\documentclass[12pt]{iopart}

\usepackage{hyperref} 
\usepackage{graphicx}
\usepackage{color}
\usepackage{dcolumn}
\usepackage{bm}
\usepackage{verbatim} 
\usepackage{cite} 
\usepackage[normalem]{ulem}
\begin{document}


    \long\def \cblu#1{\color{blue}#1}
    \long\def \cred#1{\color{red}#1}
    \long\def \cgre#1{\color{green}#1}
    \long\def \cpur#1{\color{purple}#1}

\newcommand{\eric}[1]{{\color{blue}#1}}
\newcommand{\guido}[1]{{\color{violet}#1}}
\newcommand{\matthias}[1]{{\color{blue}#1}}
\newcommand{\fabian}[1]{{\color{blue}#1}}
\newcommand{\di}[1]{{\color{blue}#1}}
\newcommand{\ericC}[1]{{\color{red}\textit{\textbf{Eric:} #1}}}
\newcommand{\guidoC}[1]{{\color{red}\textit{\textbf{Guido:} #1}}}
\newcommand{\matthiasC}[1]{{\color{red}\textit{\textbf{Matthias:} #1}}}
\newcommand{\fabianC}[1]{{\color{red}\textit{\textbf{Fabian:} #1}}}
\newcommand{\diC}[1]{{\color{red}\textit{\textbf{Di:} #1}}}

\def\FileRef{
\input FName
{
\newcount\hours
\newcount\minutes
\newcount\min
\hours=\time
\divide\hours by 60
\min=\hours
\multiply\min by 60
\minutes=\time
\
\advance\minutes by -\min
{\small\rm\em\the\month/\the\day/\the\year\ \the\hours:\the\minutes
\hskip0.125in{\tt\FName}
}
}}

\mathchardef\muchg="321D
\let\na=\nabla
\let\pa=\partial

\let\muchg=\gg

\let\t=\tilde
\let\ga=\alpha
\let\gb=\beta
\let\gc=\chi
\let\gd=\delta
\let\gD=\Delta
\let\ge=\epsilon
\let\gf=\varphi
\let\gg=\gamma
\let\gh=\eta
\let\gj=\phi
\let\gF=\Phi
\let\gk=\kappa
\let\gl=\lambda
\let\gL=\Lambda
\let\gm=\mu
\let\gn=\nu
\let\gp=\pi
\let\gq=\theta
\let\gr=\rho
\let\gs=\sigma
\let\gt=\tau
\let\gw=\omega
\let\gW=\Omega
\let\gx=\xi
\let\gy=\psi
\let\gY=\Psi
\let\gz=\zeta

\let\lbq=\label
\let\rfq=\ref
\let\na=\nabla
\def\daI{{\dot{I}}}
\def\dsq{{\dot{q}}}
\def\dgj{{\dot{\phi}}}

\def\bgs{\bar{\sigma}}
\def\bgh{\bar{\eta}}
\def\bgg{\bar{\gamma}}
\def\bgF{\bar{\Phi}}
\def\bgY{\bar{\Psi}}

\def\baF{\bar{F}}
\def\bsj{\bar{j}}
\def\baJ{\bar{J}}
\def\bsp{\bar{p}}
\def\baP{\bar{P}}
\def\bsx{\bar{x}}

\def\hgj{\hat{\phi}}
\def\hgq{\hat{\theta}}

\def\HaT{\hat{T}}
\def\HaR{\hat{R}}
\def\Hsb{\hat{b}}
\def\Hsh{\hat{h}}
\def\Hsz{\hat{z}}

\let\gG=\Gamma
\def\taA{{\tilde{A}}}
\def\taB{{\tilde{B}}}
\def\taG{{\tilde{G}}}
\def\tsp{{\tilde{p}}}
\def\tsv{{\tilde{v}}}
\def\tgF{{\tilde{\Phi}}}

\def\wgx{{\bm{\xi}}}
\def\wgz{{\bm{\zeta}}}

\def\wse{{\bf e}}
\def\wsk{{\bf k}}
\def\wsi{{\bf i}}
\def\wsj{{\bf j}}
\def\wsl{{\bf l}}
\def\wsn{{\bf n}}
\def\wsr{{\bf r}}
\def\wsu{{\bf u}}
\def\wsv{{\bf v}}
\def\wsx{{\bf x}}

\def\vaB{\vec{B}}
\def\vse{\vec{e}}
\def\vsh{\vec{h}}
\def\vsl{\vec{l}}
\def\vsv{\vec{v}}
\def\vgn{\vec{\nu}}
\def\vgk{\vec{\kappa}}
\def\vgt{\vec{\gt}}
\def\vgx{\vec{\xi}}
\def\vgz{\vec{\zeta}}

\def\waA{{\bf A}}
\def\waB{{\bf B}}
\def\waD{{\bf D}}
\def\waE{{\bf E}}
\def\waJ{{\bf J}}
\def\waV{{\bf V}}
\def\waX{{\bf X}}

\def\R#1#2{\frac{#1}{#2}}
\def\btbl{\begin{tabular}}
\def\etbl{\end{tabular}}
\def\bqbl{\begin{eqnarray}}
\def\eqbl{\end{eqnarray}}
\def\ebox#1{
  \begin{eqnarray}
    #1
\end{eqnarray}}


\def \cred#1{{\color{red}(#1)}}
\def \cblu#1{{\color{blue}#1}}

\title[MHD response and radiation asymmetry during the TQ after impurity SPI]{Radiation asymmetry and MHD destabilization during the thermal quench after impurity Shattered Pellet Injection}
\author{D. Hu$^{1,2}$, E. Nardon$^3$, M. Hoelzl$^4$, F. Wieschollek$^4$, M. Lehnen$^2$, G.T.A. Huijsmans$^{3,5}$, D. C. van Vugt$^5$\footnote{While visiting at ITER Organization, France}, S.-H. Kim$^2$, JET contributors$^6$\footnote{See the author list of \href{https://doi.org/10.1088/1741-4326/ab2276}{E. Joffrin et al. 2019 Nucl Fusion {\bf 59} 112021}.} \& JOREK Team\footnote{See https://www.jorek.eu for a list of current team members}}
 \ead{hudi2@buaa.edu.cn}
\address{
$^1$Beihang University, No. 37 Xueyuan Road, Haidian District, 100191 Beijing, China.
}
\address{
$^2$ITER Organization, Route de Vinon sur Verdon, CS 90 046,13067 Saint Paul-lez-Durance, Cedex, France.
}
\address{
$^3$CEA, IRFM, F-13108 Saint-Paul-Lez-Durance, France
}
\address{
$^4$Max Planck Institute for Plasma Physics, Boltzmannstr. 2, 85748 Garching b. M., Germany
}
\address{
$^5$Eindhoven University of Technology, De Rondom 70 5612 AP Eindhoven, the Netherlands.
}
\address{
$^6$EUROfusion Consortium, JET, Culham Science Centre, Abingdon, OX14 3DB, UK.
}
\ead{hudi2@buaa.edu.cn}

\vspace{10pt}
\begin{indented}
\item[]\today
\end{indented}

\begin{abstract}
The radiation response and the MHD destabilization during the thermal quench after a mixed species Shattered Pellet Injection (SPI) with impurity species neon and argon are investigated via 3D non-linear MHD simulation using the JOREK code. Both the $n=0$ global current profile contraction and the local helical cooling at each rational surface caused by the pellet fragments are found to be responsible for MHD destabilization after the injection. Significant current driven mode growth is observed as the fragments cross low order rational surfaces, resulting in rapidly inward propagating stochastic magnetic field, ultimately causing the core temperature collapse. The Thermal Quench (TQ) is triggered as the fragments arrive on the $q=1$ or $q=2$ surface depending on the exact $q$ profile and thus mode structure. When injecting from a single toroidal location, strong radiation asymmetry is found before and during the TQ as a result of the unrelaxed impurity density profile along the field line and asymmetric outward heat flux. Such asymmetry gradually relaxes over the course of the TQ, and is entirely eliminated by the end of it. Simulation results indicate that the aforementioned asymmetric radiation behavior could be significantly mitigated by injection from toroidally opposite locations, provided that the time delay between the two injectors is shorter than $1ms$. It is also found that the MHD response are sensitive to the relative timing and injection configuration in these multiple injection cases.
\end{abstract}

%
%
%
\maketitle
%
%

\section{Introduction}
\label{s:Introduction}

The Disruption Mitigation System (DMS) is instrumental for the sustainable operation of future high performance tokamaks such as ITER, without which the disruptive damage to the device would be intolerable \cite{Lehnen2015JNM}. Currently, the main candidates of the DMS are all based on the concept of injecting massive amounts of materials, either hydrogen isotopes, impurities or a mixture of both \cite{Hollmann2015POP,Baylor2015FST}. Among these candidates, apart from the newly developed Shell Pellet Injection scheme \cite{Hollmann2019PRL}, the Massive Gas Injection (MGI) and Shattered Pellet Injection (SPI) are presently two mainstream designs to achieve such a massive injection \cite{Lehnen2015JNM,Baylor2019NF}. Of the two, SPI shows the advantage of better injection penetration and more efficient assimilation during Thermal Quench (TQ) mitigation both in experimental \cite{Commaux2010NF,Commaux2011NF,Commaux2016NF} and numerical investigations \cite{Di2018NF}, while both schemes perform comparably for the dissipation of runaway electron beams \cite{Shiraki2018NF}. Due to its advantages, the SPI scheme is the reference concept for the ITER DMS as for now \cite{Lehnen2015JNM}.

Both, experimental and numerical investigations have been carried out regarding the performance of SPI. As the first device to implement a SPI system, DIII-D has published extensive results on the SPI efficiency for both the TQ and the Current Quench (CQ) mitigation \cite{Commaux2016NF,Shiraki2016POP,Shiraki2018NF,Raman2020NF,Sweeney2020IAEA,Sweeney2020NF}, while several other devices around the world, including JET \cite{Baylor2019NF,Gerasimov2020IAEA,Sweeney2020IAEA}, K-STAR \cite{Park2020FED}, HL-2A/2M \cite{Xu2018FST}, J-TEXT \cite{Li2018RSI} and others, have been working intensively on the SPI scheme. On the other hand, 3D non-linear simulations has been used to look into the penetration, assimilation and radiation for both deuterium and impurity SPIs \cite{Di2018NF,Kim2019POP,Hoelzl2020POP,Nardon2020NF} or massive material injections in general \cite{Izzo2013POP,Lyons2019PPCF,Ferraro2019NF}, providing insights into the MHD destabilization, transport and impurity radiation dynamics after the injection.

Despite the aforementioned results, recent numerical investigations only provide limited understanding on the evolution and mitigation of transient radiation asymmetry, both in the poloidal and the toroidal directions, during the injection. Such asymmetry could be detrimental in achieving sufficient TQ mitigation efficiency, as the injected impurities are meant to uniformly radiate away the stored pre-TQ thermal energy so as to avoid localized heat loads onto the Plasma Facing Components (PFC) which could cause localised surface melting of the first wall during mitigated disruptions in ITER \cite{Lehnen2015JNM}. Recent reports from DIII-D have shown clear evidence of poloidal radiation peaking \cite{Sweeney2020NF}, and their mitigation by dual injection \cite{Herfindal2019NF}. To extrapolate such behavior to future high performance tokamaks such as ITER, however, requires more explicit understanding of both the evolution of the impurity spatial distribution and the MHD response which is responsible for the thermal confinement destruction. The complicated interplay between the cooling induced by impurity radiation and the MHD modes necessitates which motivates the present study.
As a clarification, the total radiated energy during the CQ is generally larger than that during the TQ simply due to the larger energy reservoir in the former. However, the radiation asymmetry is expected to be much more relaxed during the CQ compared with that during the TQ as supported by our simulation results in Section \ref{ss:AsymmDualSPI}, hence the CQ is less concerning in terms of localized radiative heat flux onto the PFCs.
Due to this, we would only concern the TQ in this study.

The rest of the paper arranged as follows. In Section \ref{s:System}, our system of interest is shown, and the basic assumptions as well as their impacts are discussed. In Section \ref{s:MHD}, the MHD activity as a result of full impurity SPI is shown using a JET-like L-mode target plasma, and, after comparison with previous pure deuterium SPI results \cite{Di2018NF}, the relationship between the cooling strength and the dominant MHD destabilization mechanisms is discussed. In Section \ref{s:RadAsyMono}, the evolution of the radiation asymmetry for single injection location SPI (mono-SPI) throughout the TQ will be studied by ITER L-mode simulation, and the mitigation of such asymmetry by injecting from opposite toroidal locations (dual-SPI) will be shown in Section \ref{s:RadAsyDual}. Finally, discussion and conclusion regarding the MHD and the radiation behavior, along with the implications for future SPI operation will be presented in Section \ref{s:Conclusion}.

\section{The basic assumptions and the system of interest}
\label{s:System}

In this section, we introduce our assumptions and governing equations as well as the target equilibria for both the JET-like and the ITER SPI scenario. We use the non-linear 3D reduced MHD version of the JOREK code \cite{Huysmans2007NF,Czarny2008JCP,Hoelzl2020NF} to carry out the simulations. We introduce the governing equations and assumptions in Section \ref{ss:Equations} and Section \ref{ss:Assumptions}, and the target equilibria in Section \ref{ss:Equilibria}.

\subsection{The governing equations}
\label{ss:Equations}

We use the reduced MHD equations with impurity radiation included to consider the system evolution. We consider an ideal wall in this study, but since we are mostly concerned with the strongly driven core modes, we don't expect this would qualitatively change the dynamic of the TQ after SPIs. We assume strong charge exchange and inter-species friction so that all species and all impurity charge states share a common velocity field.
Additionally, we consider two different models for the temperatures in the simulations. The one temperature model assumes the immediate thermalization of electrons and ions such that Te = Ti. The two-temperature model assumes both species thermalize on the collisional timescale. Both models assume all ion species share the same ion temperature.
This separation of electron and ion temperature has been shown to have remarkable impact on the electron temperature evolution \cite{Ferraro2019NF}. We will further compare the impact to the ablation and radiation of the two treatments in 3D simulation here.

In the tokamak coordinates $\left(R,Z,\gj\right)$, the magnetic field and velocity field are taken to have the following form:
\bqbl
\lbq{eq:MagField}
\waB
=
F_0\na\gj
+\na\gy\times\na\gj
,\eqbl
\bqbl
\lbq{eq:VelField}
\wsv
=
v_\|\waB
-R^2\na u\times\na \gj
.\eqbl
Here, $F_0/R$ is the toroidal magnetic field and $F_0$ is taken to be constant in time and space in our study, while $\gy$ is the poloidal magnetic flux. Further, $u$ is the flow potential for the $\waE\times\waB$ flow, $v_\|$ is the parallel velocity scaled by the absolute value of the magnetic field. This ansatz-based reduced MHD model used here is not making any assumptions on the geometry, is energy conserving~\cite{Franck2015M2AN} and has been benchmarked intensely against full MHD~\cite{Pamela2020POP,Krebs2020POP}.

\begin{figure*}
\centering
\noindent
\btbl{cc}
\parbox{2.5in}{
    \includegraphics[scale=0.40]{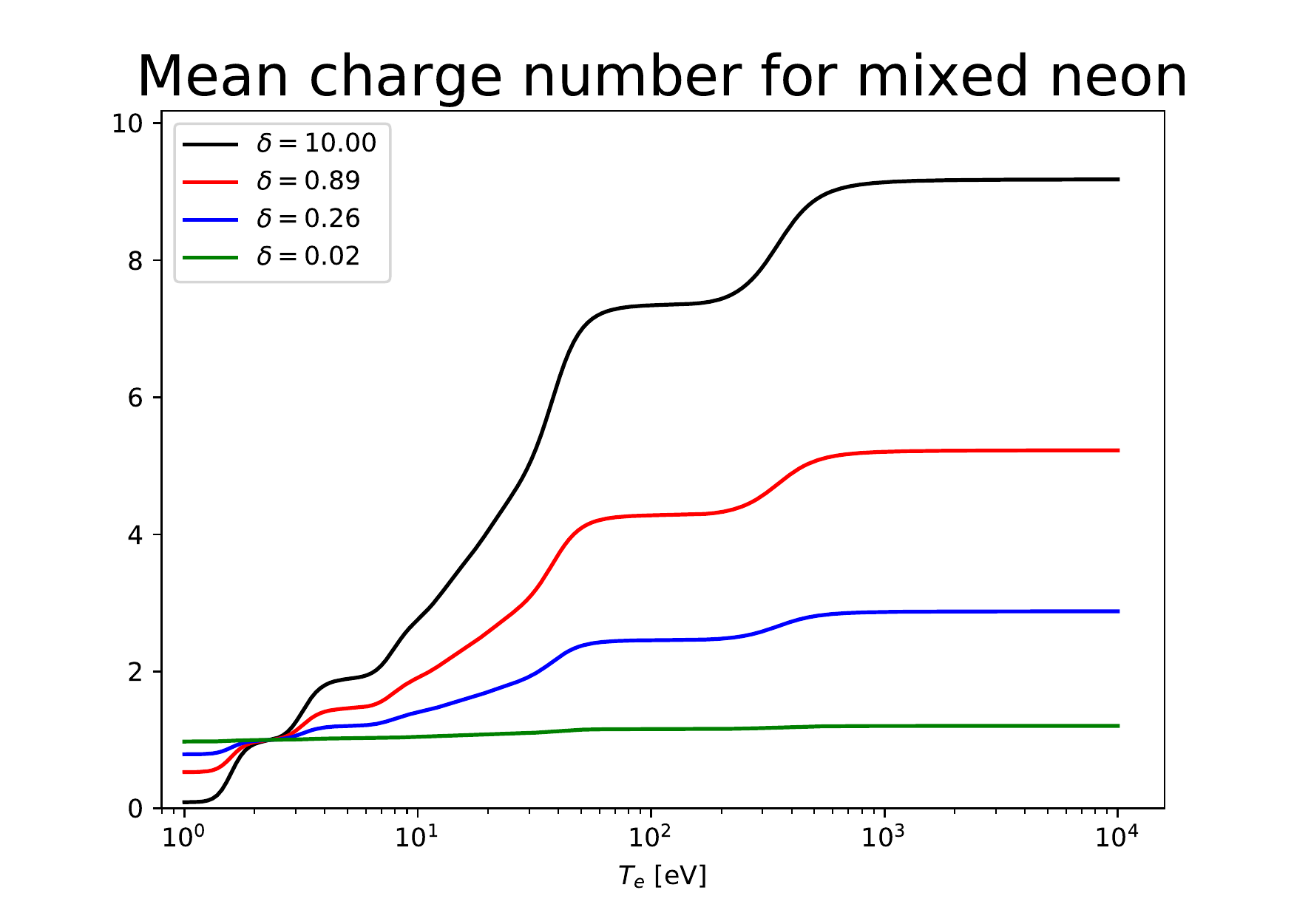}
}
&
\parbox{2.5in}{
	\includegraphics[scale=0.40]{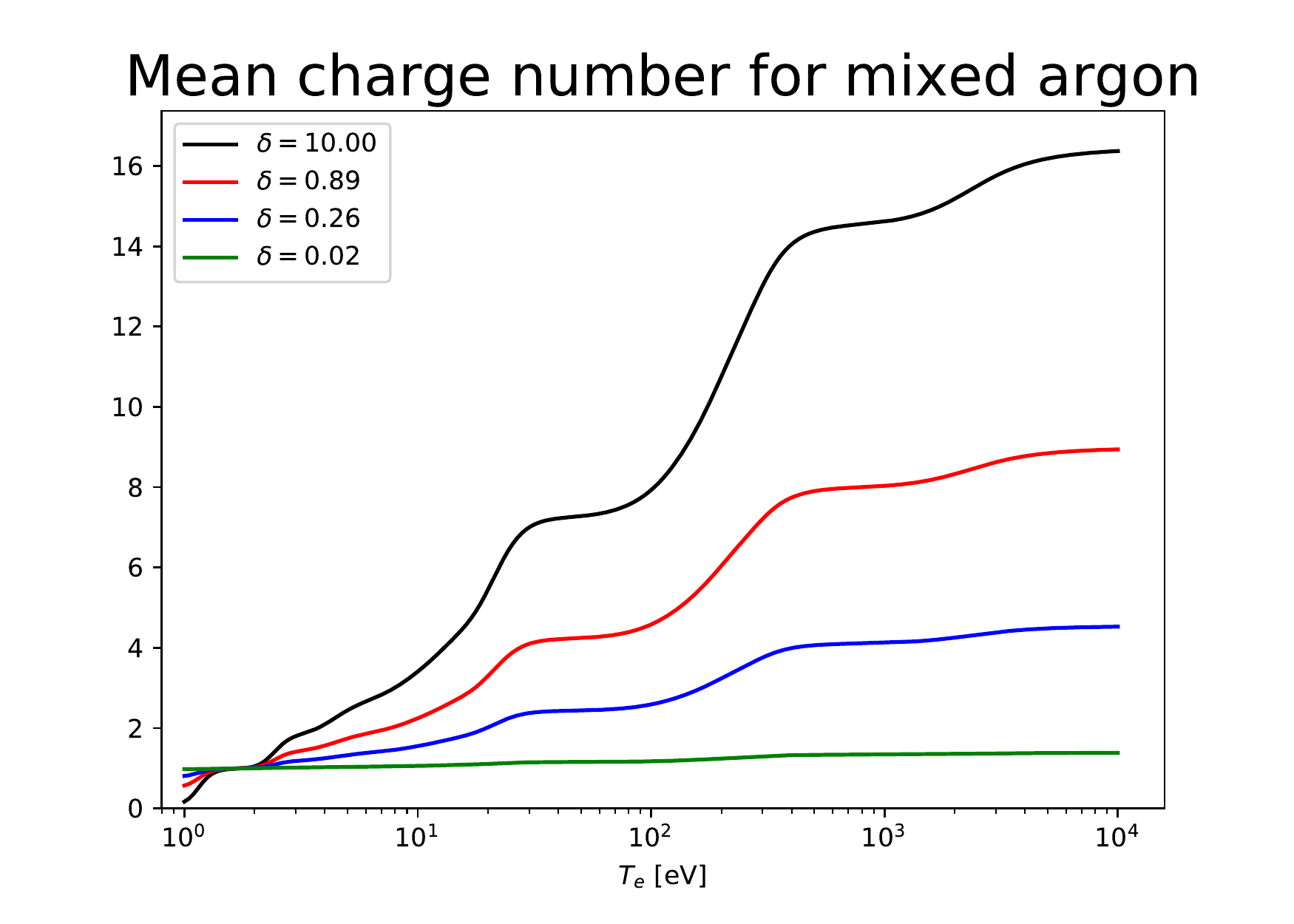}
}
\\
(a)&(b)
\\
\parbox{2.5in}{
    \includegraphics[scale=0.40]{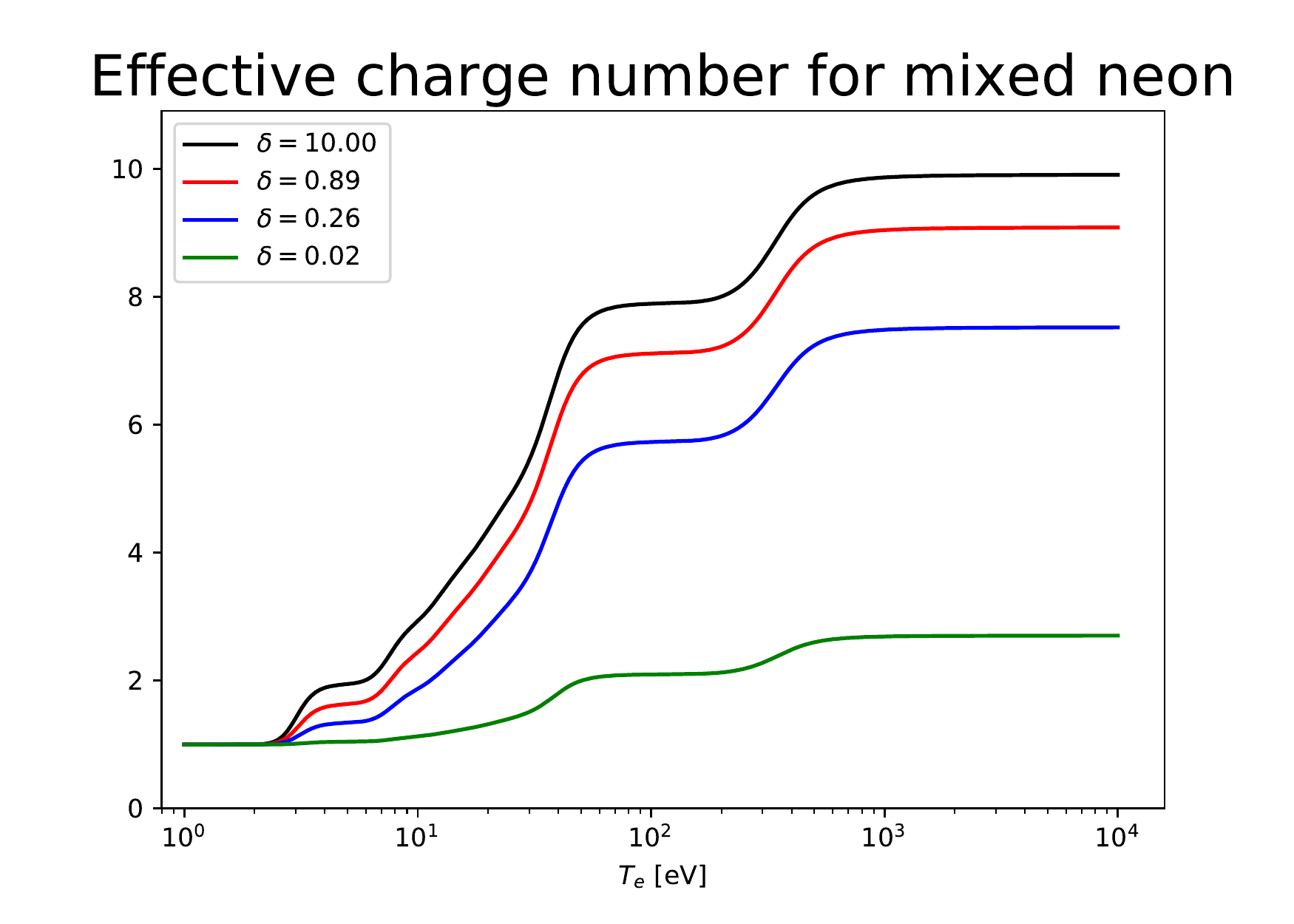}
}
&
\parbox{2.5in}{
	\includegraphics[scale=0.40]{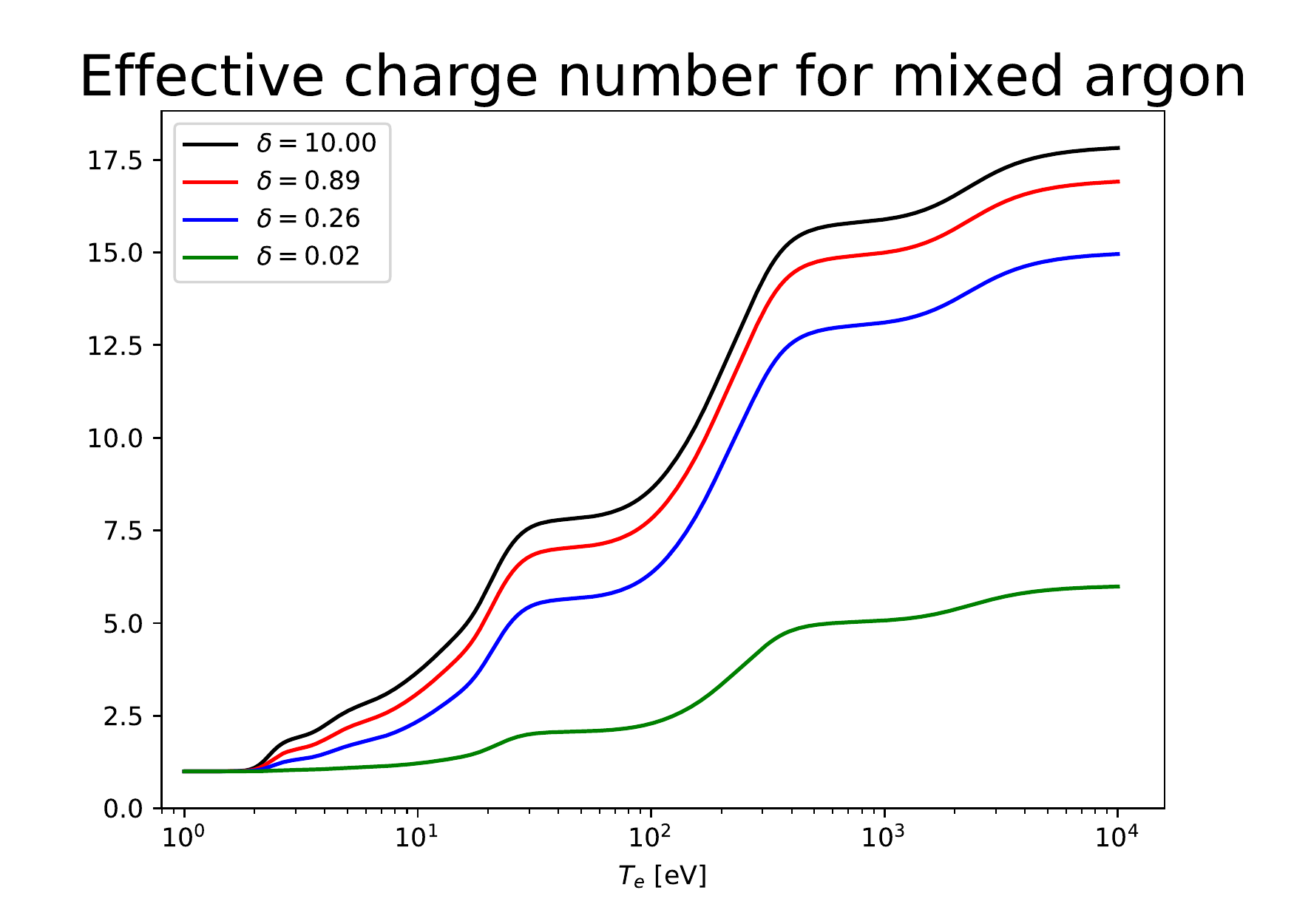}
}
\\
(c)&(d)
\etbl
\caption{The mean charge number and effective charge number for mixed hydrogen isotopes and impurity plasmas under the CE assumption, with $\gd$ defined as the number density ratio between the impurity and the background species.}
\label{fig:01}
\end{figure*}

The governing equations are then as follows. First, we have the induction equation:
\bqbl
\lbq{eq:InductionEq}
\R{\pa\gy}{\pa t}
&
=
&
\gh\left(T_e\right)\gD^*\gy
-R\left\{u,\gy\right\}
-F_0\R{\pa u}{\pa\gj}
,\eqbl
\bqbl
\lbq{eq:AmpereEq}
j
&
=
&
\gD^*\gy
,\quad
j_\gj
=
-j/R
.\eqbl
The Poisson bracket is defined as $\left\{f,g\right\}\equiv R\left(\na f\times\na g\right)\cdot\na\gj$.
Here we Spitzer resistivity considering the effective charge contribution as well as the passing ratio of the electrons \cite{Hirshman1978POF}:
\bqbl
\gh
=
\R{1}{R_{pass}}\R{\sqrt{2m_e}\,Z_\textrm{eff}\,e^2\,\ln{\gL}}{12\gp^{3/2}\,\ge_0^2\,\max{\left(T_e,T_{thres}\right)}^{3/2}}\times
\R{1+1.198\,Z_\textrm{eff}+0.222\,Z_\textrm{eff}^2}{1+2.966\,Z_\textrm{eff}+0.753\,Z_\textrm{eff}^2}
\nonumber
.\eqbl
Here we take the passing ratio $R_{pass}\sim 0.5$ for simplicity. The effective charge is
\bqbl
Z_\textrm{eff}
\equiv
\R{\sum_i{n_iZ_i^2}}{\sum_i{n_iZ_i}}
\nonumber
,\eqbl
with $n_i$ and $Z_i$ the number density and charge number of each ion species (counting each charge state separately) respectively. Note that here we use the Coronal Equilibrium (CE) assumption. This corresponds to an initial core resistivity on the order of $\gh\,[\gW\cdot m]\simeq 5.6\times 10^{-8}\left(T_{e0}\,[keV]\right)^{-3/2}$. To ensure resolution of the resistive skin current, for the ITER case in this study we introduce a threshold at $T_{thres}=2keV$, beyond which the resistivity does not decrease further. The impact of this artificial threshold on the overall post-injection dynamics is limited since the resistive time in the hot core region is much longer than our timescale of interest, and our focus is on the cooling induced MHD instability in this study. The dependence on the effective charge makes even a tiny amount of impurity species affect the resistivity remarkably, as is shown in Fig.\,\ref{fig:01} for the CE case where $\gd$ is the number density ratio between the impurity and hydrogen isotopes (the ``background species''). It can be seen that the resistivity change is $\mathcal{O}\left(1\right)$ even for $\gd\sim\mathcal{O}\left(10^{-2}\right)$.
In Eq.\,{\rfq{eq:InductionEq}}, we also have the operator $\gD^*\equiv R^2\na\cdot\left(R^{-2}\na\right)$. Eq.\,(\rfq{eq:AmpereEq}) is a consequence of Amp\`{e}re's law with the permeability absorbed into the current density.

Second, the continuity equation for the total plasma mass density is
\bqbl
\lbq{eq:ContinuityEq}
\R{\pa \gr}{\pa t}
&
=
&
-\na\cdot\left(\gr\wsv\right)
+\na\cdot\left(D\na\gr\right)
+S_{bg}
+S_{imp}
,\eqbl
and the mass density continuity equation for the whole impurity species is
\bqbl
\lbq{eq:ImpEq}
\R{\pa \gr_{imp}}{\pa t}
=
-\na\cdot\left(\gr_{imp}\wsv\right)
+\na\cdot\left(D\na\gr_{imp}\right)
+S_{imp}
.\eqbl
Here $S_{bg}$ and $S_{imp}$ are the density sources for the background species and impurity respectively. The diffusion coefficient $D$ consists of $D_\bot$ and $D_\|$ which are the parallel and perpendicular diffusion coefficients respectively, but in this study we choose $D_\|$ to be small since the parallel density relaxation is expected to be strongly dominated by the convective transport.

Then we have the perpendicular and parallel momentum equations:
\bqbl
\lbq{eq:VorticityEq}
R\na\cdot\left[R^2\R{\pa}{\pa t}\left(\gr\na_{pol}u\right)\right]
&
=
&
\R{1}{2}\left\{R^2\left|\na_{pol}u\right|^2,R^2\gr\right\}
+\left\{R^4\gr\gw,u\right\}
\nonumber
\\
&&
-R\na\cdot\left[R^2\na_{pol}u\na\cdot\left(\gr\wsv\right)\right]
+\left\{\gy,j\right\}
\nonumber
\\
&
&
-\R{F_0}{R}\R{\pa j}{\pa\gj}
+\left\{P,R^2\right\}
+R\gm_\bot\left(T_e\right)\na_{pol}^2\gw
,\eqbl
\bqbl
\lbq{eq:VorticityDef}
\gw
&
=
&
\R{1}{R}\R{\pa}{\pa R}\left(R\R{\pa u}{\pa R}\right)
+\R{\pa^2 u}{\pa Z^2}
,\eqbl
\bqbl
\lbq{eq:ParallelEq}
B^2\R{\pa}{\pa t}\left(\gr v_\|\right)
&
=
&
-\R{1}{2}\gr\R{F_0}{R^2}\R{\pa}{\pa \gj}\left(v_\|B\right)^2
-\R{\gr}{2R}\left\{B^2v_\|^2,\gy\right\}
-\R{F_0}{R^2}\R{\pa P}{\pa \gj}
\nonumber
\\
&
&
+\R{1}{R}\left\{\gy,P\right\}
-B^2\na\cdot\left(\gr\wsv\right)v_\|
+B^2\gm_\|\na_{pol}^2v_\|
.\eqbl
The vorticity equation Eq.\,(\rfq{eq:VorticityEq}) is acquired by applying $\na\gj\cdot\na\times\left(R^2\cdots\right)$ on both sides of the momentum equation. We have $\gm_\bot\propto T_e^{-3/2}$ is the perpendicular viscosity. We choose this scaling law for the perpendicular viscosity simply to keep the magnetic Prandtl number approximately fixed in space and time in these simulations. $P$ is the total pressure from all species. Further, Eq.\,(\rfq{eq:VorticityDef}) is the vorticity definition. The parallel momentum equation Eq.\,(\rfq{eq:ParallelEq}) is obtained by taking the dot product with $\waB$ on both sides of the momentum equation, and $\gm_\|$ is the parallel viscosity which we take to be a constant.

Lastly, we arrive at the pressure equation. For the single temperature treatment, we have:
\bqbl
\lbq{eq:PressureEq}
\R{\pa P}{\pa t}
&
=
&
-\wsv\cdot\na P
-\gg P\na\cdot\wsv
+\R{\gg-1}{R^2}\gh\left(T_e\right)j^2
+\na\cdot\left(\gk_\bot\na_\bot T
+\gk_\|\na_\| T\right)
\nonumber
\\
&
&
+\left(\gg-1\right)\gm_\|\left[\na_{pol}\left(v_\|B\right)\right]^2
-n_e n_{imp}L_{rad}\left(T_e\right)
\nonumber
\\
&&
+\R{\gg-1}{2}\wsv\cdot\wsv\left(S_D+S_{imp}\right)
-\pa_t E_{ion}
-\na\cdot\gG_{ion}
.\eqbl
Here the total pressure is defined as $P\equiv n_e T_e + \left(n_{bg}+n_{imp}\right)T_i$ with $n_e$, $n_{bg}$ and $n_{imp}$ representing the number density of electrons, background ions and impurity ions respectively. The Braginskii parallel thermal conduction $\gk_\|\propto T_e^{5/2}$ is used \cite{Braginskii1965RPP}. We have neglected the viscous heating by perpendicular motion due to its smallness. Also, $L_{rad}$ represents the radiation power function acquired via the CE assumption \cite{Summers1979JPB} which includes contribution from line radiation, recombination radiation and bremsstrahlung radiation. The third last term on the right hand side (RHS) corresponds to a frictional heating as the newly ablated particles are accelerated to the background plasma velocity. We choose to represent the ionization energy as a potential energy here, as can be seen from the last two terms on the RHS. The reason and impact of such choice is discussed in Section \ref{ss:AssumptionsDiscussion}. Similarly, for the two temperature treatment, we have
\bqbl
\lbq{eq:i_TemperatureEq502}
\R{\pa}{\pa t}P_i
&
=
&
-\wsv\cdot \na P_i
-\gg P_i\na\cdot\wsv
+\na\cdot\left(\gk_\bot\na_\bot T_i
+\gk_{i,\|}\na_\| T_i\right)
\nonumber
\\
&&
+\R{\gg-1}{2}\wsv\cdot\wsv\left(S_D+S_{imp}\right)
+\left(\gg-1\right)\gm_\|\left[\na_{pol}\left(v_\|B\right)\right]^2
\nonumber
\\
&&
+\left(n_{bg}+n_{imp}\right)\left(\pa_t T_i\right)_{c,e}
,\eqbl
\bqbl
\lbq{eq:e_TemperatureEq502}
\R{\pa}{\pa t}P_e
&
=
&
-\wsv\cdot \na P_e
-\gg P_e\na\cdot\wsv
+\na\cdot\left(\gk_\bot\na_\bot T_e
+\gk_{e,\|}\na_\| T_e\right)
\nonumber
\\
&&
+\R{\gg-1}{R^2}\gh\left(T_e\right)j^2
-\left(\gg-1\right)n_e n_{imp}P_{rad}\left(T_e\right)
+n_e\left(\pa_t T_e\right)_{c,i}
\nonumber
\\
&&
-\pa_t E_{ion}
-\na\cdot\gG_{ion}
.\eqbl
Here we have the electron and ion pressures $P_i\equiv \left(n_{bg}+n_{imp}\right)T_i$, $P_e\equiv n_e T_e$. The Braginskii parallel heat conduction is calculated for each species respectively. The ion-electron collisional energy exchange terms $\left(\pa_t T_i\right)_{c,e}$ and $\left(\pa_t T_e\right)_{c,i}$ are calculated considering the thermal equilibration time:
\bqbl
\left(\pa_t T_e\right)_i
=
\left(\gn_{e/imp}+\gn_{e/bg}\right)\left(T_i-T_e\right)
,\quad
\left(\pa_t T_i\right)_e
=
-\R{n_e}{n_{imp}+n_{bg}}\left(\pa_t T_e\right)_i
\nonumber
.\eqbl
Here $\gn_{e/imp}$ and $\gn_{e/bg}$ are the electron-ion collision rate for the impurity and background species respectively.

Together, Eq.\,(\rfq{eq:InductionEq}) to Eq.\,(\rfq{eq:e_TemperatureEq502}) form our governing equations. To close the equations, however, we need to specify the ablation density source and the radiation energy loss terms.

\subsection{The ablation and radiation model}
\label{ss:Assumptions}

To close the equations, we consider an ablation scaling law which resembles the ``strongly shielded'' Neutral Gas Shielding (NGS) model in a Maxwellian plasma as our ablation density source \cite{Sergeev2006PPR,Parks2020XXX,Bosviel2019EPS}. Here ``strongly shielded'' means the neutral gas cloud dissipate almost all the incoming heat flux before they reach the pellet surface due to the smallness of the material's ablation energy. The physics of such model is that, in a quasi-steady state, the ablation rate must be such as to maintain a sufficient line integrated neutral density to substantially deplete the incident heat flux along the field line, so that the actual flux arriving at the fragment surface vanishes. The exact ablation rate is subject to the pellet species as well as details in the model. For historical reasons, we used two different ablation models in our study: Sergeev's model for full impurity pellets \cite{Sergeev2006PPR} and Parks's model which is capable of dealing with truly mixed pellets \cite{Parks2020XXX}. We will specify which model has been used for each case later on.

When dealing with mixed species SPI, caution has to be paid to the exact combination of species. For example, for a neon and deuterium mixed pellet, the two species are uniformly mixed together in the pellet, and the corresponding mass ablation rate for a given molecular mixture ratio $X\equiv\R{N_{D2}}{N_{Ne}+N_{D2}}$ and pellet radius $r_p$ is then \cite{Parks2020XXX,Bosviel2019EPS}
\bqbl
\lbq{eq:NeD2AblRate}
G\left[g/s\right]
&
=
&
\gl\left(X\right)\left(\R{T_e\left[eV\right]}{2000}\right)^{5/3}
\left(\R{r_p\left[cm\right]}{0.2}\right)^{4/3}n_e\left[10^{14}cm^{-3}\right]^{1/3}
,\eqbl
while the parameter $\gl$ is a function of the mixture ratio:
\bqbl
\lbq{eq:glNeD2}
\gl\left(X\right)
&
\equiv
&
27.08
+\tan{\left(1.49X\right)}
.\eqbl
The particle number ablation rate for each species is then
\bqbl
\lbq{eq:NeAblFrac}
\R{d}{dt}N_{Ne}\left[mol/s\right]
&
=
&
\R{\left(1-X\right)G}{\left(1-X\right)W_{Ne}+XW_{D2}}
,\eqbl
\bqbl
\lbq{eq:D2AblFrac}
\R{d}{dt}N_{D2}\left[mol/s\right]
&
=
&
\R{XG}{\left(1-X\right)W_{Ne}+XW_{D2}}
,\eqbl
where the mole mass weight is
\bqbl
W_{Ne}\left[g/mol\right]
=
20.18
,\quad
W_{D2}\left[g/mol\right]
=
4.028
\nonumber
.\eqbl
Here $N_{Ne}$ and $N_{D2}$ are the ablation rate in $mol/s$ for neon atom and deuterium molecular respectively. On the other hand, in the case of argon and deuterium mixed pellets, since the two species essentially form separately due to the separation of their respective triple points, we assume that each pellet fragment consists of either pure argon or pure deuterium. Their respective mass ablation rate take the form of Eq.\,(\rfq{eq:NeD2AblRate}), but instead of Eq.\,(\rfq{eq:glNeD2}) we have \cite{Parks2020XXX}
\bqbl
\gl_{Ar}
\equiv
36.63
,\quad
\gl_{D2}
\equiv
39.00
.\eqbl
In our simulations, we generate the fragments from each species separately according to our fragment size distribution.

In both kinds of mixed pellet, we assume the fragment size $r_p$ follows the Statistical Fragmentation model \cite{ParkDistribution}:
\bqbl
\lbq{eq:ParkP}
P\left(r_p\right)
=
\R{r_pK_0\left(\gk_p r_p\right)}{I}
,\quad
I
\equiv
\int_0^\infty{r_pK_0\left(\gk_p r_p\right)dr}
=
\gk_p^{-2}
,\eqbl
where $K_0$ is the modified Bessel function of the second kind, and $\gk_p$ is the inverse of the characteristic fragment size.
In our simulation, the ablated atoms are then deposited around the fragment with a poloidally and toroidally gaussian distribution:
\bqbl
S_n
\propto
\exp{\left(-\R{\left(R-R_{f}\right)^2+\left(Z-Z_{f}\right)^2}{\gD r^{2}_{NG}}\right)}
\times\exp{\left(-\left(\R{\gj-\gj_{f}}{\gD\gj_{NG}}\right)^2\right)}
.\eqbl
In the above equation, $R_{f}$, $Z_{f}$ and $\gj_{f}$ are the spatial coordinates of a fragment. In this work, we choose the neutral cloud parameter $\gD r_{NG}=2cm$ and $\gD\gj_{NG}=1\,\textrm{rad}$, leading to a toroidally elongated shape. Such a shape is partly justified due to the fast expansion of the ablation cloud in a way that cannot be modelled by fluid equations anyway and is unavoidable in our simulations due to limited resolution in toroidal harmonics.
Thus we only meant to represent an already somewhat toroidally relaxed plasma cloud in our source term, although such treatment may result in artificially mitigated toroidal peaking in the impurity density as well as in the radiation power density.
Instead of the currently used gaussian shape, it is possible to provide a more accurate toroidally relaxed source term by considering detailed modelling of the parallel expansion of the ablation cloud \cite{Aleynikov2019JPP}, but that would be left for future works.
The pellet radius is consistently evolved according to the ablation rate and the mass conservation.

In this study, we assume the plasma is transparent to all radiations for simplicity, meaning there is no absorption of the radiated power. Although this is generally true for the recombination and the bremsstrahlung radiations, it is not strictly valid for the line radiation, especially in the close vicinity of the fragments where the impurity density is large \cite{Morozov2007PPR}. Away from the ablation cloud, however, two effect would mitigate this opacity. The first is the general relaxation of the total impurity density, the second is the change of the dominant ionization state influenced by the electron temperature which removes the resonant ions absorbing the line radiation photons. Indeed, for recent DIII-D experiments \cite{Raman2020NF}, distinctive ``radiation band'' could be seem within the plasma, suggesting the plasma as a whole is more or less transparent even if the ablation cloud itself could not be described as such. A more detailed study of the photon mean-free-path would be pursued in future works.

The last important component in our model is the use of the CE model, which requires the plasma to both be dominated by radiative recombination rather than three-body recombination, and to exist in an ionization equilibrium \cite{Mosher1974PRA}. The validity of this is not guaranteed in a massive injection scenario since the timescale of reaching ionization equilibrium is comparable with that of ablation and transport during such a scenario \cite{Smith2010APJ}. However, we are mostly concerned with the accuracy of the impurity radiation function as it dominates the behavior of the radiation power density. Hence, so long as there is no significant deviation in the radiative power loss $P_{rad}=n_en_{imp}L_{rad}\left(T_e,n_e\right)$ between the CE result and that of more detailed non-equilibrium models, we would consider the use of the CE assumption satisfactory. We now test whether or not that is indeed satisfied in the considered scenario.

\subsection{Discussion on the impact of the CE assumption}
\label{ss:AssumptionsDiscussion}

\begin{figure*}
\centering
\noindent
\btbl{cc}
\parbox{2.5in}{
    \includegraphics[scale=0.40]{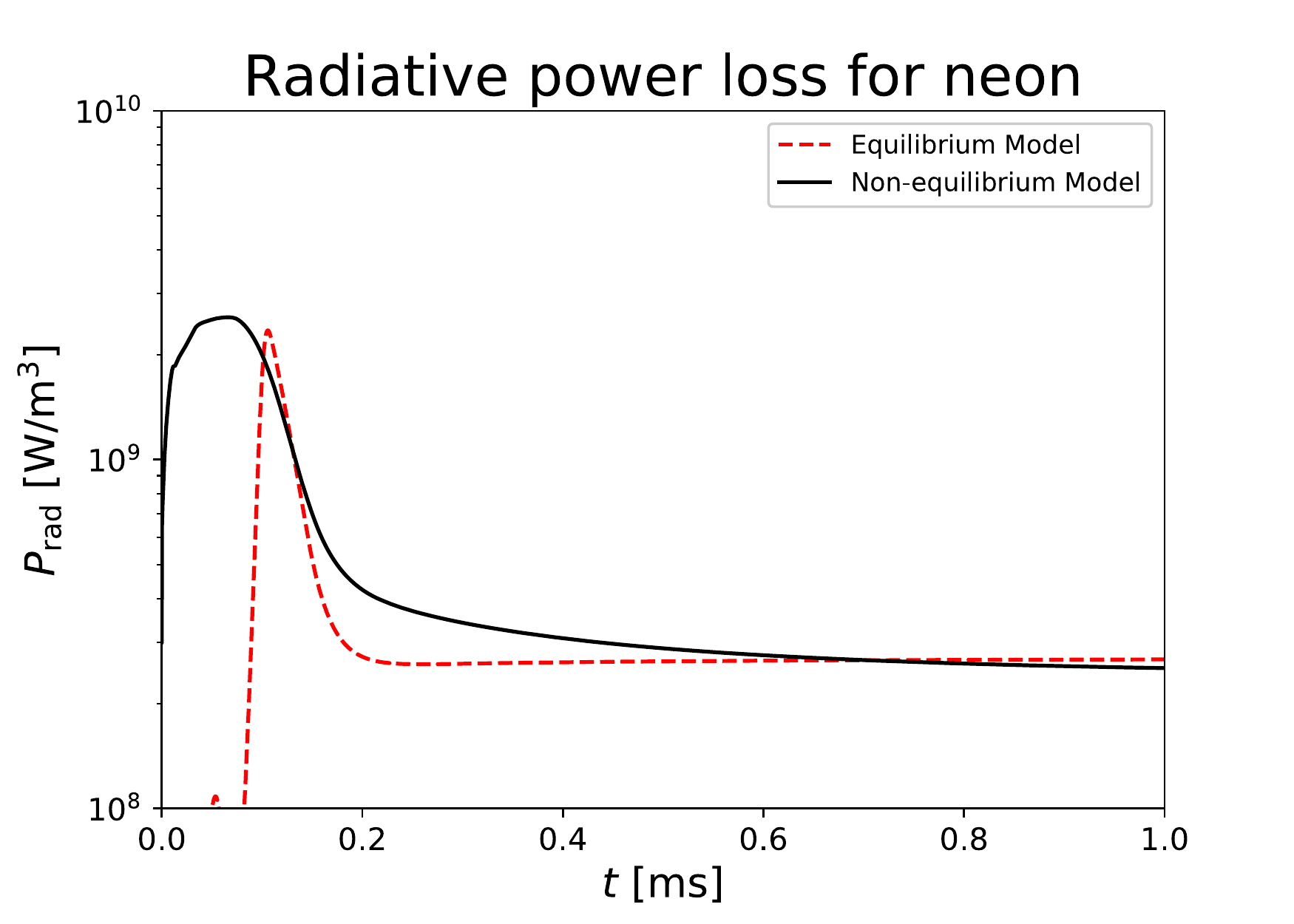}
}
&
\parbox{2.5in}{
	\includegraphics[scale=0.40]{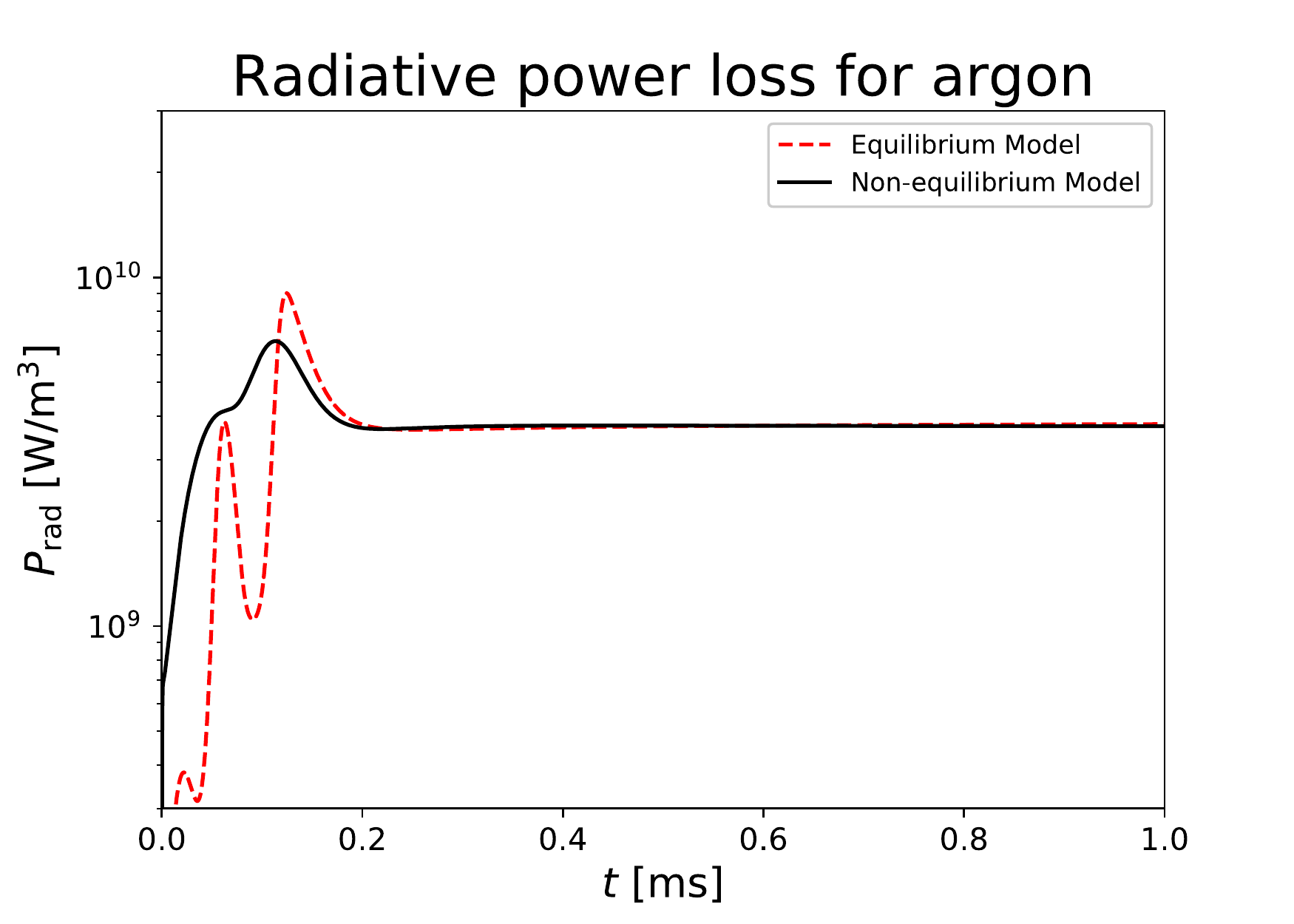}
}
\\
(a)&(b)
\etbl
\caption{The comparison of the radiative power density between the CE model and the non-equilibrium time evolving model for (a) neon and (b) argon.}
\label{fig:02}
\end{figure*}

The validity of the CE model can be tested in 0D by comparing the CE model with a model that consistently evolves the individual charge states. We consider the ambient electron temperature falls exponentially from $5keV$ to $10eV$ with cooling timescale of $20\gm s$, which is consistent with our numerical observation that the fragments take about $80\gm s$ to cool the plasma down from $5keV$ to $100eV$. For the CE case, the charge state distribution is a function of temperature and (weakly) density. For the non-equilibrium case, let the impurity charge state distribution begin with all neutral, then evolve according to the ionization and recombination probability \cite{OPENADAS}. Further, let all charge states experience an exponential number density loss to represent impurity transport, which is balanced by an equal neutral source representing pellet ablation, so that the total impurity number density is kept constant. The comparison of the resulting radiative power loss is shown in Fig.\,\ref{fig:02}. It is shown that despite significant initial discrepancy, when the temperature cools down below $\mathcal{O}\left(100eV\right)$ (corresponds to time on the order of $80 \gm s$), the difference in the radiative power loss is acceptable. On the other hand, in case the dilution cannot cool the plasma fast enough, there could be some deviation between the CE result and that of a self-evolving model. As a result, in those cases, it could be expected that the CE assumption produces a milder radiation collapse than that is actually the case.

Another question related to the CE assumption is the treatment of the ionization energy. In a realistic model, part of the electron thermal energy would be used for ionization, cooling down the plasma. Upon recombination, in the absence of three-body recombination, the ionization energy released would be radiated away as recombination radiation. Such recombination radiation is in accordance with the recombination probability, so that in a self-evolving treatment the self-consistency is ensured. However, under CE assumption, rapid temperature change or strong density source may cause artificially fast evolution of charge state distribution, thus result in artificially high recombination radiation. To avoid this, we treat the ionization energy as a potential energy (meaning it will feedback to the electron thermal energy) in our study while using the recombination radiation function at CE from the open ADAS data to model the corresponding radiation power. Such treatment ensures the energy conservation and prevents artificially large recombination radiation. A coronal non-equilibrium impurity model is presently under development for JOREK, and will be published at a later stage.

\subsection{The target equilibria and the injection configurations}
\label{ss:Equilibria}

We consider both a JET-like L-mode and an ITER L-mode plasma as our target equilibria. The JET-like equilibrium is used to show the MHD excitation during a pure argon injection, which will be compared with the pure deuterium SPI case investigated in Ref. \cite{Di2018NF}. Here, argon is chosen instead of neon for its stronger line radiation, so that the JET-like case would be used to represent the extremely strong radiation cooling limit, as opposed to the above mentioned mildly cooling limit of the deuterium case. Such a comparison will help to demonstrate the characteristic MHD destabilization mechanism, namely the axisymmetric and the helical current redistribution, for each respective limit. In general cases, the current density and MHD response would show a combination of the characteristics from these two limits. Later on, the ITER equilibrium will be used to investigate the radiation asymmetry for both mono- and dual-SPI with hydrogen/neon mixed pellets, as well as the MHD response in both cases.

\begin{figure*}
\centering
\noindent
\btbl{cc}
\parbox{2.7in}{
    \includegraphics[scale=0.27]{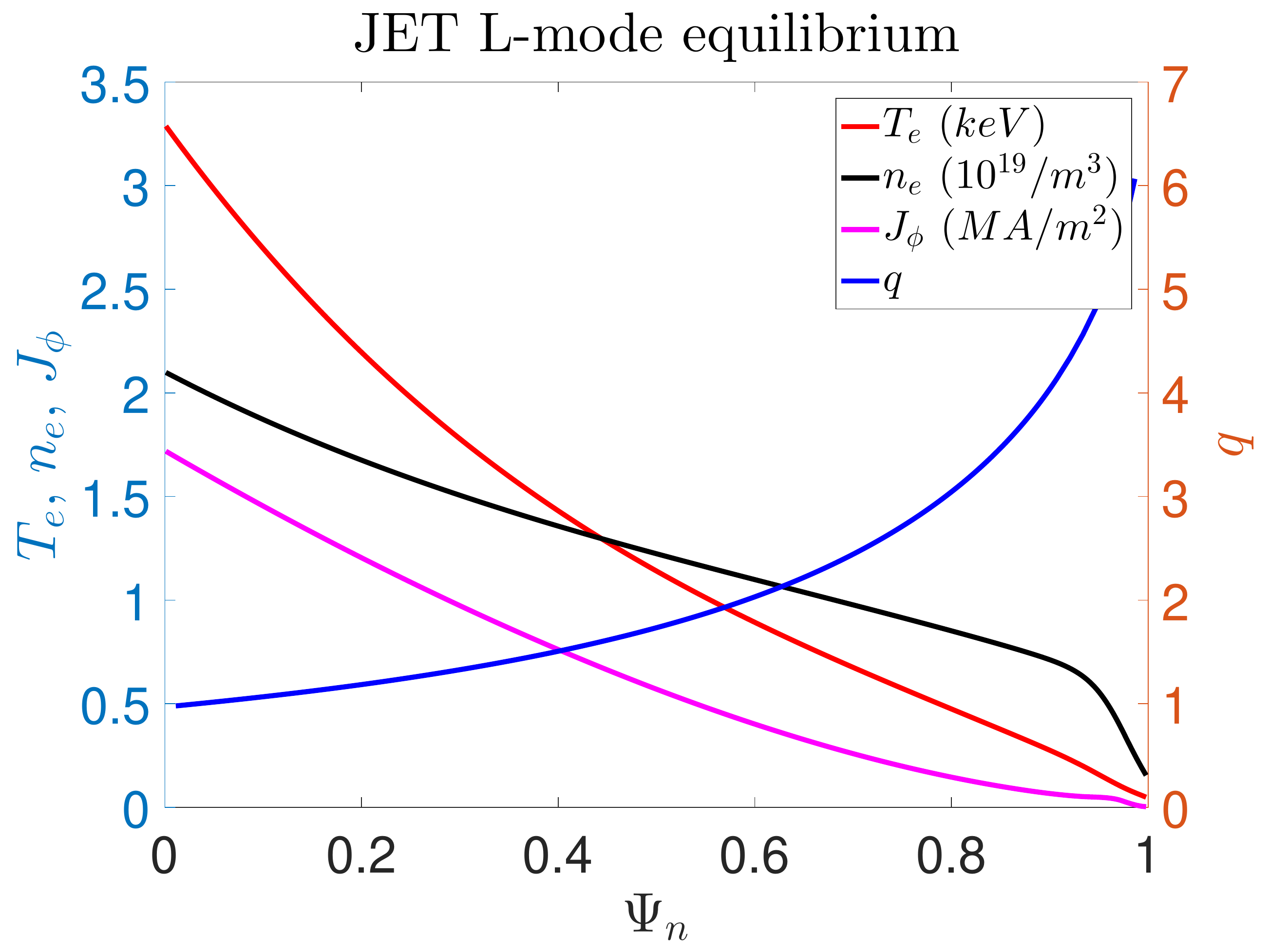}
}
&
\parbox{2.7in}{
	\includegraphics[scale=0.27]{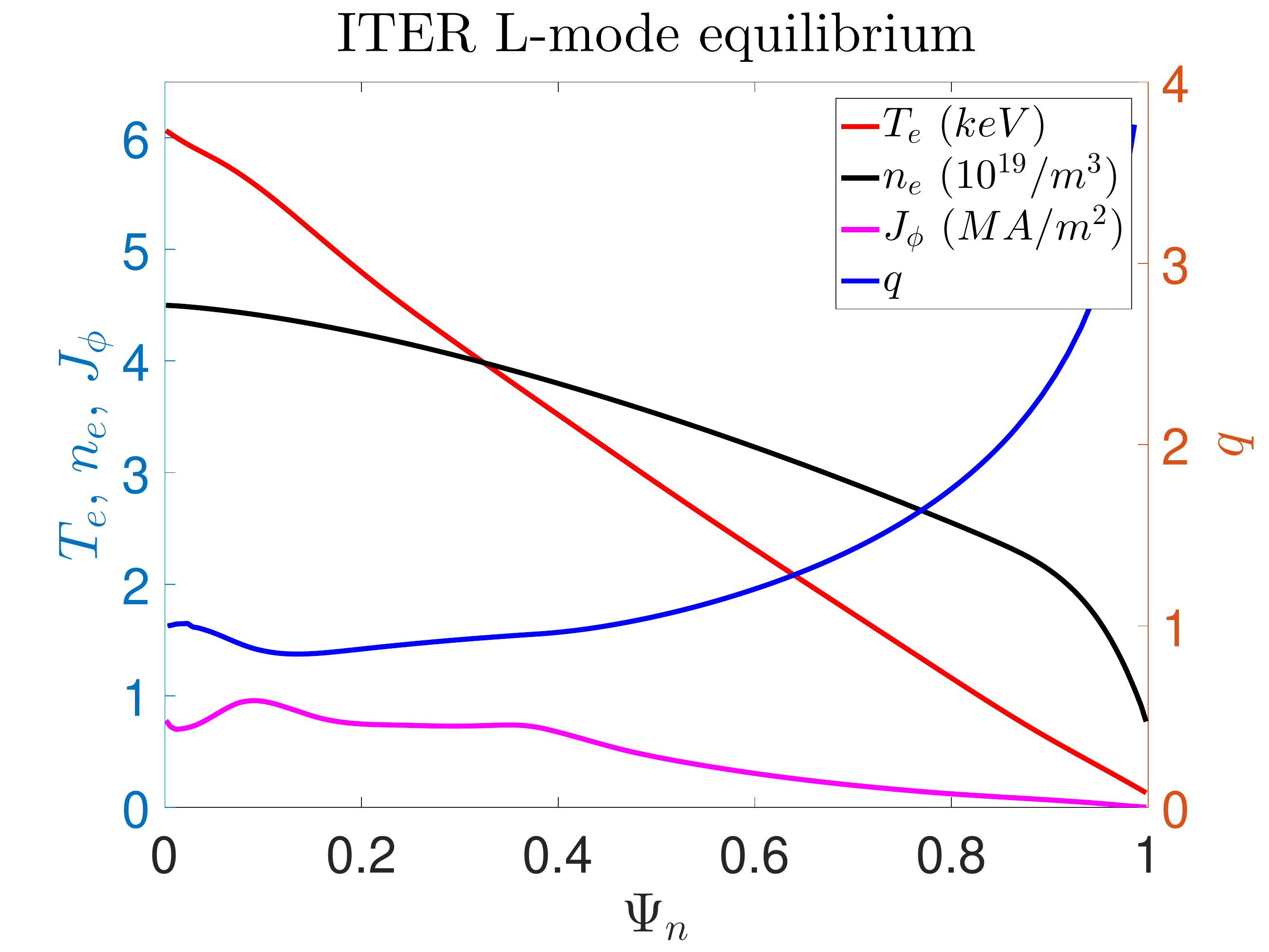}
}
\\
(a)&(b)
\etbl
\caption{The target equilibrium for (a) JET-like case and (b) ITER case.}
\label{fig:03}
\end{figure*}

The initial equilibria used for the injection studies are shown in Fig.\,\ref{fig:03}, where the safety factor $q$, electron temperature and density, as well as the current density profile are plotted as function of the normalized flux $\gY_n$. For the JET-like case, we choose the our target plasma template to be resemble that of JET pulse No. 85943 at time $62.4s$. For historical reasons, we were not able to use the exact same equilibrium of the much more recent JET SPI discharges \cite{Gerasimov2020IAEA,Sweeney2020IAEA}. However, it would be seen in Section \ref{s:MHD} that the MHD response and the radiation structure of our simulation agree well with the experimental observations. There is a very small region of the plasma within the $q=1$ surface, with axis safety factor $q_0=0.98$. The toroidal field $B_t\simeq 3 T$, and the total plasma current is $I_p\simeq 2MA$. The core electron temperature is $T_e(0)\simeq 3.28keV$ and the electron density is $n_e(0)\simeq 2.1\times 10^{19}/m^3$. We have assumed that electrons and ions share the same temperature.
As for the ITER case, we chose a hydrogen L-mode scenario, which has $B_t\simeq 2.65 T$, $I_p\simeq 7.5 MA$. The core electron temperature and density is $T_e(0)\simeq 6.05keV$ and $n_e(0)\simeq 4.5\times 10^{19}/m^3$ respectively, the thermal energy content before the injection is about $33.3MJ$. We also assumed the equipartition of electron and ion temperature initially. This ITER equilibrium has a weakly reversed shear in the core, hence although the core safety factor is slightly above unity, there is still a significant portion of the plasma that is within the $q=1$ surface. Despite this, numerical investigation found that the $1/1$ kink mode remains negligibly small within our timescale of interest. This could be due to the relatively low $\gb$ of the L-mode plasma which makes the ideal kink stable \cite{Bussac1975PRL,WhiteBook}, while the resistive kink, although always unstable, has a long growth time compared with our time of interest.

We choose the realistic JET SPI configuration for our JET-like simulation \cite{BaylorReport,JETSPImemo}, while for the ITER case we consider a simple configuration where the injectors locate at the outer mid-plane and shoot inward along the major radial direction. For the JET-like simulation, we inject from a single location into the plasma, while the spreading cone vertex angle is 20 degree. Sergeev's ablation model \cite{Sergeev2006PPR} is used. For the ITER case, we have the option of injecting pellets with various mixture ratios at one or more toroidal locations, and in each case the spread angle is 20 degree. Since the pellet is truly mixed, we use Parks's ablation model \cite{Parks2020XXX} for the ITER case. In both scenarios, the fragment size distribution is set according to Eq.\,(\rfq{eq:ParkP}), and the fragment velocity distribution is flat within their respective range. In Table \ref{tab:1}, the set of injection configurations we will be using is shown including the configuration notation, the total impurity and hydrogen isotope injection amount, the fragment number, the velocity and the injection symmetry. In the injection symmetry, ``mono-SPI'' means we inject from a single location, ``symmetric dual-SPI'' means we inject from toroidally opposite locations and both injections are exactly identical except for the location. The setup for ``asymmetric dual-SPI'' is the same as the symmetric one apart from the fact that there is a time delay between the two injections.

\begin{table*}
\centering
\noindent
\btbl{|c|c|c|c|c|c|}
\hline
Notation & Impurity & H Isotope & Frag. & Velocity & Symmetry\\
\hline
JET shot 1 & $1.5\times 10^{22}$ Ar & - & $100$ & $200\pm 50m/s$ & Mono-SPI\\
\hline
ITER shot 1 & $4\times 10^{21}$ Ne & $3.6\times 10^{22}$ H & $400$ & $150\pm 50m/s$ & Mono-SPI\\
\hline
ITER shot 2 & $4\times 10^{21}$ Ne & - & $100$ & $300\pm 100m/s$ & Mono-SPI\\
\hline
ITER shot 3 & $2.6\times 10^{22}$ Ne & $2.1\times 10^{24}$ H & $1000$ & $150\pm 50m/s$ & Mono-SPI\\
\hline
ITER shot 4 & $2.6\times 10^{22}$ Ne & $2.1\times 10^{24}$ H & $1000$ & $150\pm 50m/s$ & Symm. dual-SPI\\
\hline
ITER shot 5 & $2.6\times 10^{22}$ Ne & $2.1\times 10^{24}$ H & $1000$ & $150\pm 50m/s$ & Asymm. dual-SPI\\
\hline
\etbl
\caption{The injection parameters for the SPI considered in this study. }
\label{tab:1}
\end{table*}

\section{The current redistribution and MHD response after full impurity SPI}
\label{s:MHD}

\begin{figure*}
\centering
\noindent
\btbl{c}
\parbox{4.5in}{
    \includegraphics[scale=0.35]{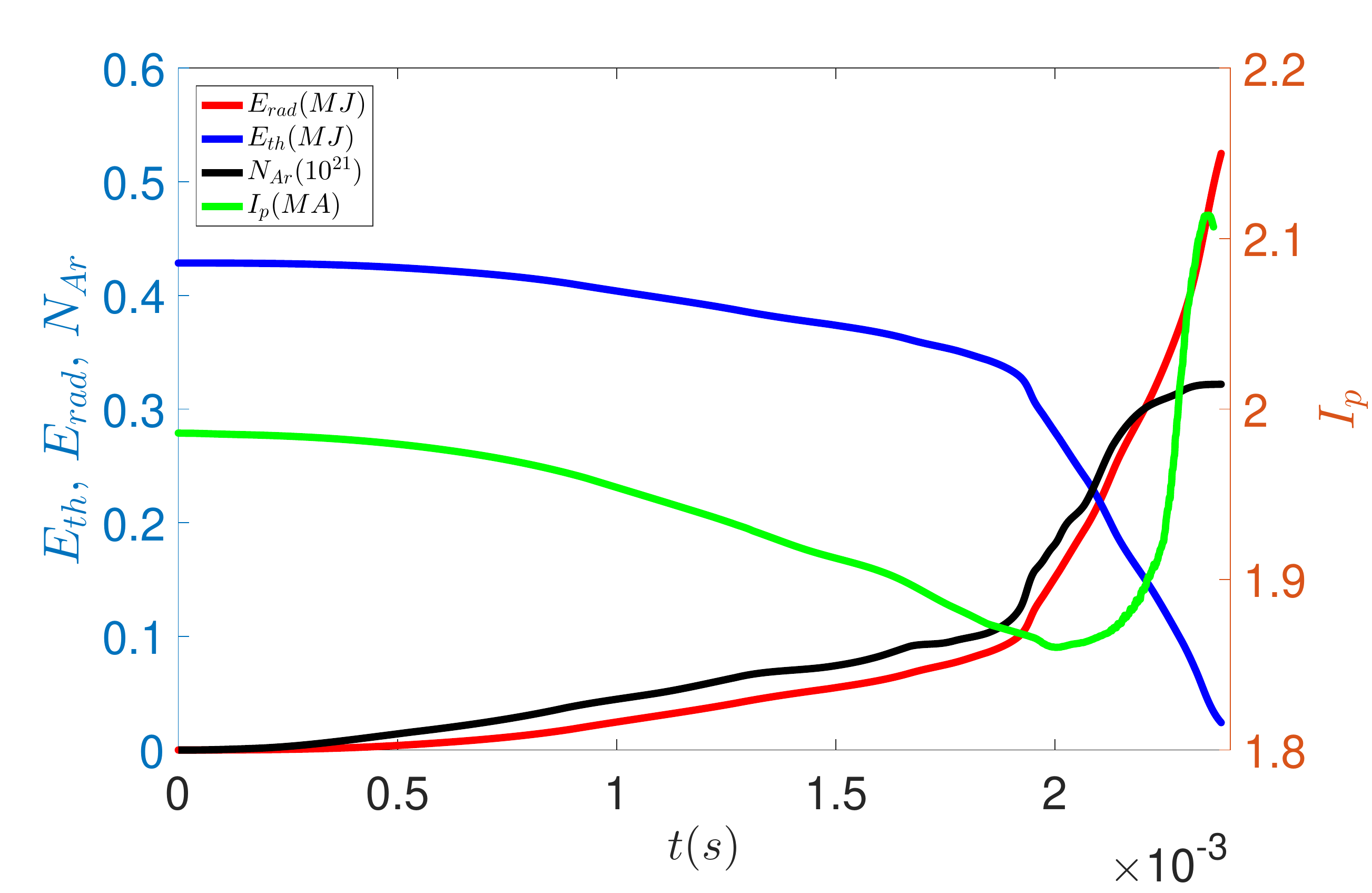}
}
\etbl
\caption{The ``JET shot 1'' evolution of the thermal energy content $E_{th}$, radiated energy $E_{rad}$, assimilated argon atom number $N_{Ar}$, and the plasma current $I_p$. The TQ is approximately triggered at $t=2.00ms$.}
\label{fig:04}
\end{figure*}

In this section we will show the characteristic MHD response for the full argon SPI ``JET shot 1'' and compare it with previous deuterium SPI simulations. Here we consider a single temperature model as described in Section \ref{ss:Equations}.
The general evolution of the thermal content, the radiated energy, the total ablated argon atoms and the total current is shown in Fig.\,\ref{fig:04}. The TQ occurs around $t=2.00ms$, as the TQ proceed there can be seen a current spike developing. Sudden increase of radiated energy and ablation rate can also be seen around the time of the TQ onset.

Two kinds of MHD excitation mechanism have been identified to play a role in the massive material injection process \cite{Di2018NF,Nardon2017PPCF}.
The first is the $n=0$ axisymmetric current contraction caused by the periphery cooling. This causes current density redistribution on a fraction of the minor radius $a$, creating an inward propagating current sheet on the cooling front as the fragments move toward the plasma core on the timescale of $\gt\propto a^2/\gh$ where $\gh$ is the resistivity in the cooled down region. The second is the helical cooling effect on each major rational surface, which causes helical current density redistribution around a small distance $\gd l$ close to the rational surface on the timescale $\gt\propto \gd l^2/\gh$.
When the $n=0$ current contraction time is faster than the fragment flying time, we expect to see the current contraction to follow the fragments closely, thus contribute significantly to the MHD destabilization. Otherwise there is no time for the $n=0$ mechanism to respond, and we expect to mostly see the contribution from the helical one.

\begin{figure*}
\centering
\noindent
\btbl{cc}
\parbox{2.8in}{
    \includegraphics[scale=0.25]{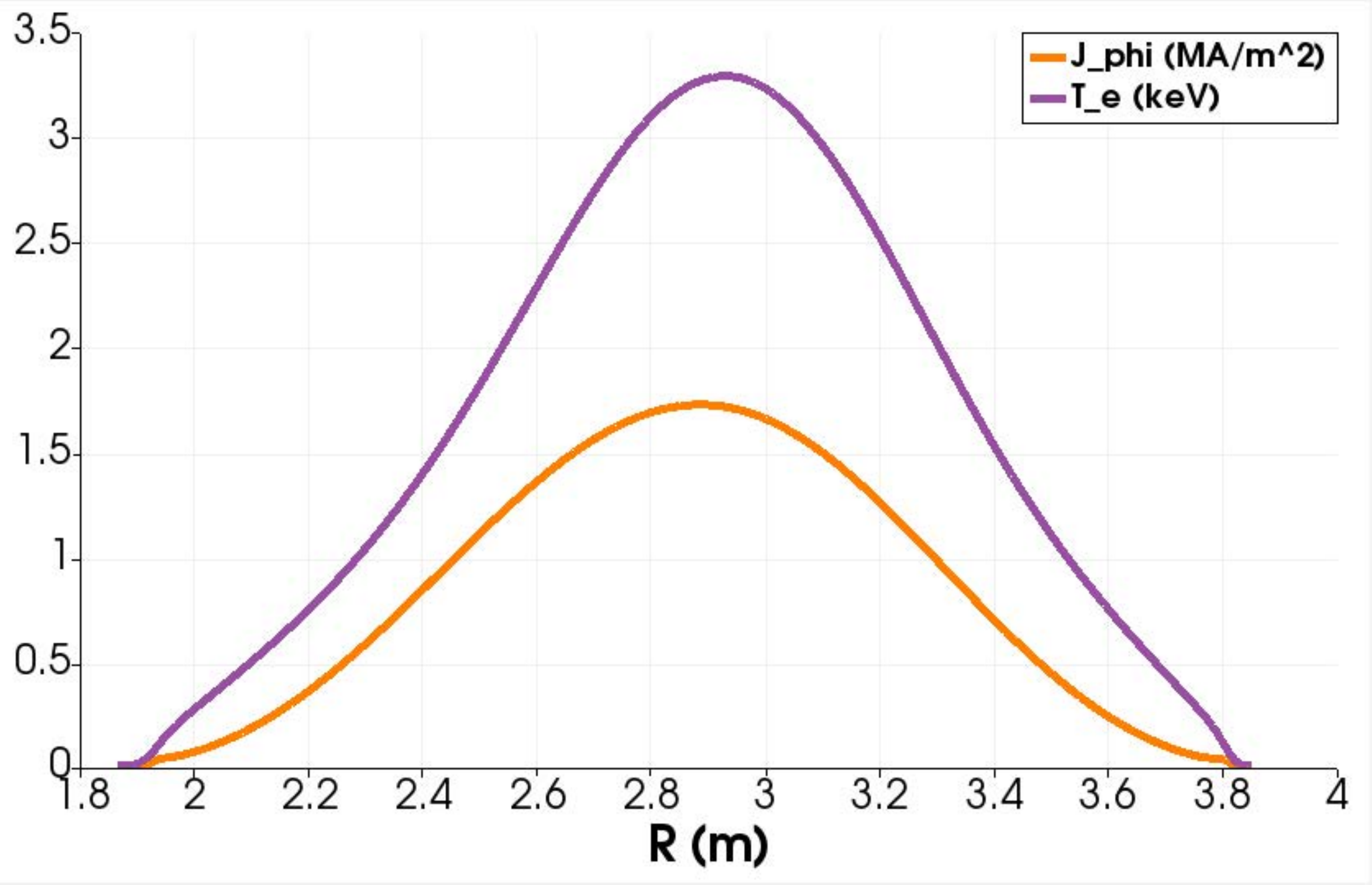}
}
&
\parbox{2.8in}{
	\includegraphics[scale=0.25]{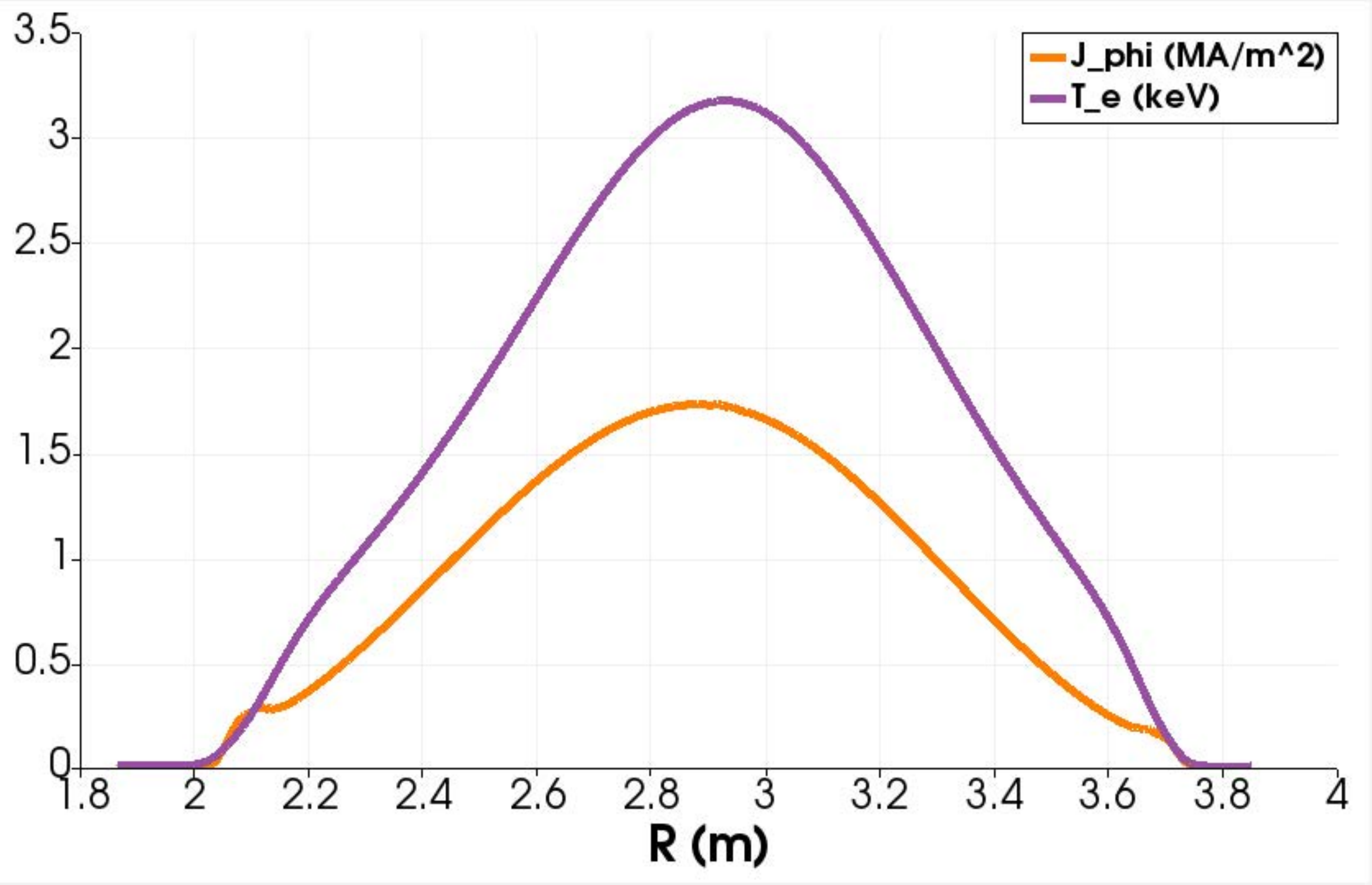}
}
\\
(a)&(b)
\\
\parbox{2.8in}{
  	\includegraphics[scale=0.25]{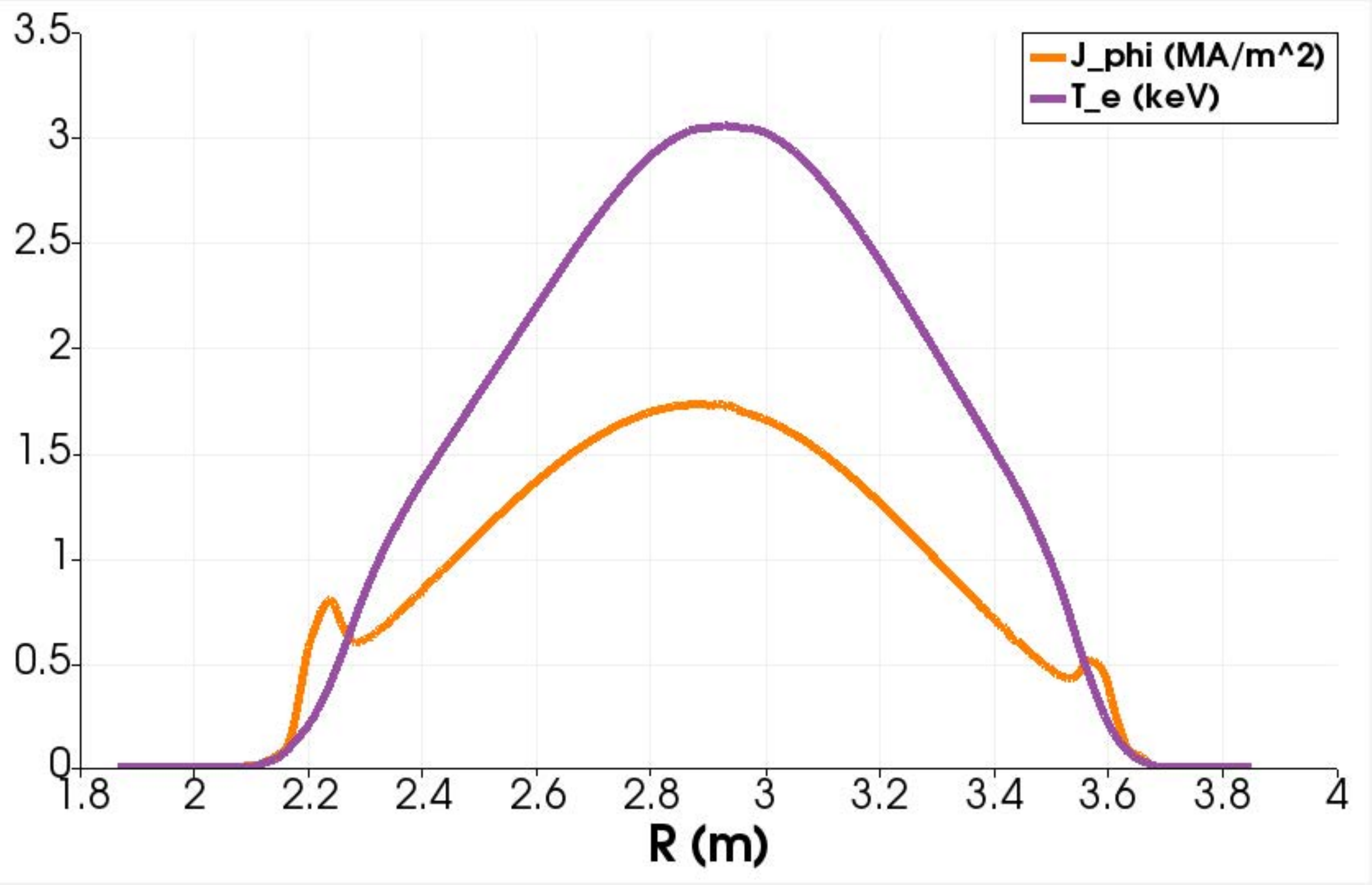}
}
&
\parbox{2.8in}{
	\includegraphics[scale=0.25]{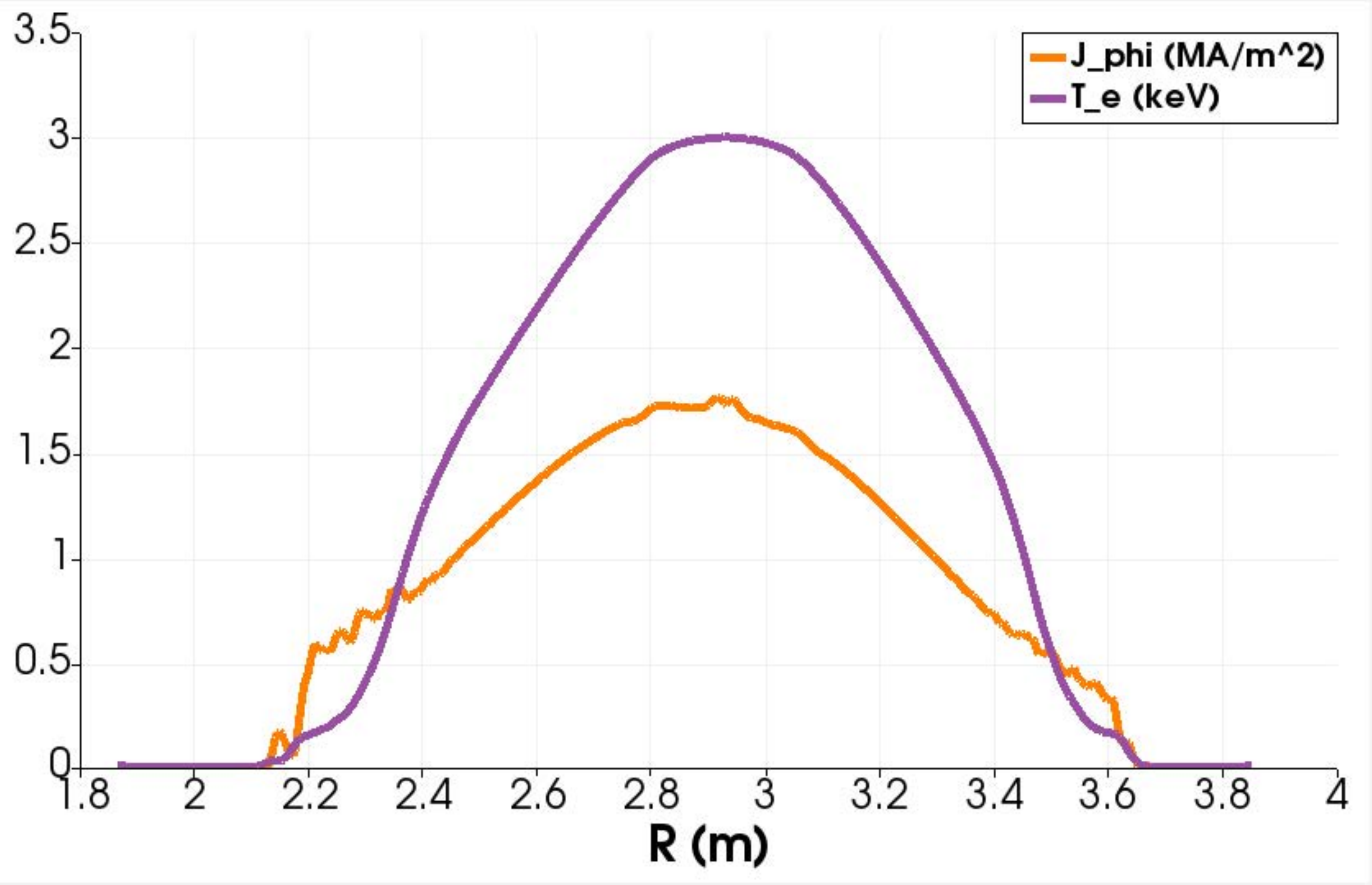}
}
\\
(c)&(d)
\etbl
\caption{The $n=0$ electron temperature and current density profile on the mid-plane at time (a) $t=0.00ms$, (b) $t=1.01ms$, (c) $t=1.72ms$ and (d) $t=1.91ms$ for pure argon SPI into JET-like L-mode. The current density profile contraction can be seen to closely follow that of the temperature, until the onset of magnetic stochasticity at $t=1.91ms$. Approximately, the $q=2$ surface corresponds to $R=2.26m$ and $R=3.55m$ on the mid-plane.}
\label{fig:05}
\end{figure*}

In the absence of any impurity, deuterium SPI causes a mild dilution cooling which leads to a comparably high post injection electron temperature on the order of $100eV$. In such a scenario there is very little axisymmetric current contraction observed, and the dominant MHD excitation is by helical cooling \cite{Di2018NF}. In drastic contrast, on the other limit that is the pure argon SPI studied here, we find the current density profile contraction follows the propagation of the cooling front closely until the onset of field line stochasticity as is shown in Fig.\,\ref{fig:05}.
The $n=0$ axisymmetric current density profile $J_\gf$ and electron temperature profile $T_e$ at the midplane at times $t=0.00ms$, $t=1.01ms$, $t=1.72ms$ and $t=1.91ms$ are shown as a function of the major radius $R$. At $t=1.72ms$, a significant current sheet can be seen forming on the cooling front, which coincides with the time at which the vanguard fragment begin to arrive on the $q=2$ surface. At $t=1.91ms$, further contraction of the current profile is prevented by the onset of the magnetic stochasticity as will be seen later, since the nonlinear $\wsv\times\waB$ term tends to act like a hyper-resistivity upon the $n=0$ component of the current density, flattening its profile \cite{Boozer1986JPP,Strauss1986,Craddock1991}. Such current contraction behavior is in accordance with our aforementioned argument of timescale comparison, and it can be expected that, in the case of mixed species injection, the exact current redistribution behavior and MHD response would be a mixture of those two limits discussed above.
Not shown in the figure, the current density profile will also be flattened in the core after the TQ as is reported in previous deuterium SPI study \cite{Di2018NF}. Such flattening is the result of the hyper-resistivity induced by the nonlinear $\wsv\times\waB$ of the strongly non-linear MHD activity during the TQ \cite{Boozer1986JPP,Strauss1986,Craddock1991}, as is mentioned above.

\begin{figure*}
\centering
\noindent
\btbl{cc}
\parbox{2.8in}{
    \includegraphics[scale=0.35]{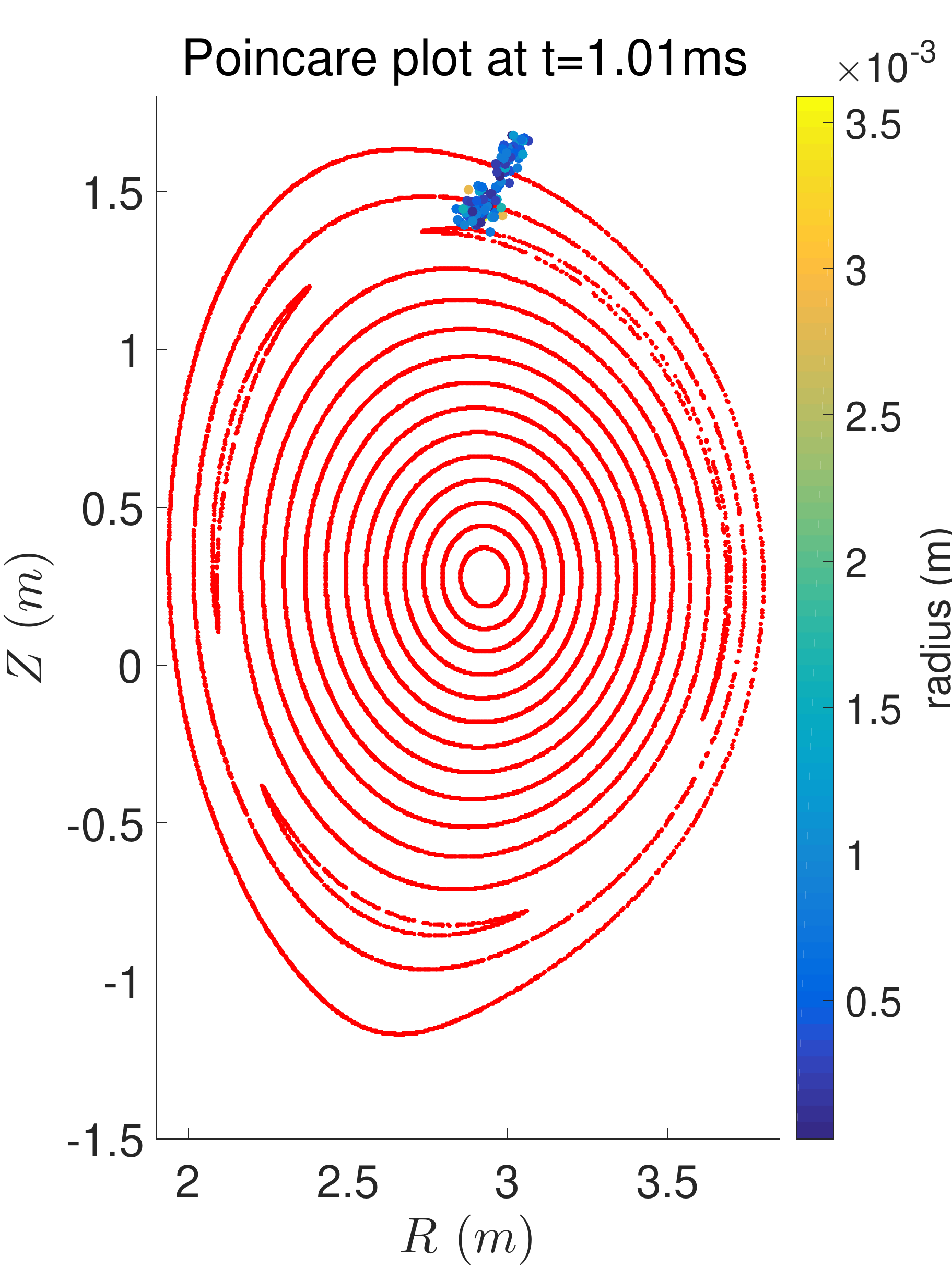}
}
&
\parbox{2.8in}{
	\includegraphics[scale=0.35]{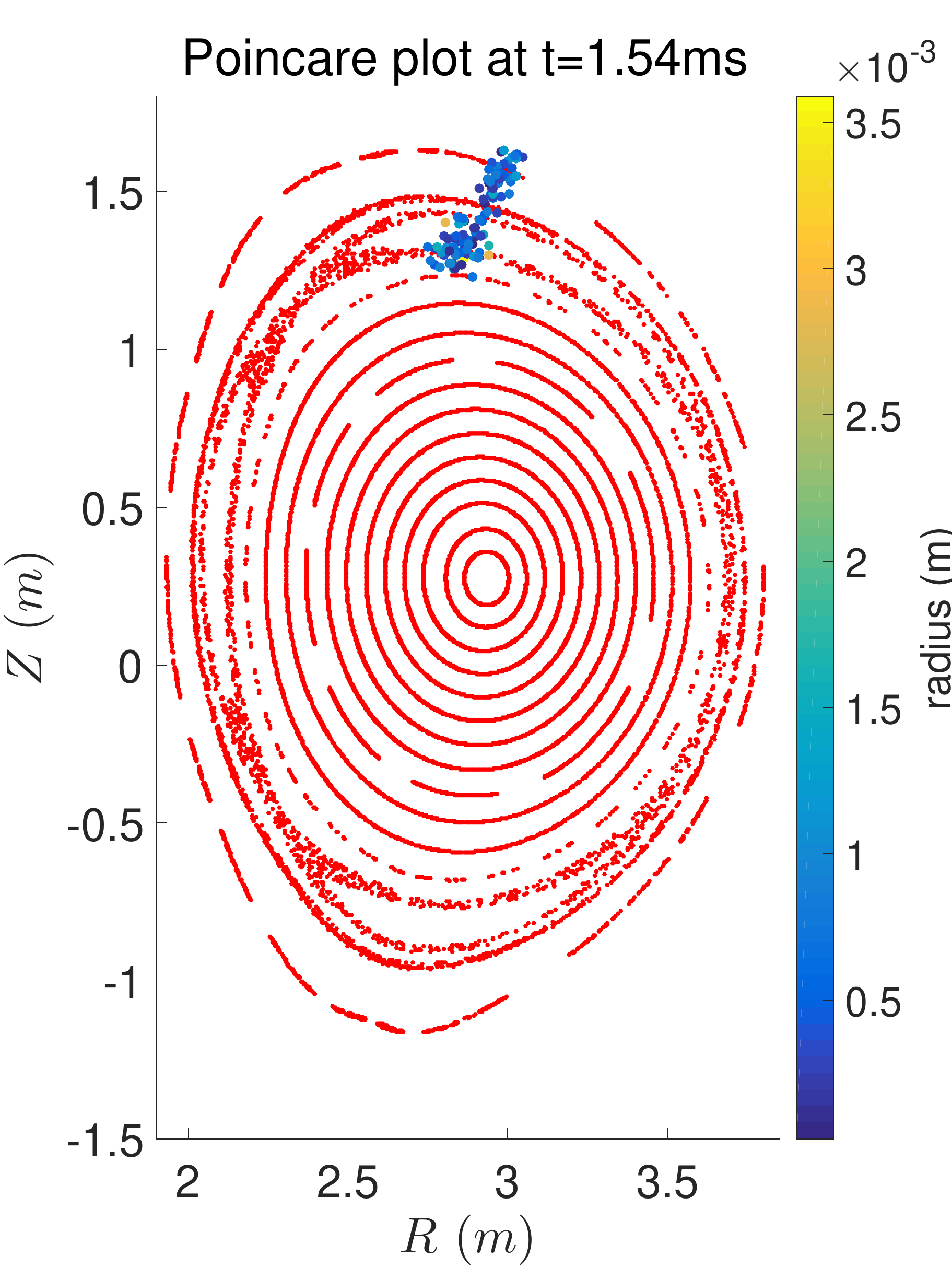}
}
\\
(a)&(b)
\\
\parbox{2.8in}{
  	\includegraphics[scale=0.35]{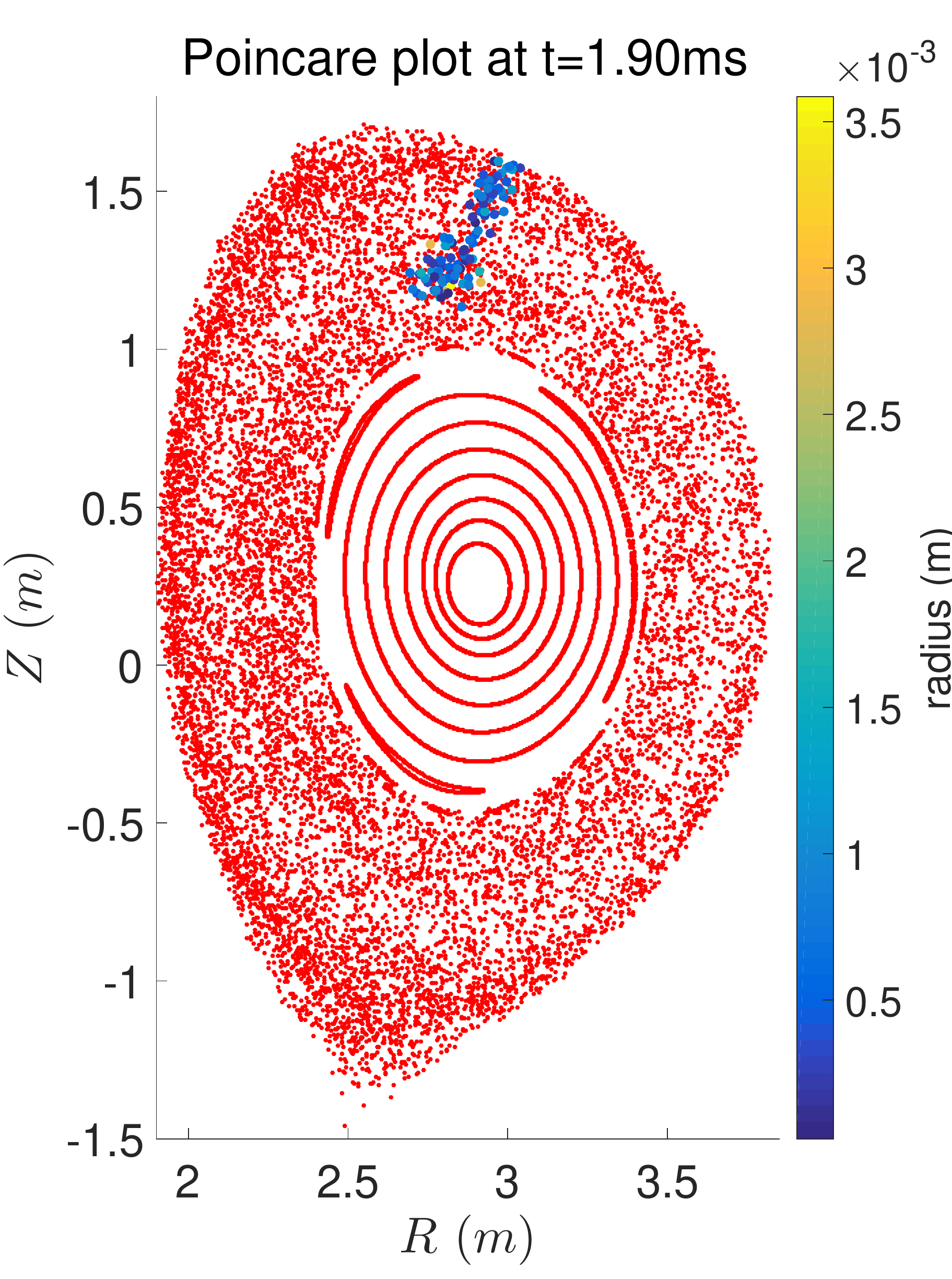}
}
&
\parbox{2.8in}{
	\includegraphics[scale=0.35]{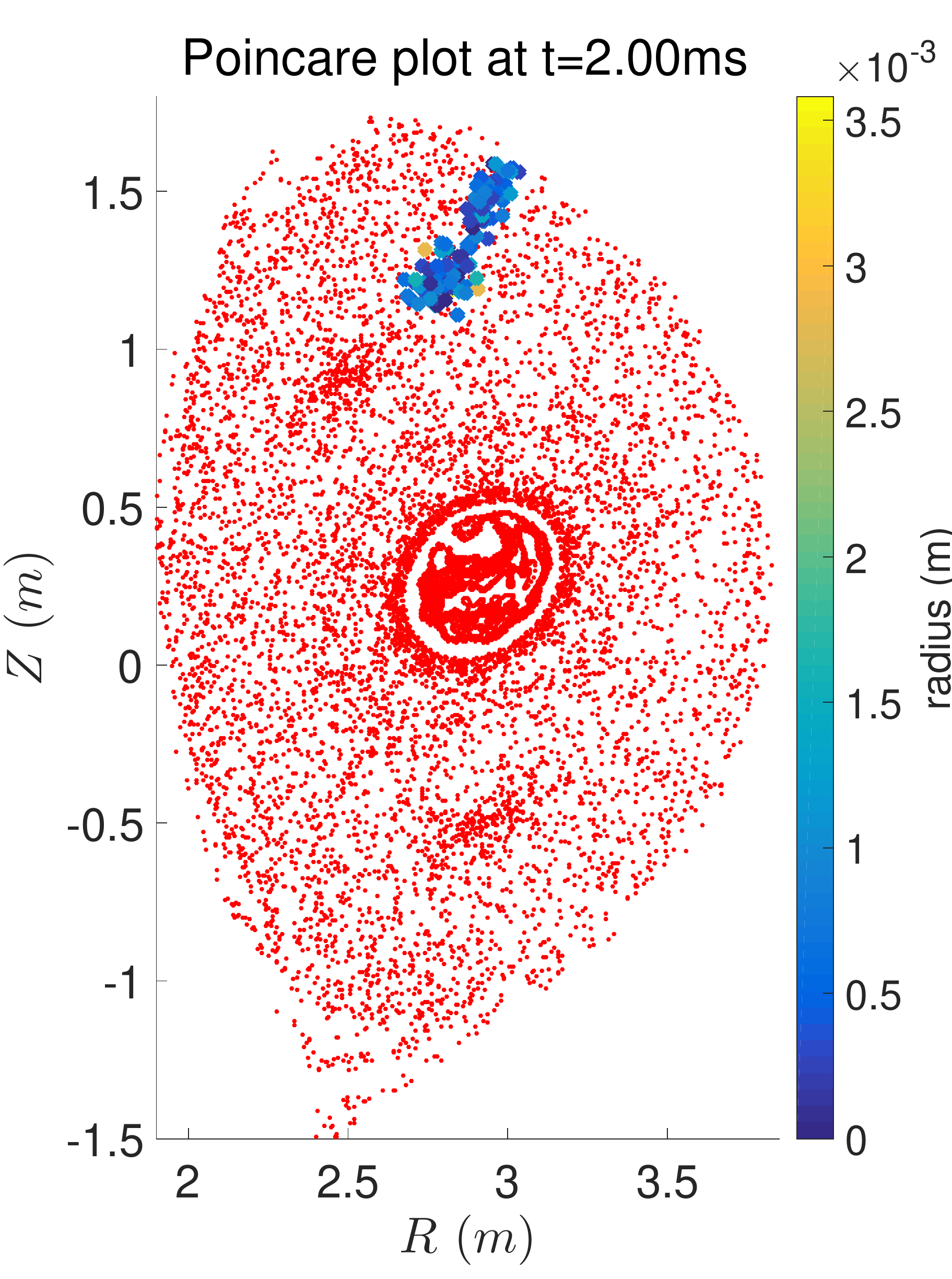}
}
\\
(c)&(d)
\etbl
\caption{The Poincar\'{e} plot of the magnetic field line for ``JET shot 1'' at (a) $t=1.01ms$, (b) $t=1.54ms$, (c) $t=1.90ms$ and (d) $t=2.00ms$, as well as the fragment size and location at each corresponding time.}
\label{fig:06}
\end{figure*}

Poincar\'{e} plots of the magnetic field topology at time $t=1.01ms$ $t=1.54ms$, $t=1.90ms$ and $t=2.00ms$, as well as the projection of fragment positions onto the poloidal plane are shown in Fig.\,\ref{fig:06}. It can be seen that at first when the fragments enter the plasma there is no immediate magnetic stochasticity. The $3/1$ islands in Fig.\,\ref{fig:06}(a) are probably mainly a consequence of the axisymmetric current contraction discussed above. Indeed, the O-point phase in Fig.\,\ref{fig:06}(a) does not entirely correspond to that of the vanguard fragments, that the drive by the helical cooling effect is not dominant, as opposed to the pure deuterium SPI case. As the fragments and the cooling front propagate inwards, the $3/1$ island continues to grow and its separatrix begins to become stochastic. Finally when the fragments arrive on the $q=2$ surface and destabilize the $2/1$ mode, the mode coupling makes the whole outer plasma stochastic. This onset of magnetic stochasticity results in the flattening of the current sheet in the cooled region as seen in Fig.\,\ref{fig:05}(d). At this time, some core modes such as the $3/2$ mode are also nonlinearly destabilized, as can be seen in Fig.\,\ref{fig:06}(c). Shortly after, when these core modes couple with the already large outer modes, a stochastization of the entire plasma domain leads to a de-confinement of the core region, the TQ.
\begin{figure*}
\centering
\noindent
\btbl{c}
\parbox{4.0in}{
    \includegraphics[scale=0.35]{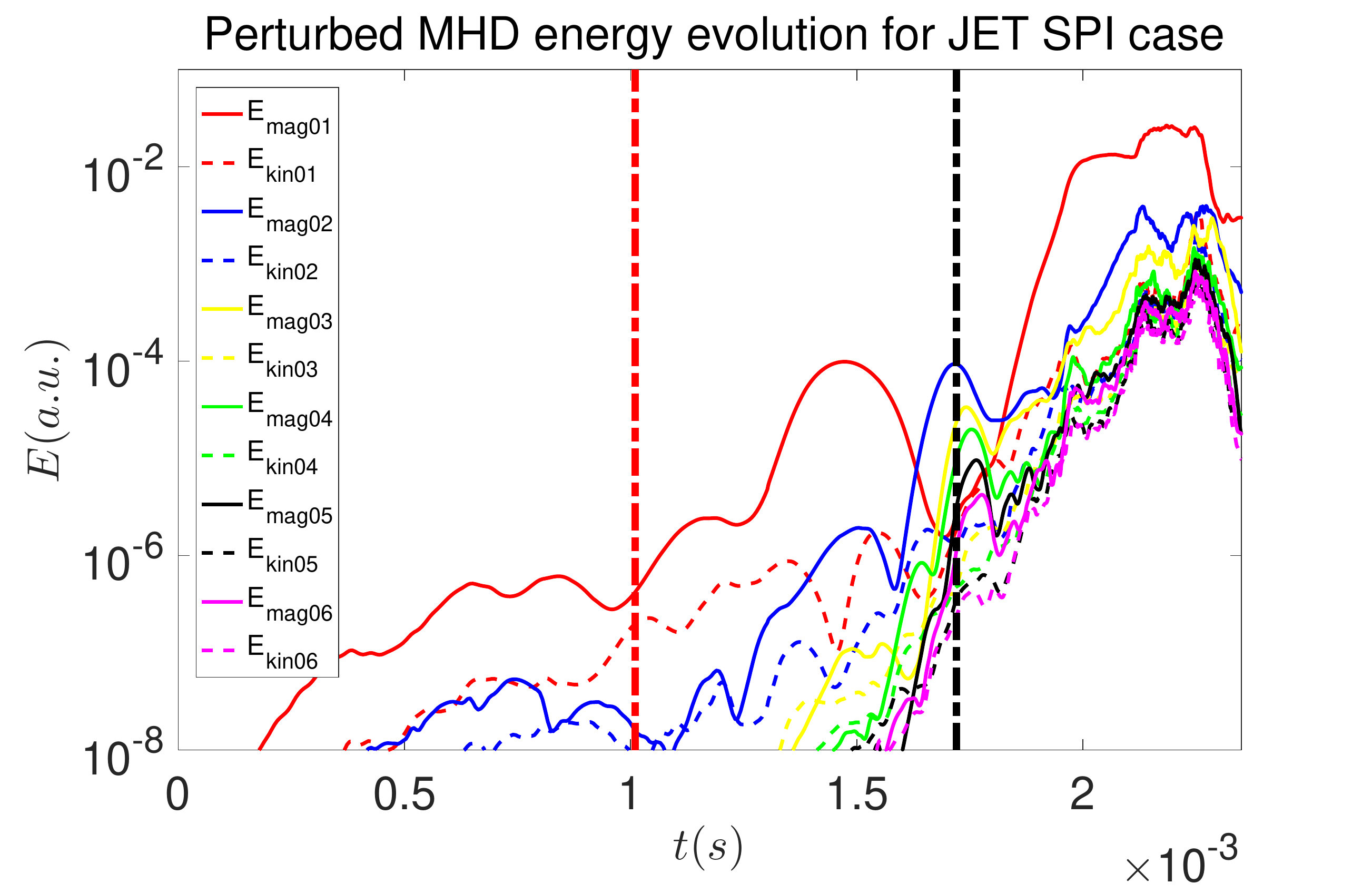}
}
\etbl
\caption{The $n=1$ to $n=6$ perturbed magnetic and kinetic energy after the injection. The red chained line corresponds to the approximate time when the vanguard fragments (thus the cooling front) arrive on the $q=3$ surface, and the black one corresponds to when they arrive on the $q=2$ surface.}
\label{fig:07}
\end{figure*}

The correlation between the cooling front propagation and the MHD excitation can also be seen in the perturbed energy spectrum as is shown in Fig.\,\ref{fig:07}, where the $n=1$ magnetic energy responds directly to the inward movement of the cooling front. The $3/1$ mode becomes unstable after the cooling front arrives at the $q=3$ surface, which is approximately marked by the vertical red chained line. Later, the $2/1$ begins to grow shortly after the cooling front approaches the $q=2$ surface around $t=1.72ms$, which is marked by the black line. This is also the time at which we see a strong current sheet in Fig.\,\ref{fig:05}(c). Afterward, the combined growth of $2/1$, $3/1$ and other core modes destroys the whole plasma confinement as is shown in Fig.\,\ref{fig:06}(c) and Fig.\,\ref{fig:06}(d), and the MHD activity remain strong during the TQ process. The above sequence of MHD developments shows remarkable agreement with recent JET SPI observation \cite{Gerasimov2020IAEA}, especially in terms of the TQ triggering location and the heightened MHD activity, despite our equilibrium is not exactly the same. This suggests a general behavior at least for the case of small or no $q=1$ surface.

\begin{figure*}
\centering
\noindent
\btbl{cccc}
\parbox{1.4in}{
    \includegraphics[scale=0.213]{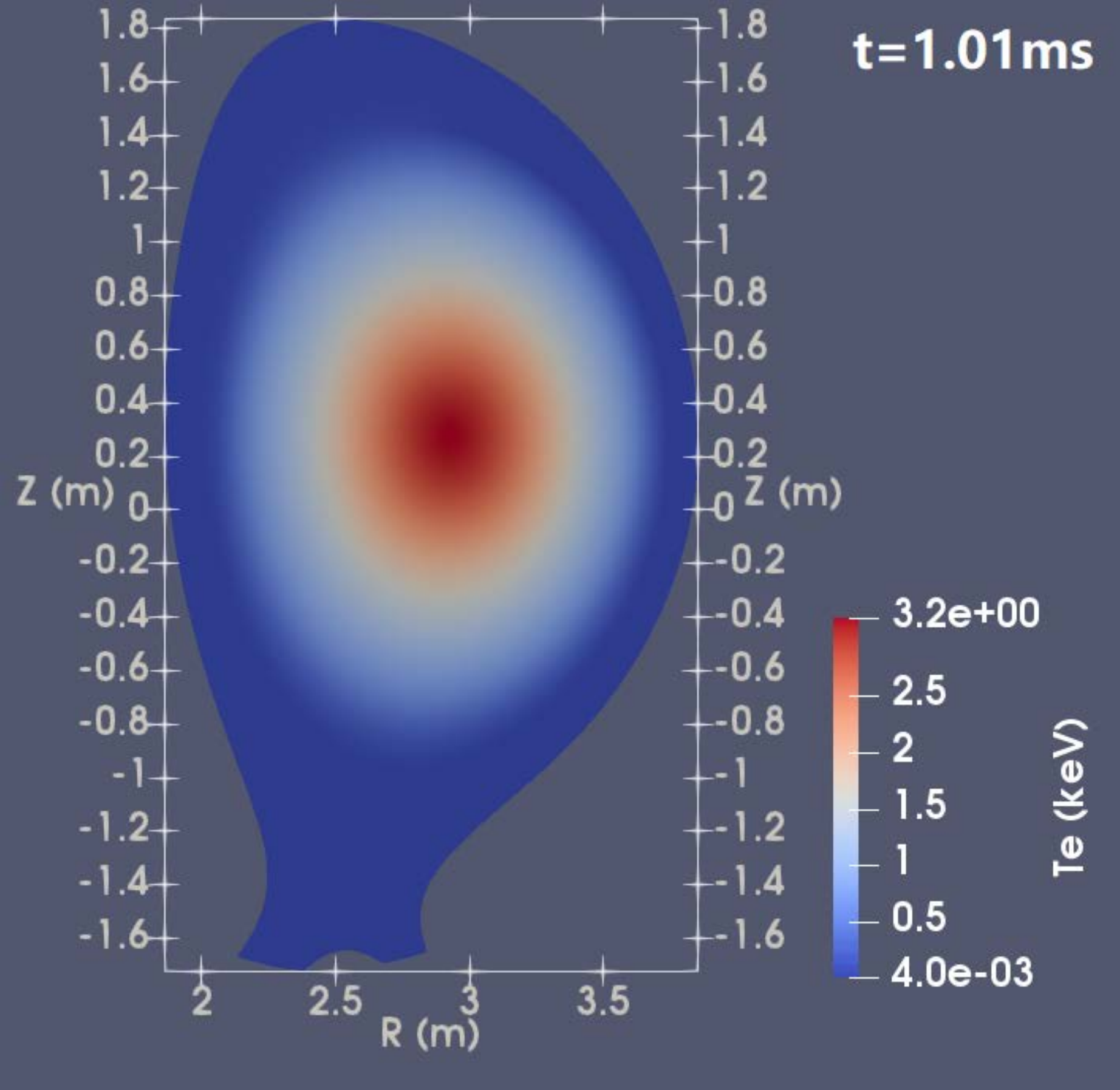}
}
&
\parbox{1.4in}{
	\includegraphics[scale=0.213]{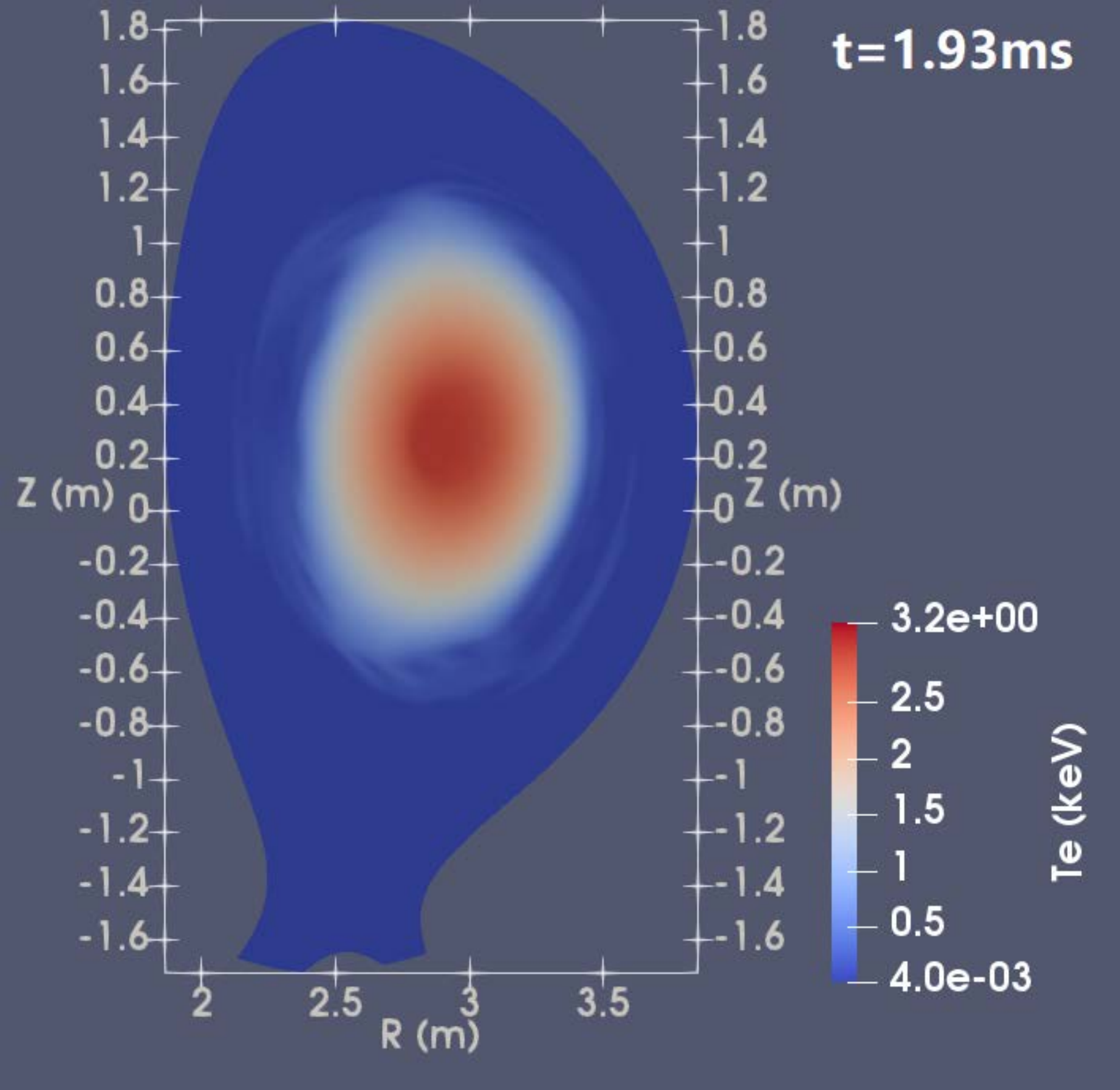}
}
&
\parbox{1.4in}{
	\includegraphics[scale=0.213]{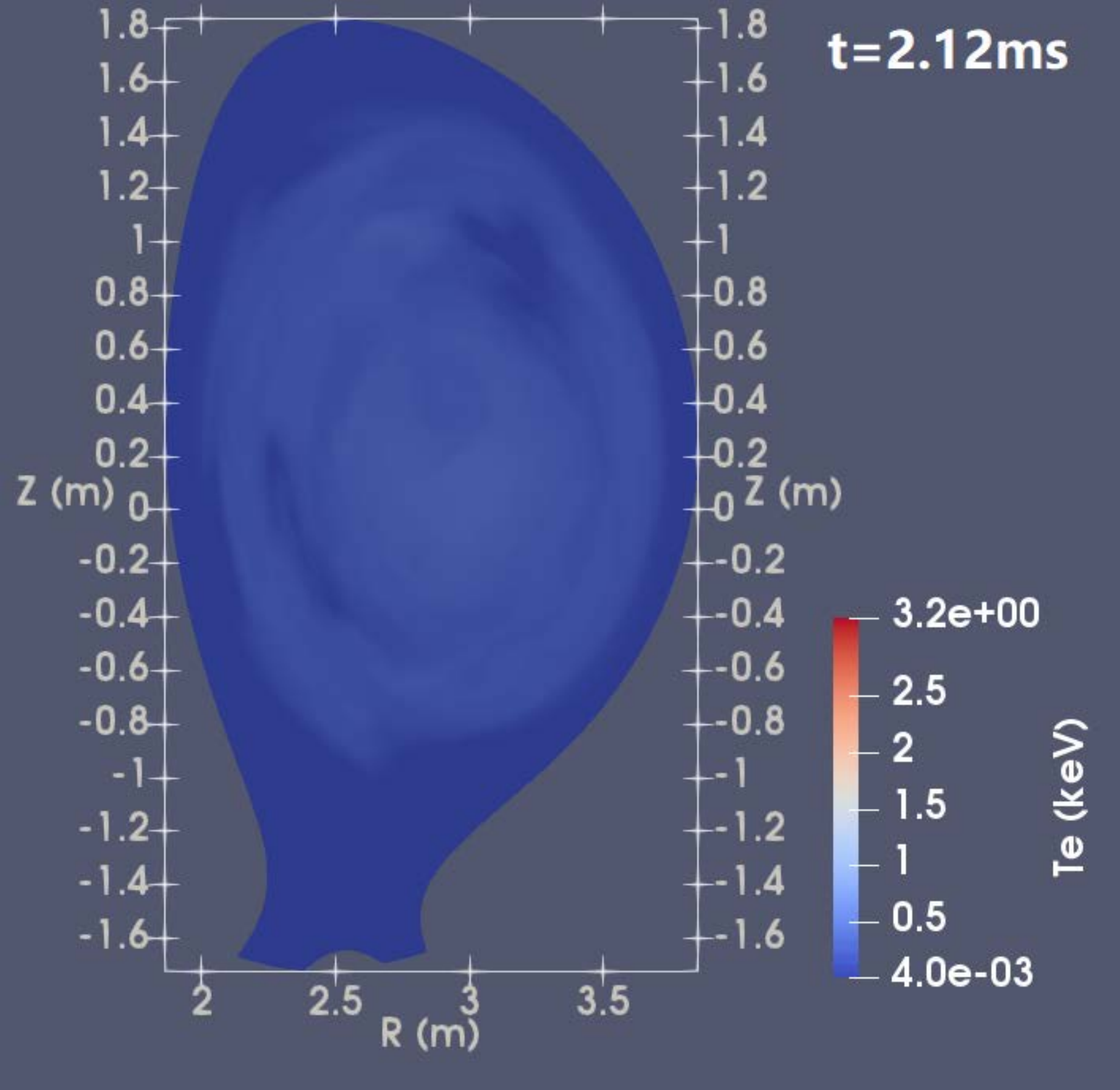}
}
&
\parbox{1.4in}{
	\includegraphics[scale=0.213]{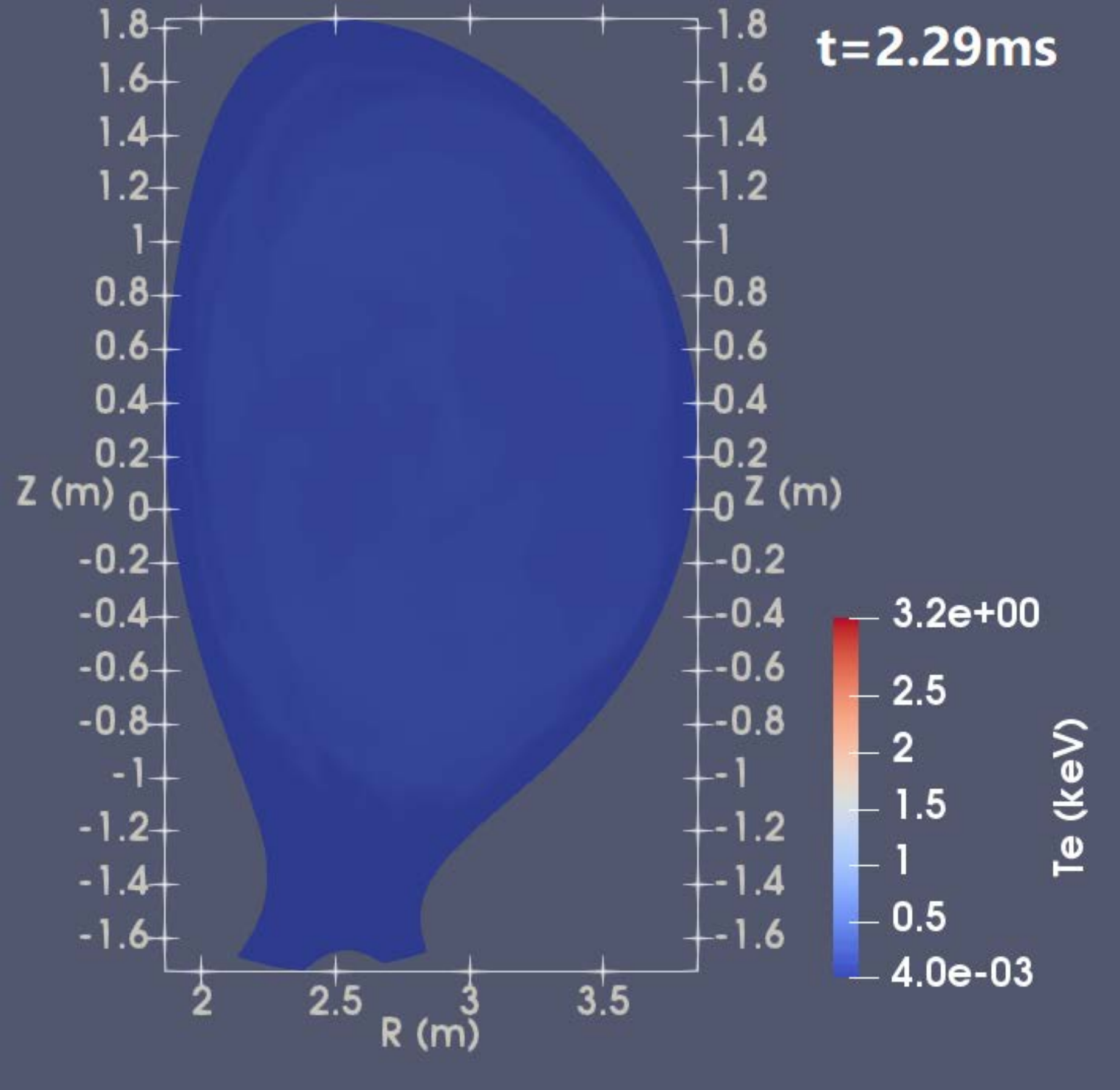}
}
\\
(a)&(b)&(c)&(d)
\\
\parbox{1.4in}{
	\includegraphics[scale=0.213]{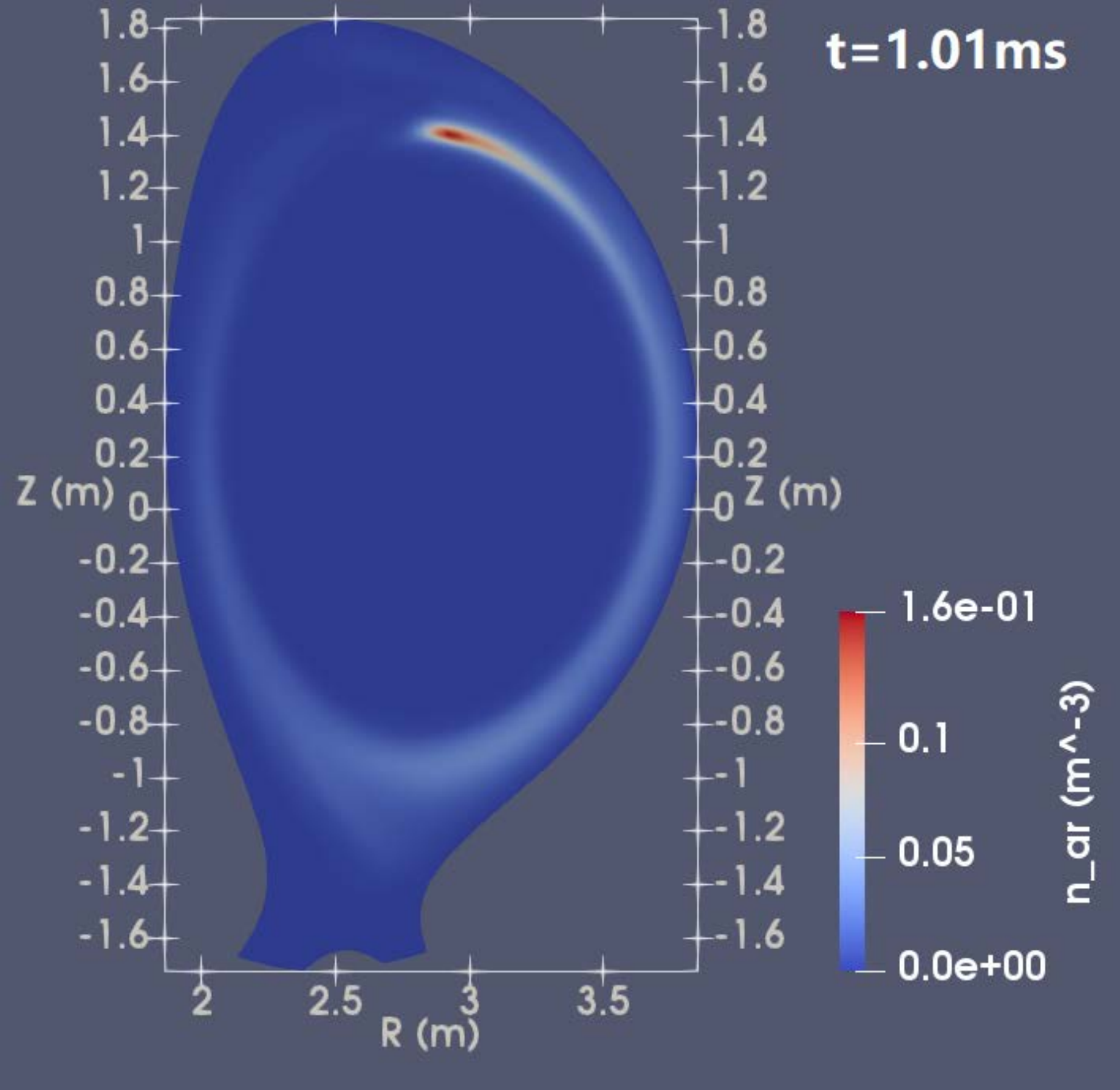}
}
&
\parbox{1.4in}{
	\includegraphics[scale=0.213]{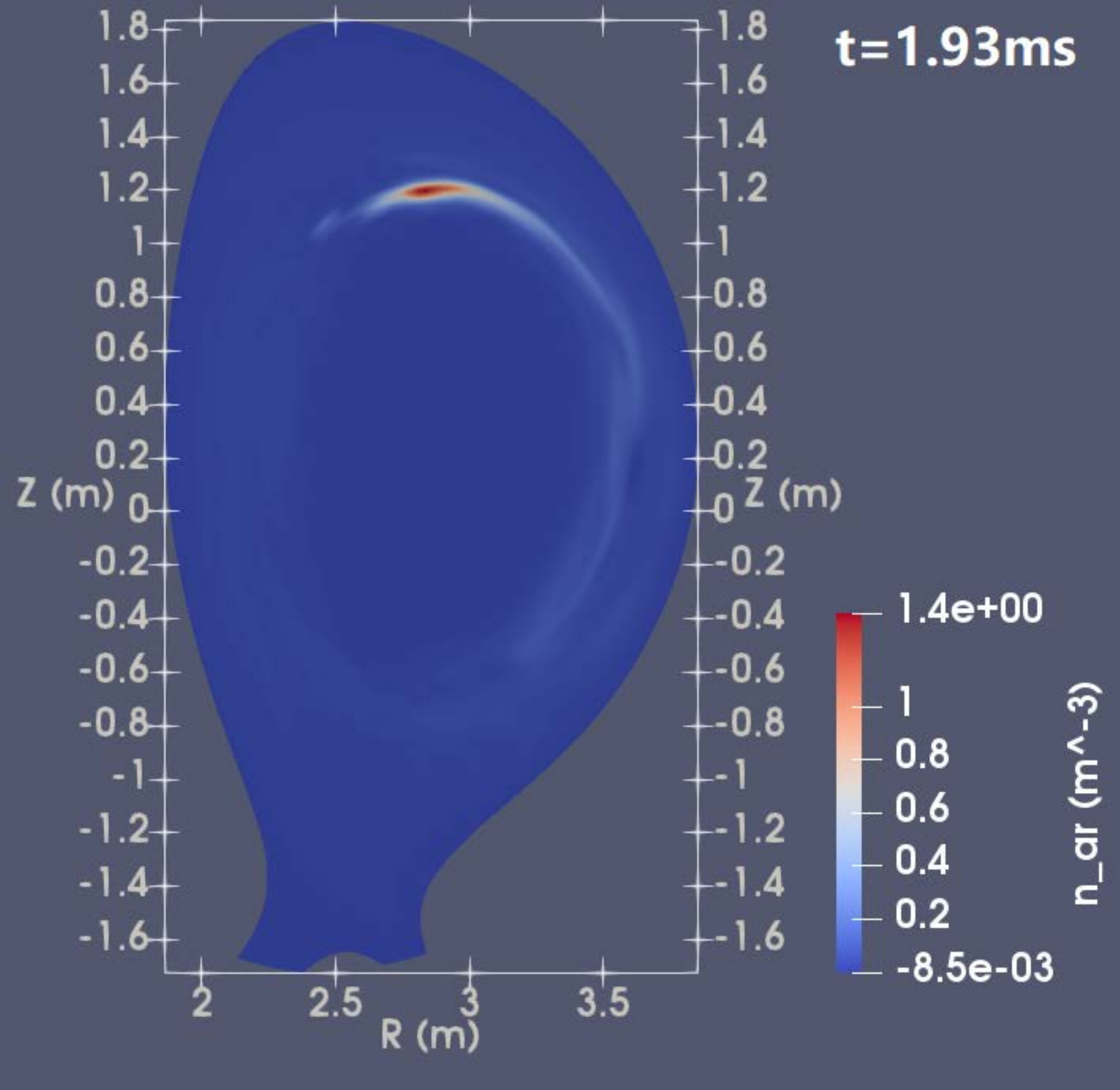}
}
&
\parbox{1.4in}{
	\includegraphics[scale=0.213]{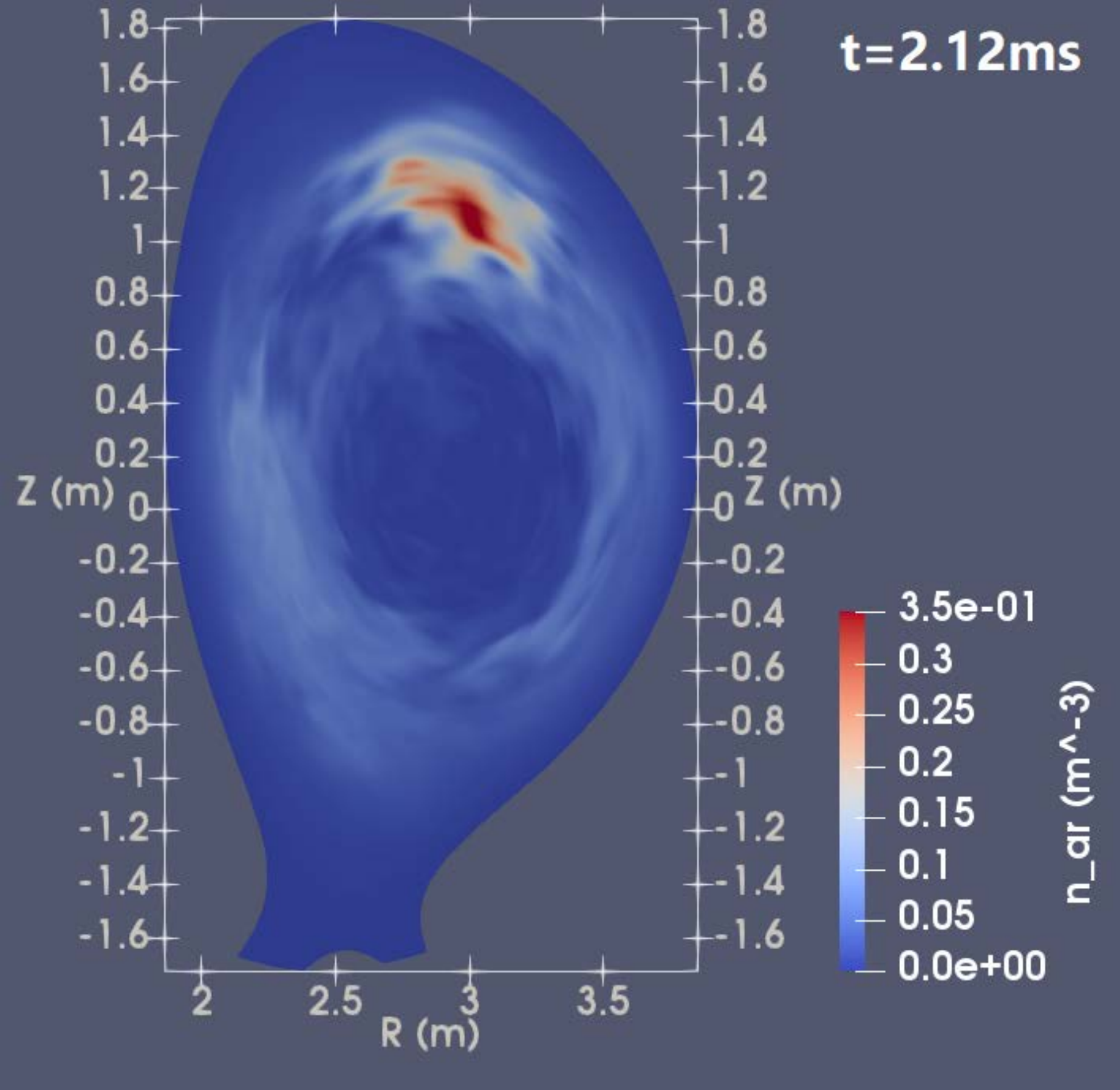}
}
&
\parbox{1.4in}{
	\includegraphics[scale=0.213]{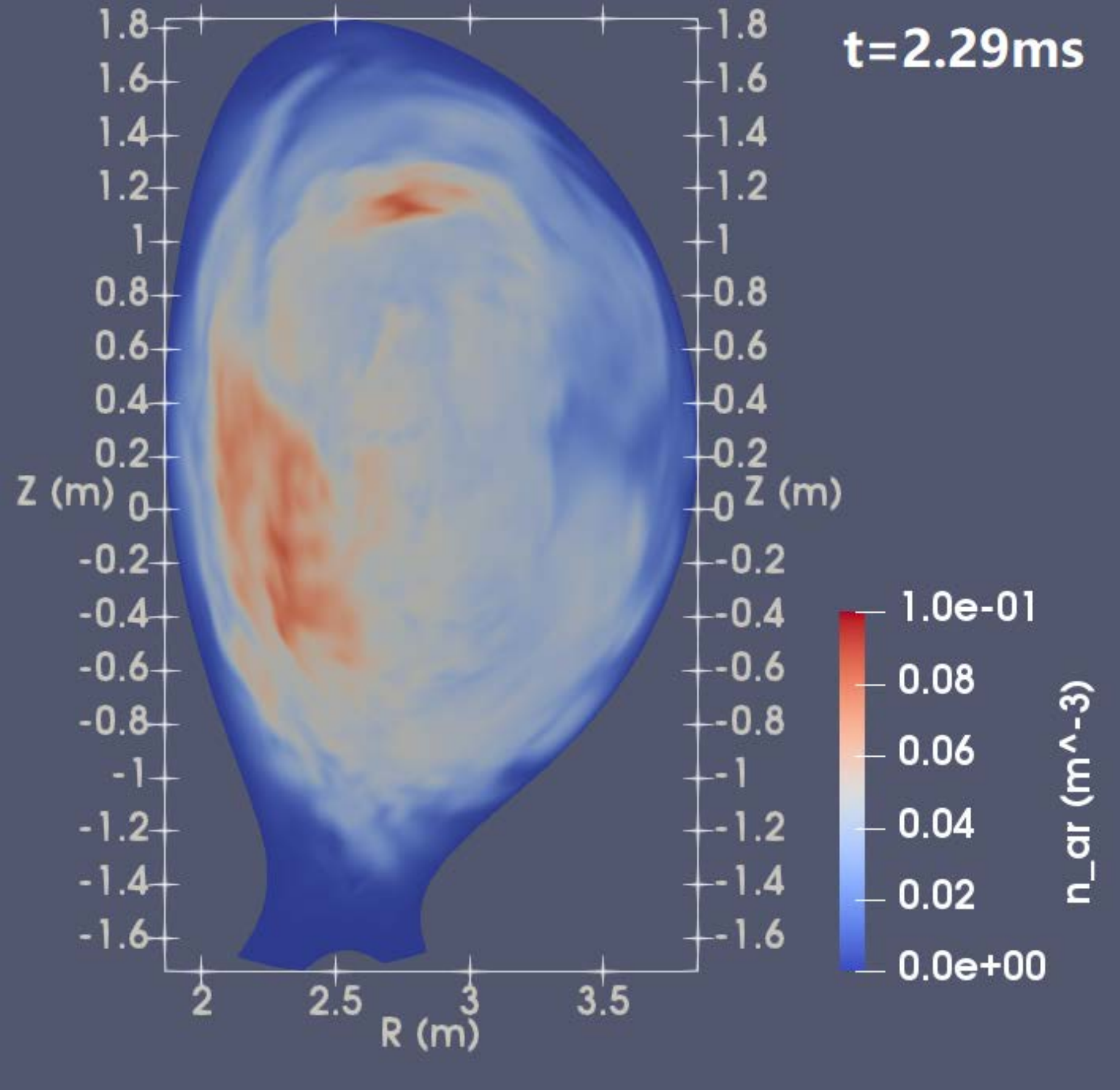}
}
\\
(e)&(f)&(g)&(h)
\etbl
\caption{Poloidal cross-sections of the electron temperature for ``JET shot 1'' at (a) $t=1.01ms$, (b) $t=1.93ms$ (approaching the TQ onset), (c) $t=2.12ms$ (shortly after the TQ onset) and (d) $2.29ms$, as well as the argon number density at the same times. There is more than $100\gm s$ delay between the core temperature collapse (around (b) and (c)) and the core impurity mixing by the MHD modes (around (h)). The high impurity density region in (e), (f) and (g) corresponds to the approximate position of the fragments.}
\label{fig:08}
\end{figure*}

\begin{figure*}
\centering
\noindent
\btbl{ccc}
\parbox{1.4in}{
    \includegraphics[scale=0.213]{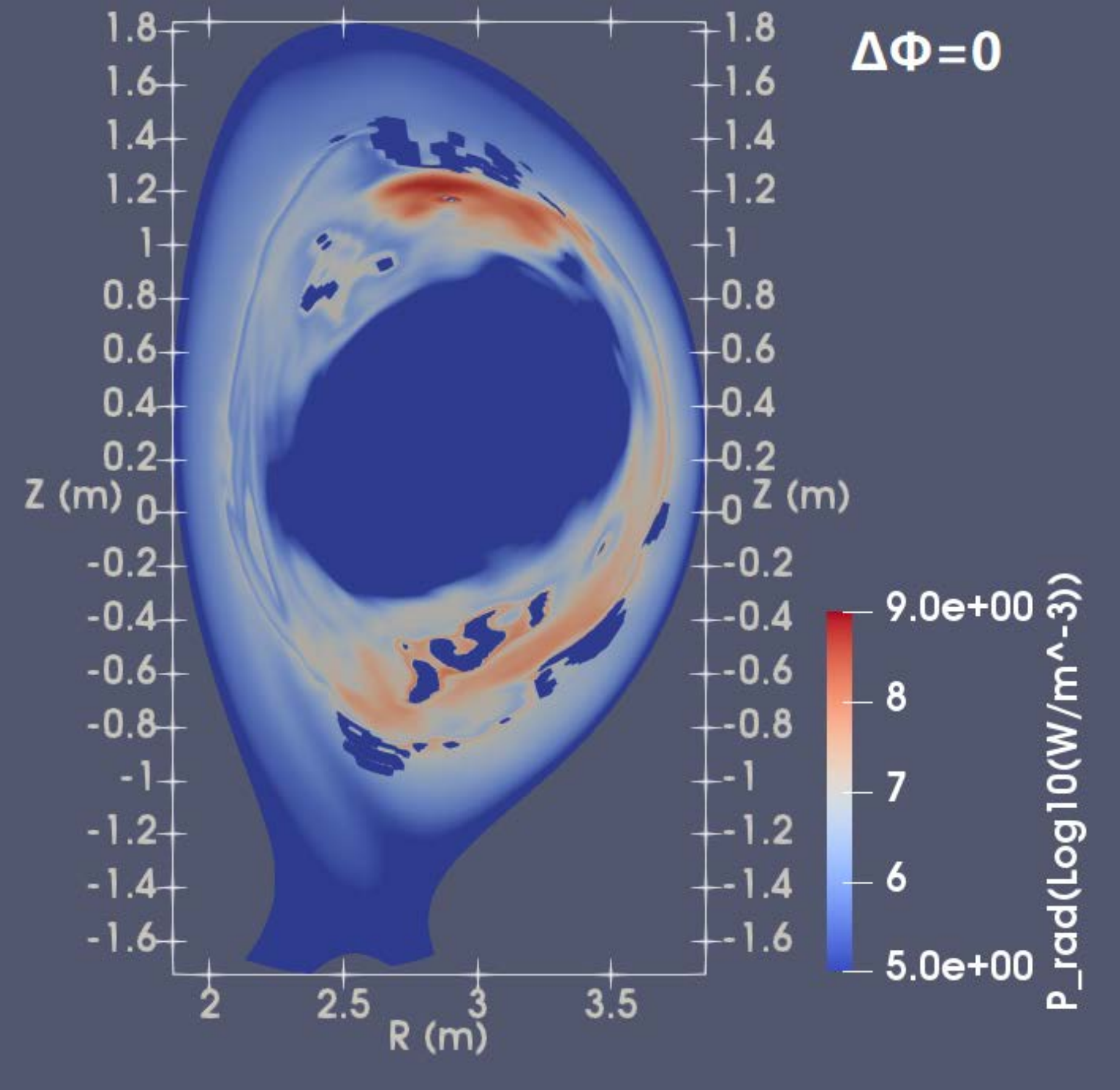}
}
&
\parbox{1.4in}{
	\includegraphics[scale=0.213]{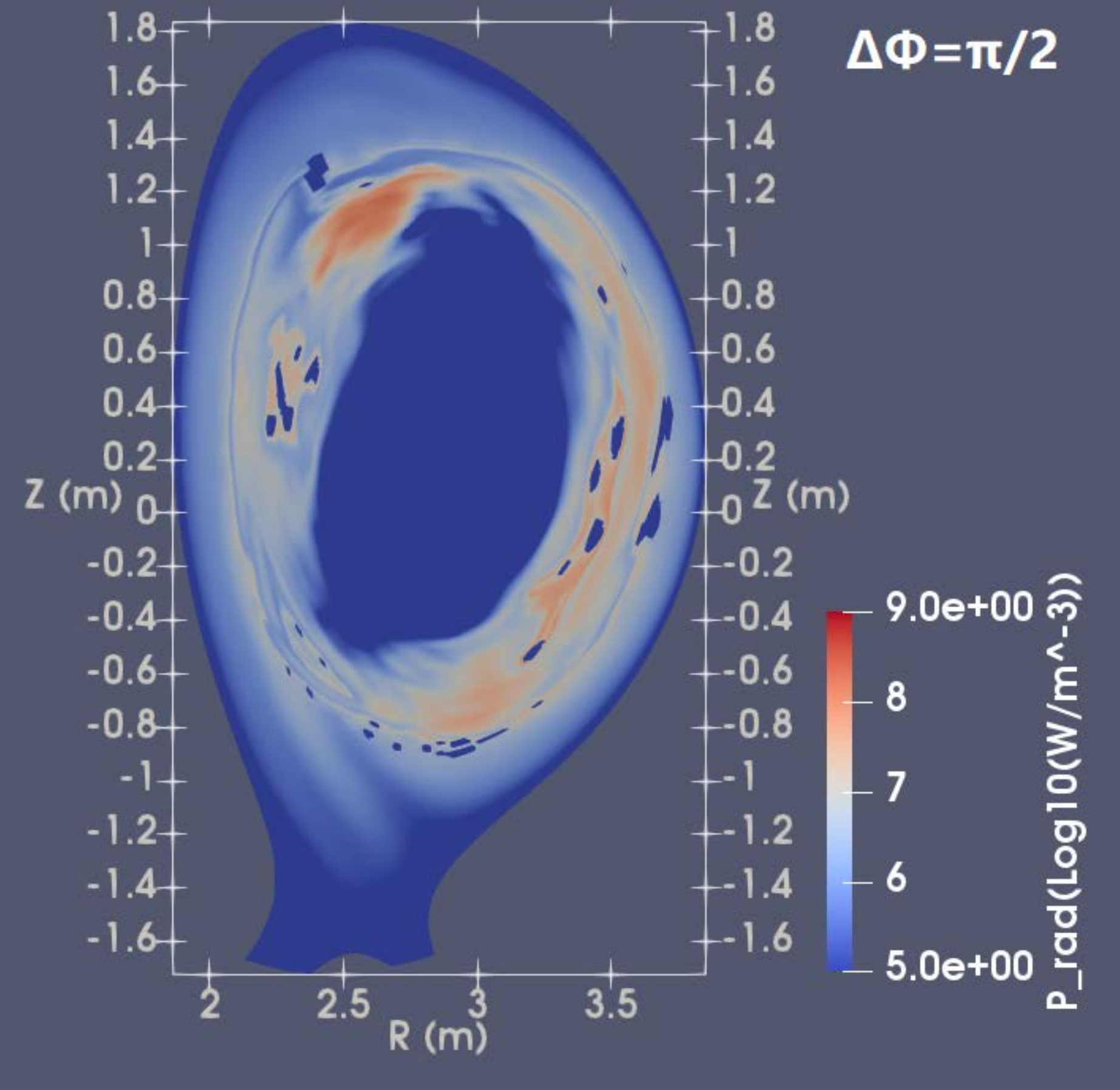}
}
&
\parbox{1.4in}{
	\includegraphics[scale=0.213]{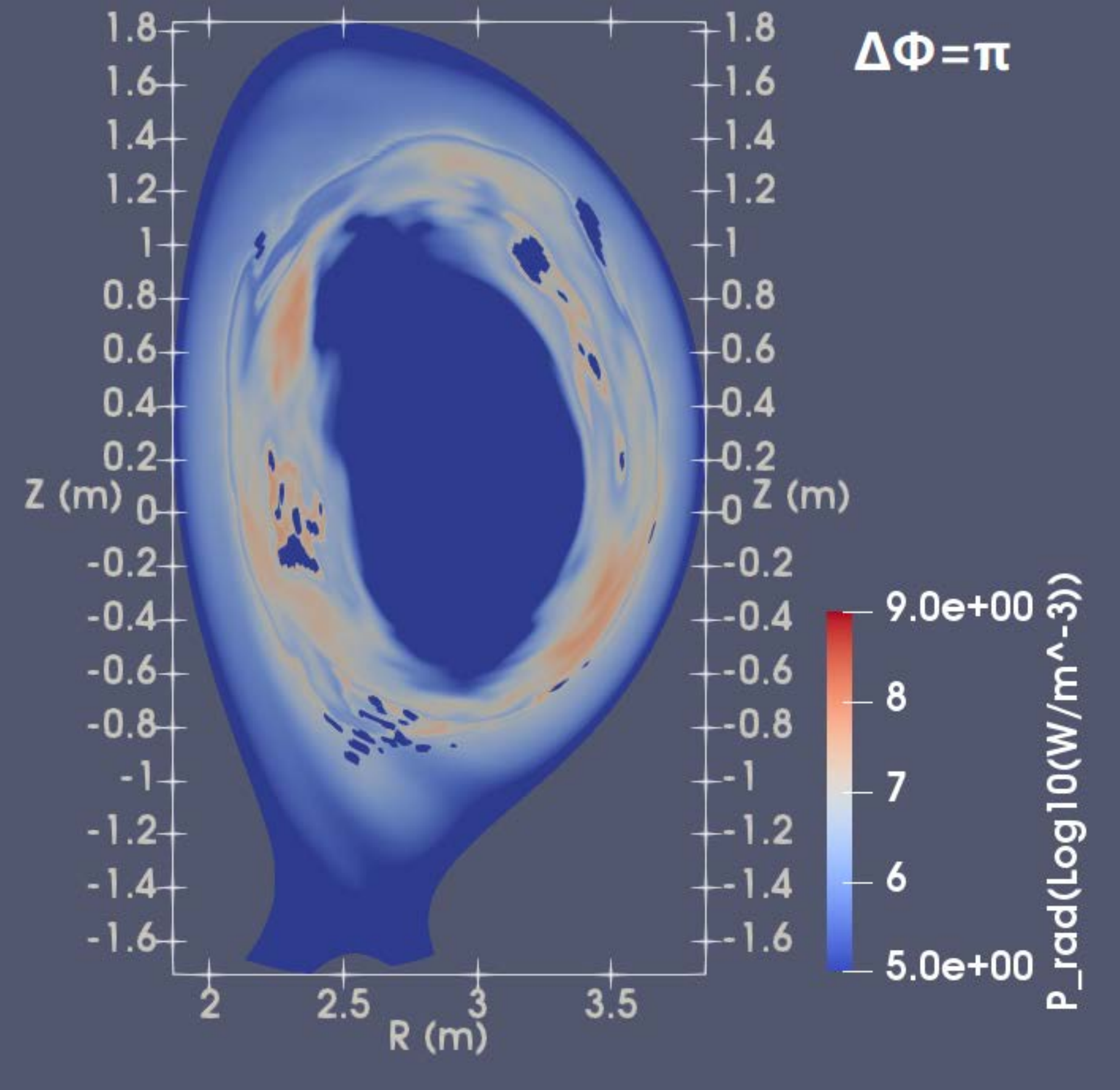}
}
\\
(a)&(b)&(c)
\etbl
\caption{The $\log_{10}$ of the radiation power density of ``JET shot 1'' at time $t=2.00ms$ for (a) the SPI toroidal location, (b) toroidally $\gp/2$ away from the SPI location and (c) toroidally $\gp$ away from the SPI location.}
\label{fig:09}
\end{figure*}

It should be noted that such full impurity SPIs tend to trigger the core collapse before they actually reach the plasma core. This is undesirable for runaway electron suppression as the core impurity density rise could suffer a time delay relative to the temperature collapse, providing a preferred region for the runaway electrons to generate. This can be seen in Fig.\,\ref{fig:08} where the electron temperature and the impurity density profile is shown for $t=1.01ms$, $t=1.93ms$, $t=2.12ms$ and $2.29ms$ respectively.
The evolution of the temperature profile shows the characteristics of ``scraping off'' before the onset of the TQ, while it is dominated by the diffusive behavior due to conduction along stochastic field line after the TQ begins at $t\simeq 2.00ms$, and it can be seen that at $t=2.12ms$ the temperature profile is almost completely flattened as is shown in Fig.\,\ref{fig:08}(c).
The argon density shows similar 3D stream as observed in recent JET SPI experiments \cite{Gerasimov2020IAEA}. Furthermore, the radiation power density shows a distinctive unrelaxed helical structure, with the peak close to the toroidal location of the injection location as is shown in Fig.\,\ref{fig:09}. Apart from the toroidal asymmetry as is shown by the order-of-magnitude difference in the radiation power density peak, the radiation also exhibits a $2/1$ helical structure, corresponding to the $q=2$ surface which triggers the TQ. The above behaviors in the TQ triggering position, enhanced MHD amplitude during the TQ, helical impurity density stream and helical unrelaxed radiation structure at the time of the TQ qualitatively agree with the recent JET experiments \cite{Sweeney2020IAEA,Gerasimov2020IAEA}.

\begin{figure*}
\centering
\noindent
\btbl{ccc}
\parbox{1.4in}{
    \includegraphics[scale=0.20]{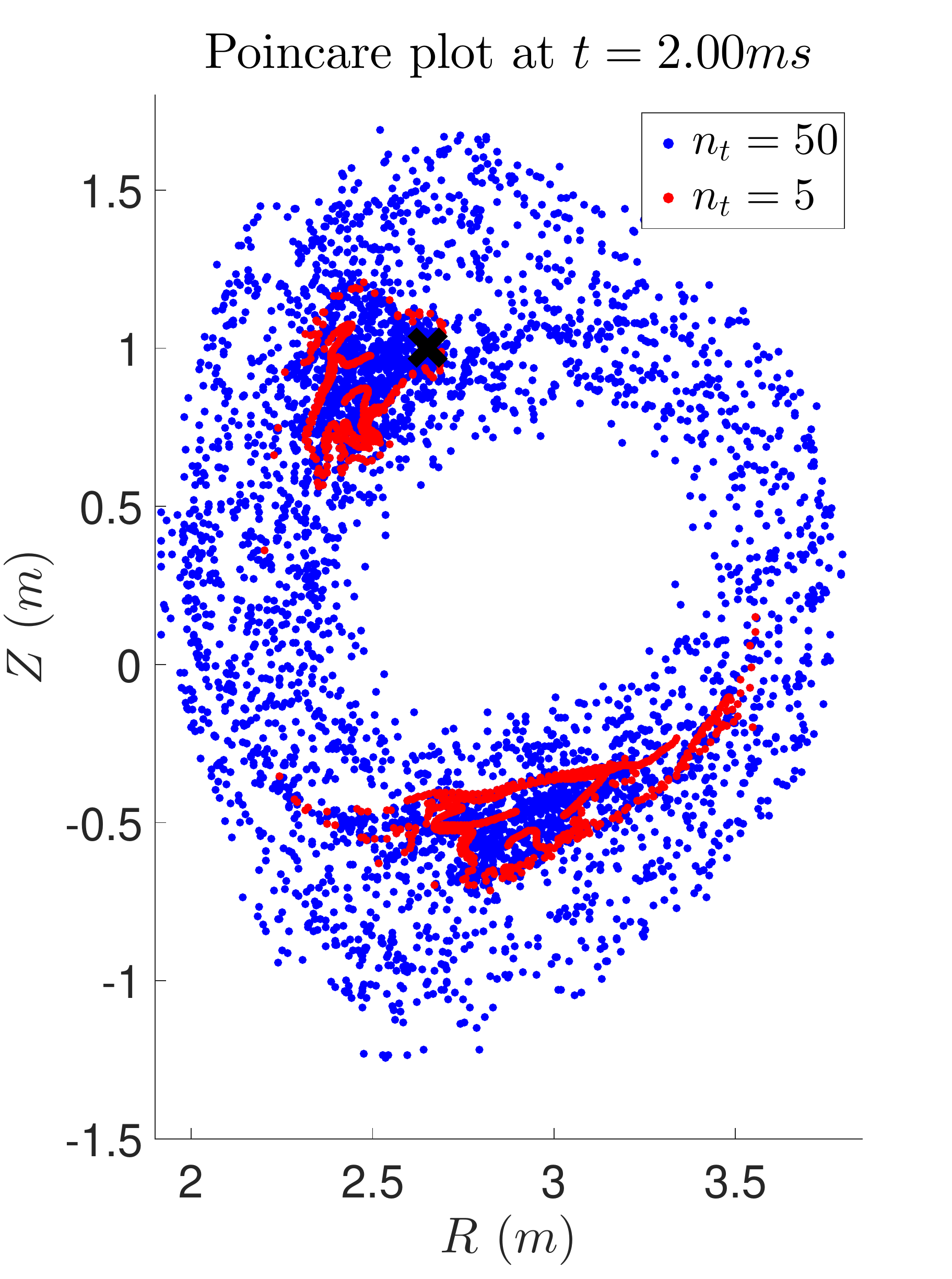}
}
&
\parbox{1.4in}{
	\includegraphics[scale=0.20]{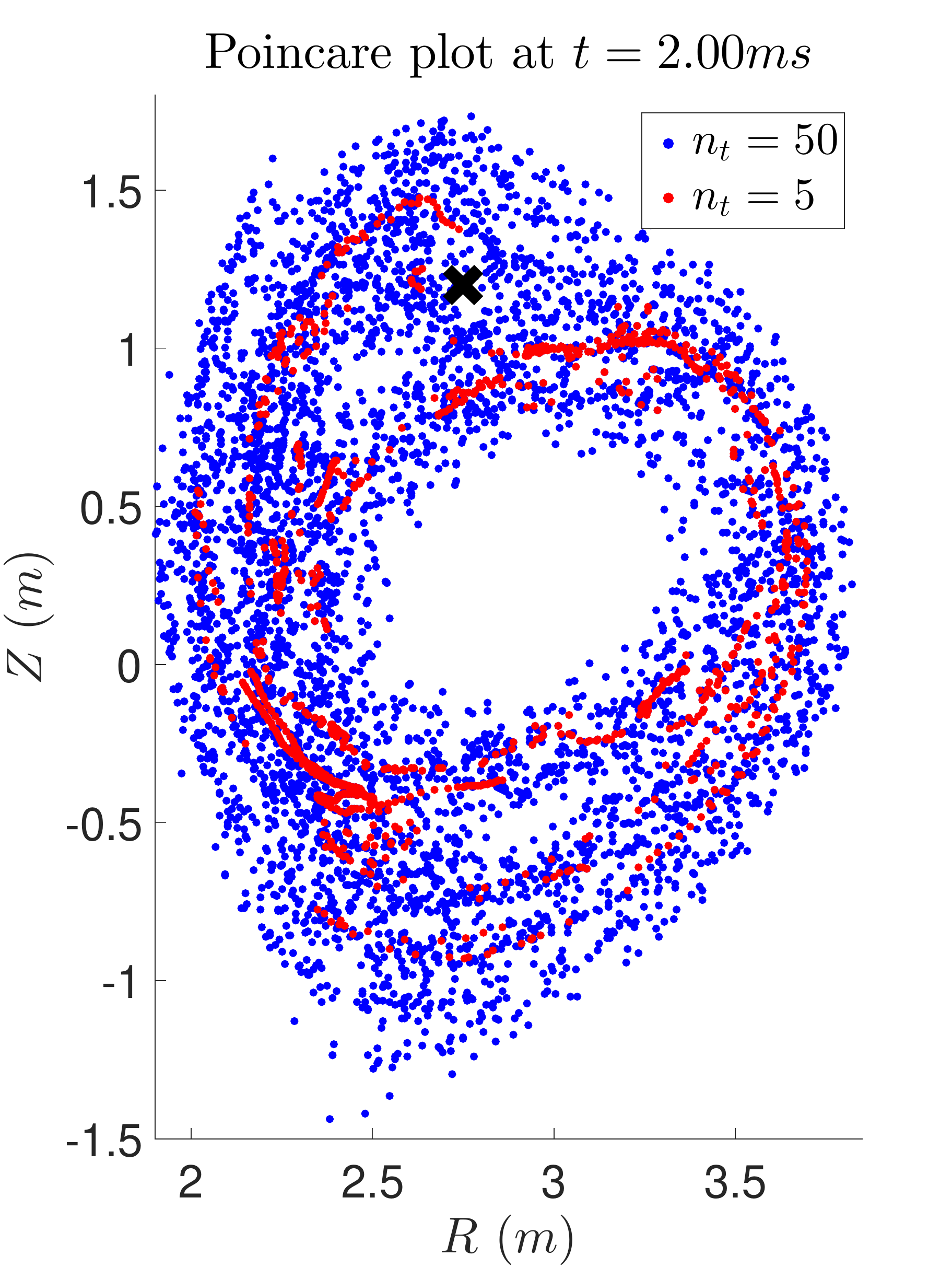}
}
&
\parbox{1.4in}{
	\includegraphics[scale=0.20]{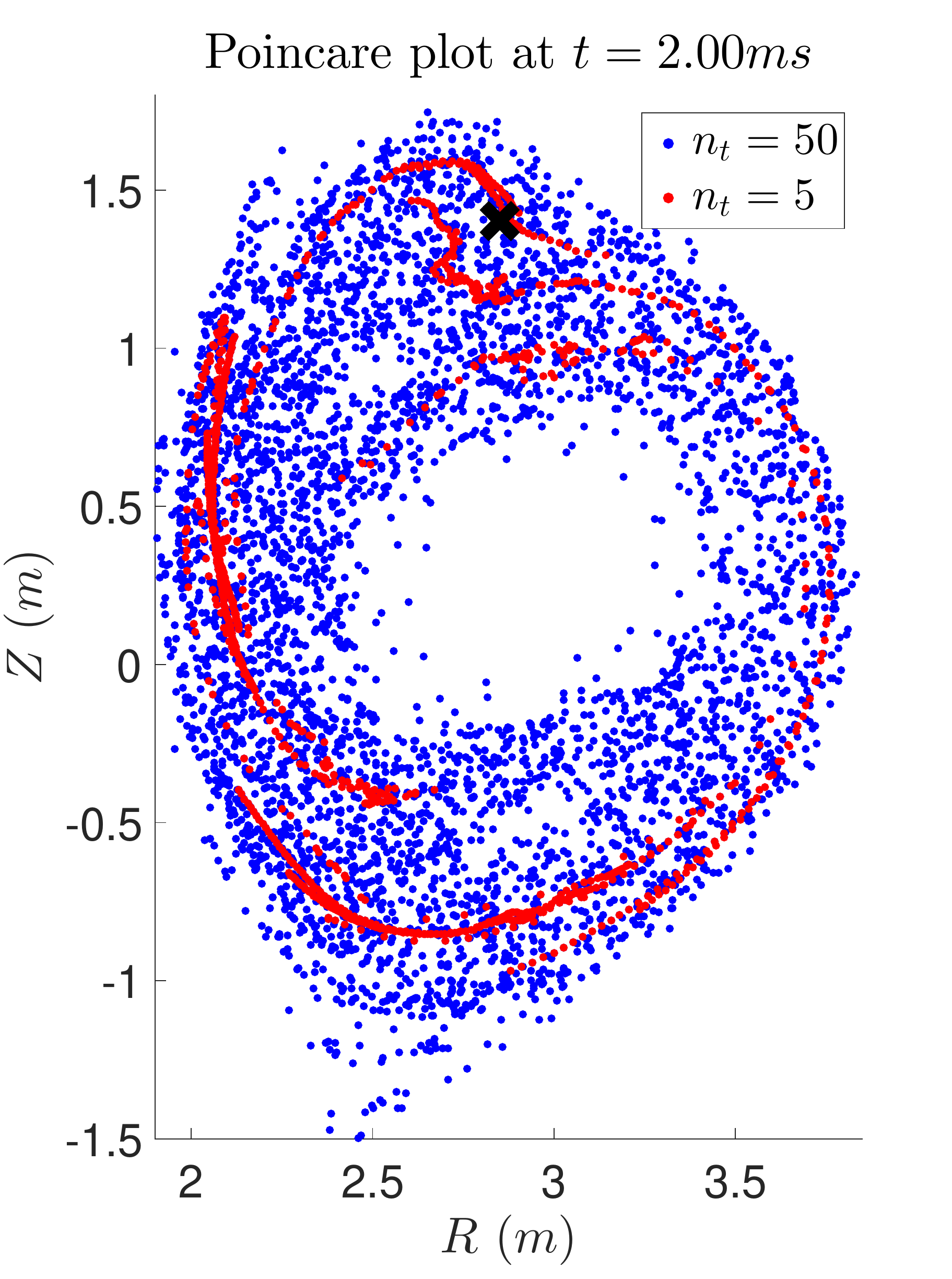}
}
\\
(a)&(b)&(c)
\\
\parbox{1.4in}{
	\includegraphics[scale=0.20]{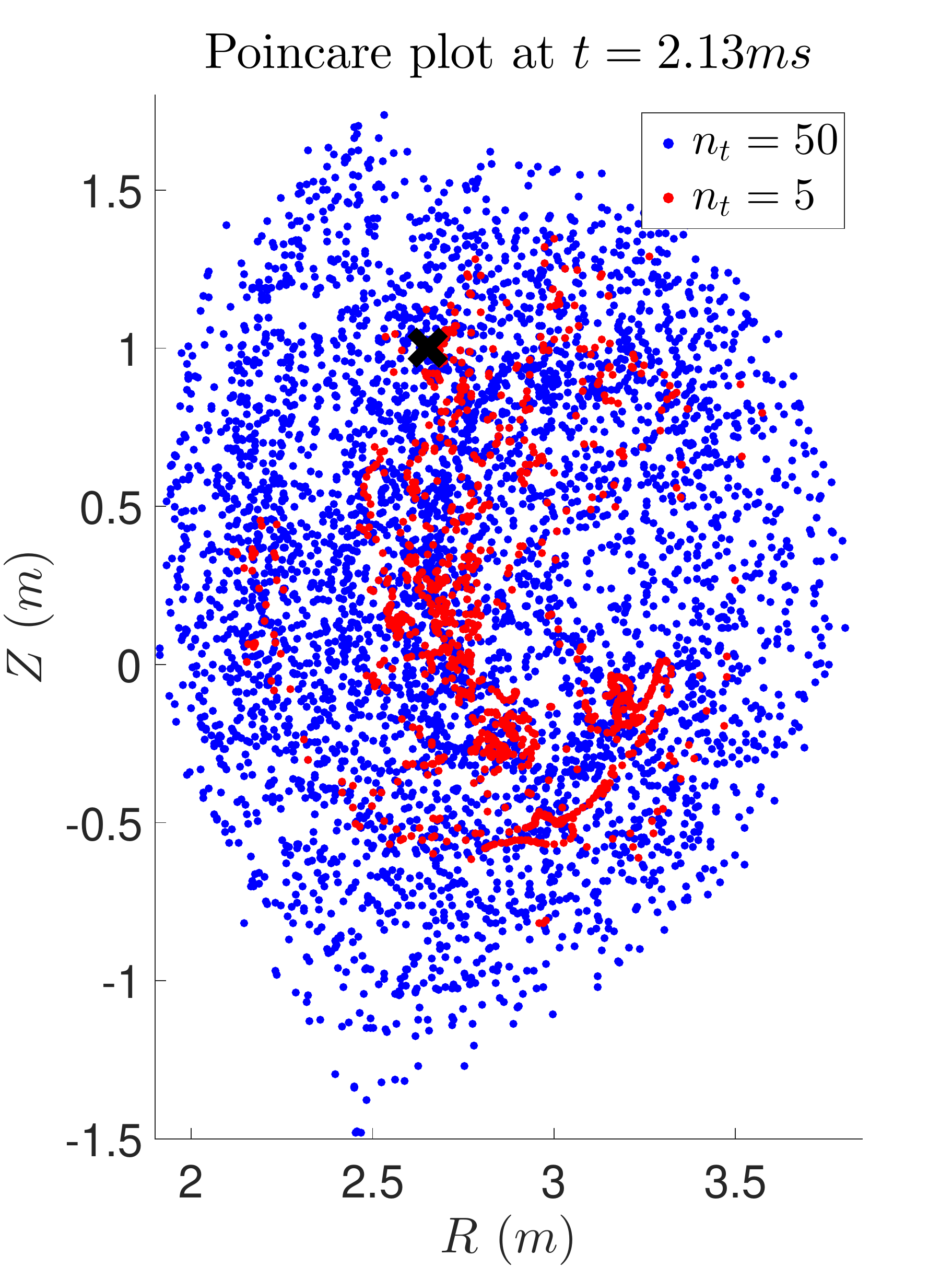}
}
&
\parbox{1.4in}{
	\includegraphics[scale=0.20]{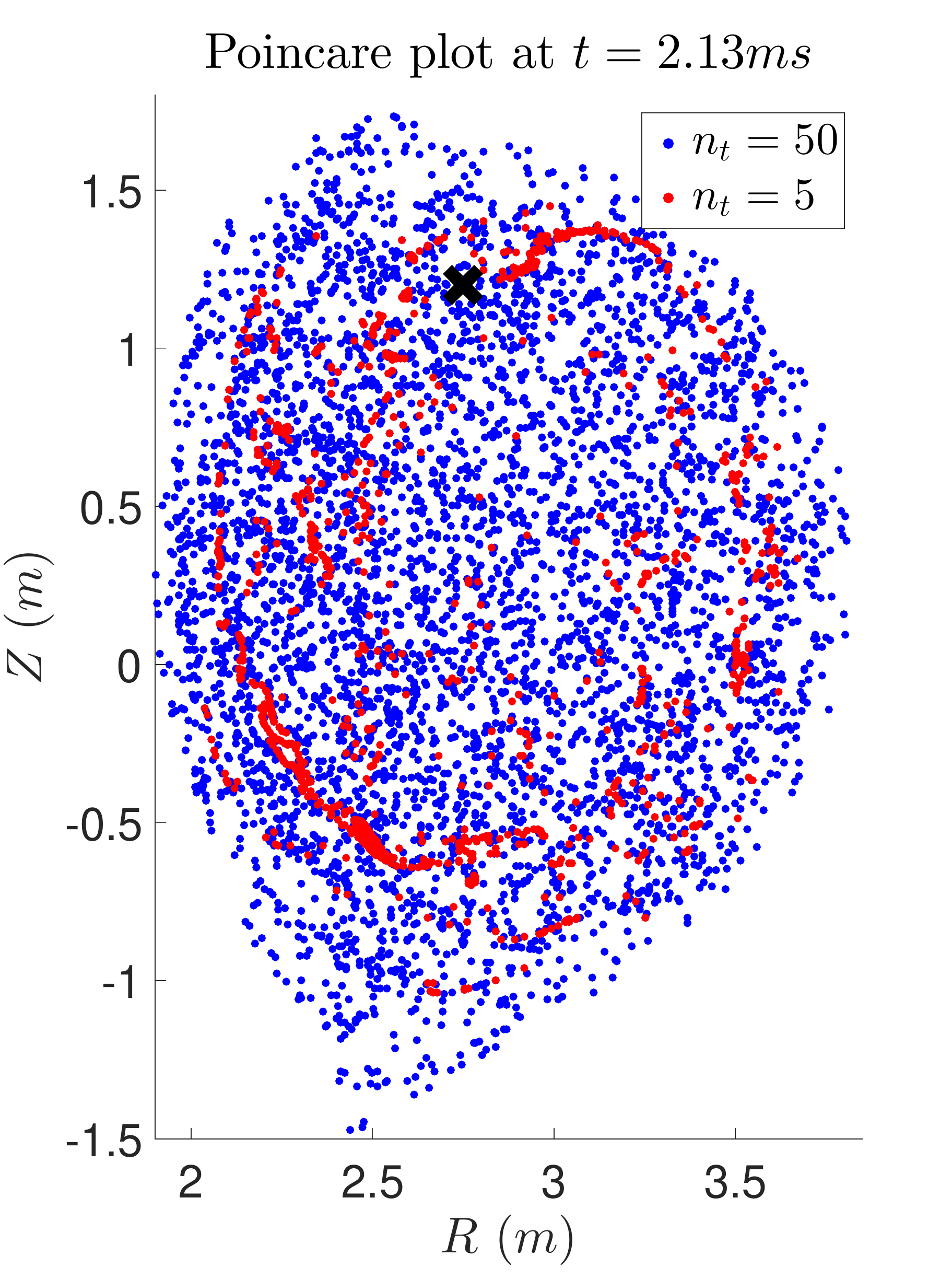}
}
&
\parbox{1.4in}{
	\includegraphics[scale=0.20]{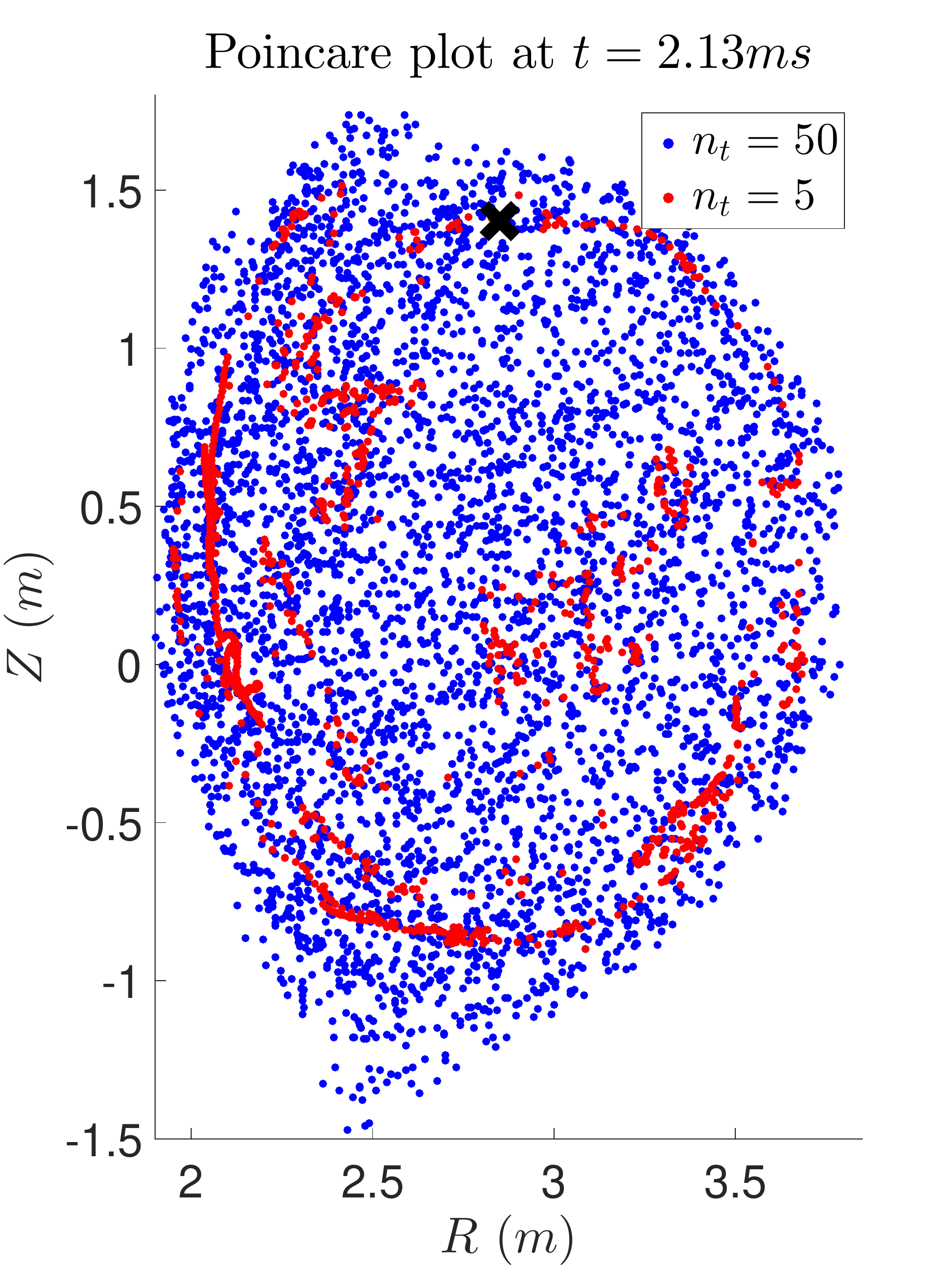}
}
\\
(d)&(e)&(f)
\etbl
\caption{The Poincar\'{e} plot of magnetic field line for ``JET shot 1'' at two different times and three different origin regions. The red points indicate the result after tracking the field lines for 5 toroidal turns, while the blue ones indicate that after 50 turns. The black X indicates the approximate origin position of the field line tracing.}
\label{fig:10}
\end{figure*}

It can be seen from Fig.\,\ref{fig:08} that there is more than $100\gm s$ delay between the core temperature collapse and the eventual core impurity mixing by MHD modes.
This delay between the temperature and density response could partly be attributed to the evolution of the stochasticity strength, as is shown in Fig.\,\ref{fig:10}, where the Poincar\'{e} plots of magnetic field at $t=2.00ms$ (onset of the TQ) and $t=2.13ms$ are compared. The red dots represent the result after tracing the field lines for 5 toroidal turns, while the blue ones represent that after tracing for 50 turns. The black crosses indicate the origin region, we begin field line tracing from several points within a small box around these crosses. It can be seen that at the time of the TQ onset, despite the global stochasticity as is shown in Fig.\,\ref{fig:06}(d), the stochastic core penetration from the edge is still limited, as even after 50 toroidal turns the field lines remain outside of the core for all three initial positions as are shown in Fig.\,\ref{fig:10}(a), (b) and (c). After $100ms$, however, the global stochasticity grows significantly, and the core region can be accessed within as few as 5 turns from the same three initial positions as are shown in Fig.\,\ref{fig:10}(d), (e) and (f) respectively. It is a few hundreds of microseconds after this time that we see significant core impurity penetration as is shown in Fig.\,\ref{fig:08}(h). Further, for the 5-turn-cases, it appears that the core accessibility of slightly inner initial positions could be much better than that of the outer initial positions, as can be seen comparing Fig.\,\ref{fig:10}(d) against Fig.\,\ref{fig:10}(e) and (f). We will discuss the significance of this property in the conclusion section.

As the electron density and the impurity density share a similar evolution, the period of delay between the temperature collapse and the density rise means the core hollow region would experience a high electric field but an only slightly increased electron and impurity density to stop runaway electrons. Such undesirable behavior provides incentive for more advanced SPI schemes, such as injecting hydrogen isotopes to mildly dilute the electron temperature before injecting the impurities \cite{Nardon2020NF}. On the other hand, the stochastic field line means the electrons would experience the averaged density throughout the stochastic region over time, mitigating the aforementioned detrimental effect. Hence more detailed analysis with JOREK runaway electron test particle model \cite{Sommariva2017NF} needs to be carried out to quantitatively determine the impact of this hollowed density region. That, however, is out of the scope of this paper.

\section{The plasma response and radiation asymmetry for ITER mono-SPI}
\label{s:RadAsyMono}

In this section, we will use both the single and the two temperature models described in Section \ref{ss:Equations} to investigate the profile evolution and radiation asymmetry after SPI from a single toroidal location. In Section \ref{ss:MonoSPIProfile}, we first consider the single temperature model to briefly show the general process of TQ triggering using ``ITER shot 1'', then compare it with ``ITER shot 2'' to demonstrate the importance of utilizing the favorable MHD mode structure in SPI core penetration. Later on, we will show the electron and ion temperature deviation during the TQ as well as the impurity radiation asymmetry after SPI using ``ITER shot 3'' with the two temperature model in Section \ref{ss:TempDevSpecies} and \ref{ss:MonoSPIAsym} respectively.

\subsection{The temperature collapse and density transport in ITER simulations}
\label{ss:MonoSPIProfile}

\begin{figure*}
\centering
\noindent
\btbl{c}
\parbox{4.5in}{
    \includegraphics[scale=0.35]{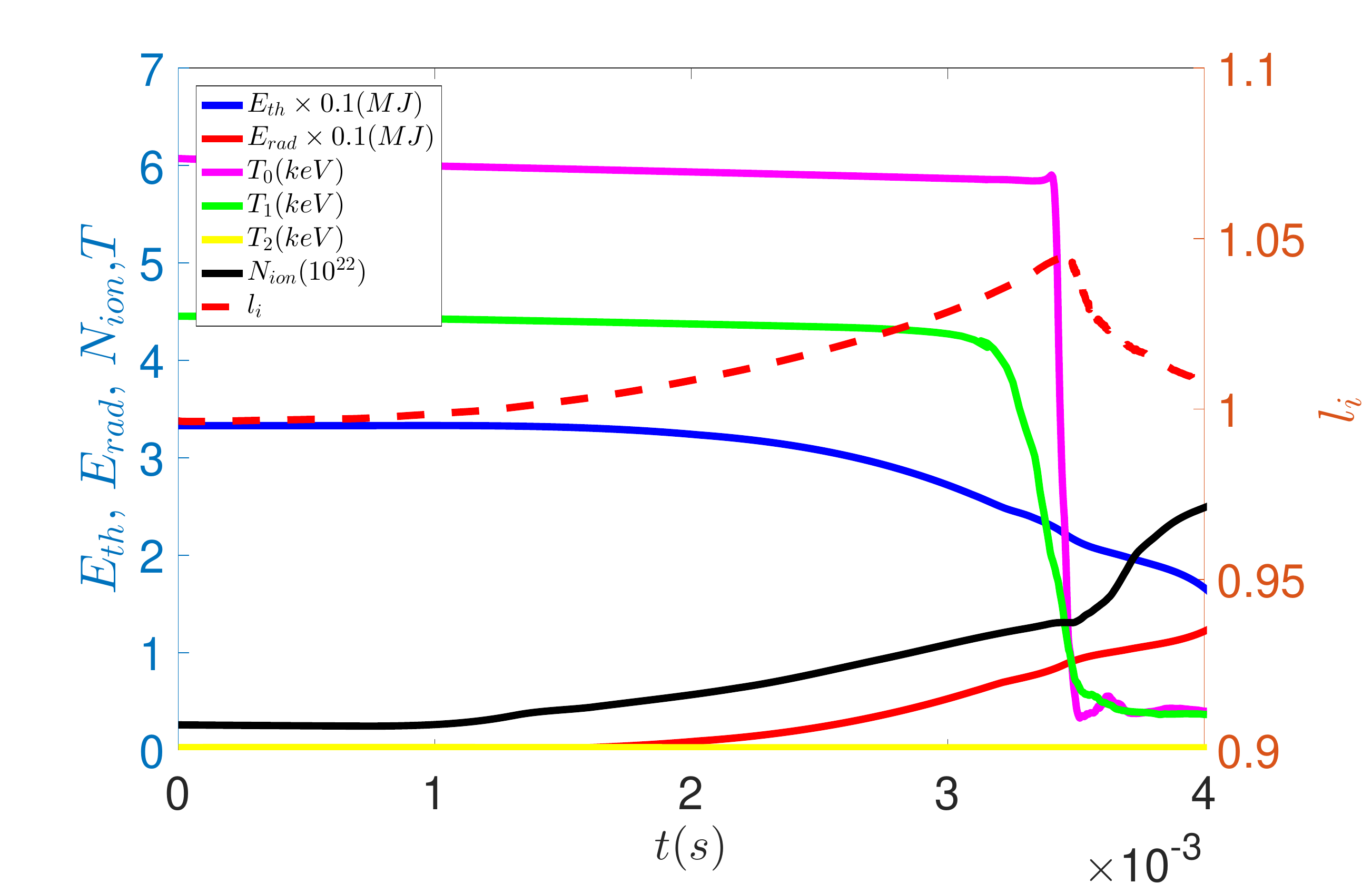}
}
\etbl
\caption{The ``ITER shot 1'' evolution of the thermal energy content $E_{th}$, radiated energy $E_{rad}$, total ion particle content $N_{ion}$, core temperature $T_0$ ($R=6.22m$ on mid-plane), edge temperature $T_2$ ($R=8.20m$ on mid-plane) and the temperature halfway between those two $T_1$ ($R=7.21m$ on mid-plane), and the internal inductance $l_i$ (the dashed red line). The TQ is approximately triggered at $t=3.50ms$.}
\label{fig:11}
\end{figure*}

\begin{figure*}
\centering
\noindent
\btbl{cc}
\parbox{2.8in}{
    \includegraphics[scale=0.32]{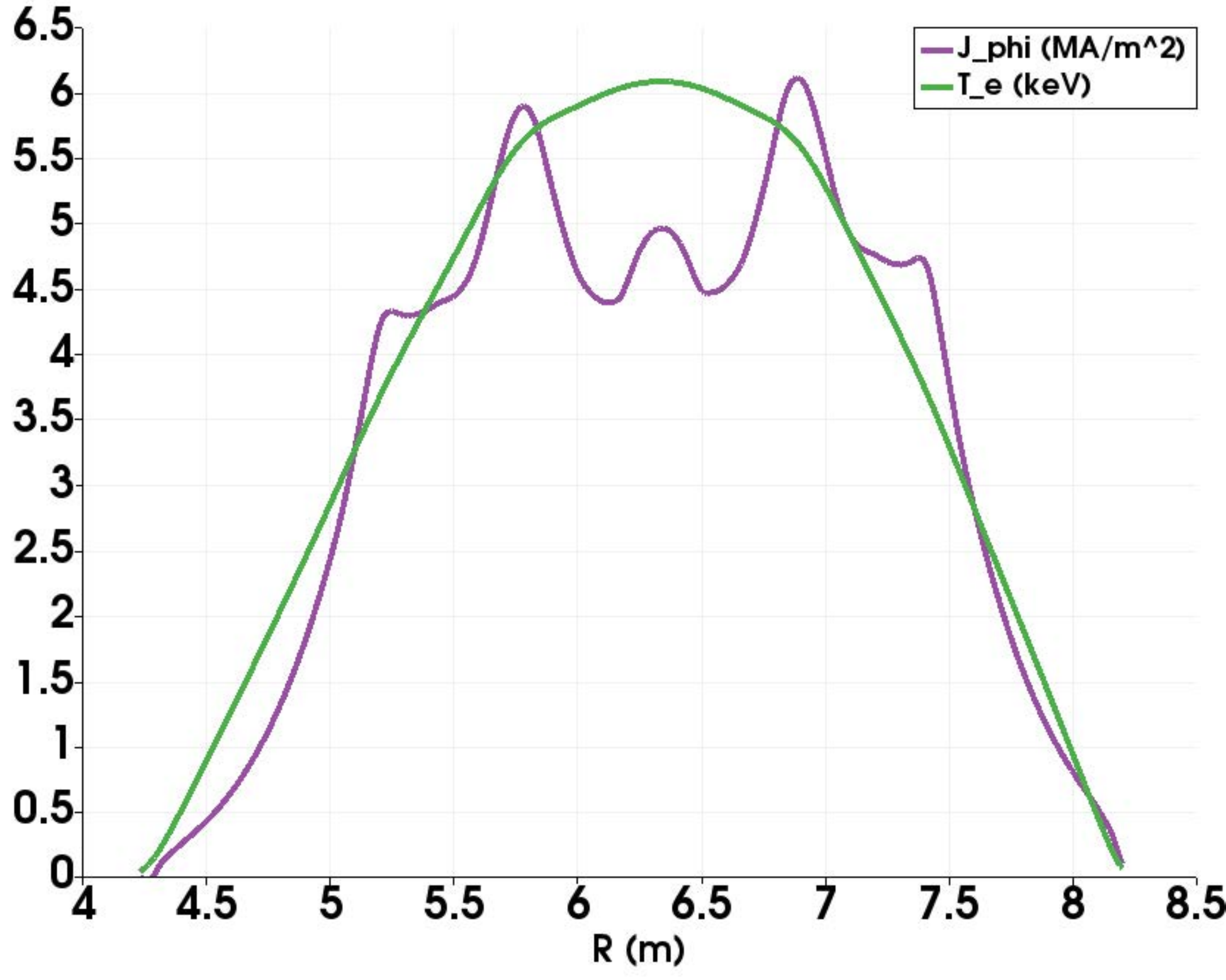}
}
&
\parbox{2.8in}{
	\includegraphics[scale=0.32]{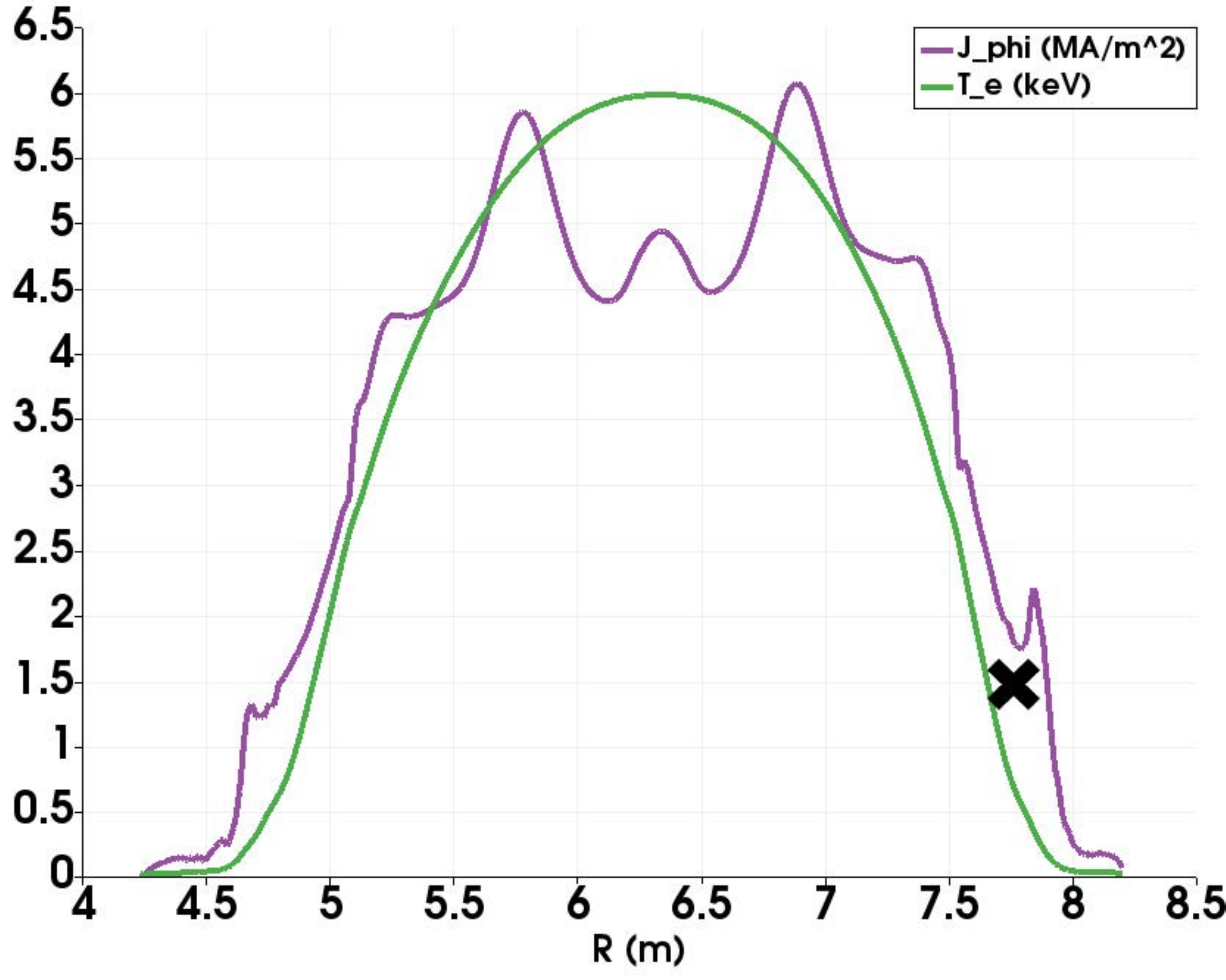}
}
\\
(a)&(b)
\etbl
\caption{The $n=0$ temperature and the $n=0$ current density profile at (a) $t=0ms$ and (b) $t=2.64$. There exists some current contraction, but it does not follow the cold front as closely as the full argon case. The black cross corresponds to the approximate position of the vanguard fragments.}
\label{fig:12}
\end{figure*}

The evolution of several plasma parameters for ``ITER shot 1'' is shown in Fig.\,\ref{fig:11} as a demonstration of the general sequence of events during our ITER SPI simulations. The fragments arrive at the plasma approximately at $t=0.8ms$, creating a slight current contraction as shown by the internal inductance $l_i$ going up, until the triggering of the TQ marked by the core temperature collapse and the relaxation of the current density profile as shown by $l_i$ going down. However, the plasma boundary temperature $T_2$ did not change drastically over the TQ, implying no significant heat flux have reached the plasma boundary. This could partly because of our ideal wall boundary condition which suppresses the boundary stochastic field. The impurity radiation power did not immediately jump up, however. This could be due to the difference between small impurity fraction mixed species SPI and full impurity SPI, and is in agreement with previous NIMROD result where mixed species SPI is found to cause a milder radiation spike compared with the full impurity case \cite{Kim2019POP}. This is also in accordance with experimental results showing the radiated energy dependence on the impurity mixture ratio \cite{Shiraki2016POP}.
The current profile response to the mixed species SPI show some similar axisymmetric contraction as is shown in Fig.\,\ref{fig:12}, but it is not following the flight of the fragments as closely as in the previous full argon case as shown in Fig.\,\ref{fig:05}. Later on in Section \ref{ss:SymmDualSPI} we will also see that the helical cooling plays a significant role in the MHD response of the mixed species SPI, hence both mechanisms are present in our ITER simulations.

\begin{figure*}
\centering
\noindent
\btbl{ccc}
\parbox{1.45in}{
    \includegraphics[scale=0.25]{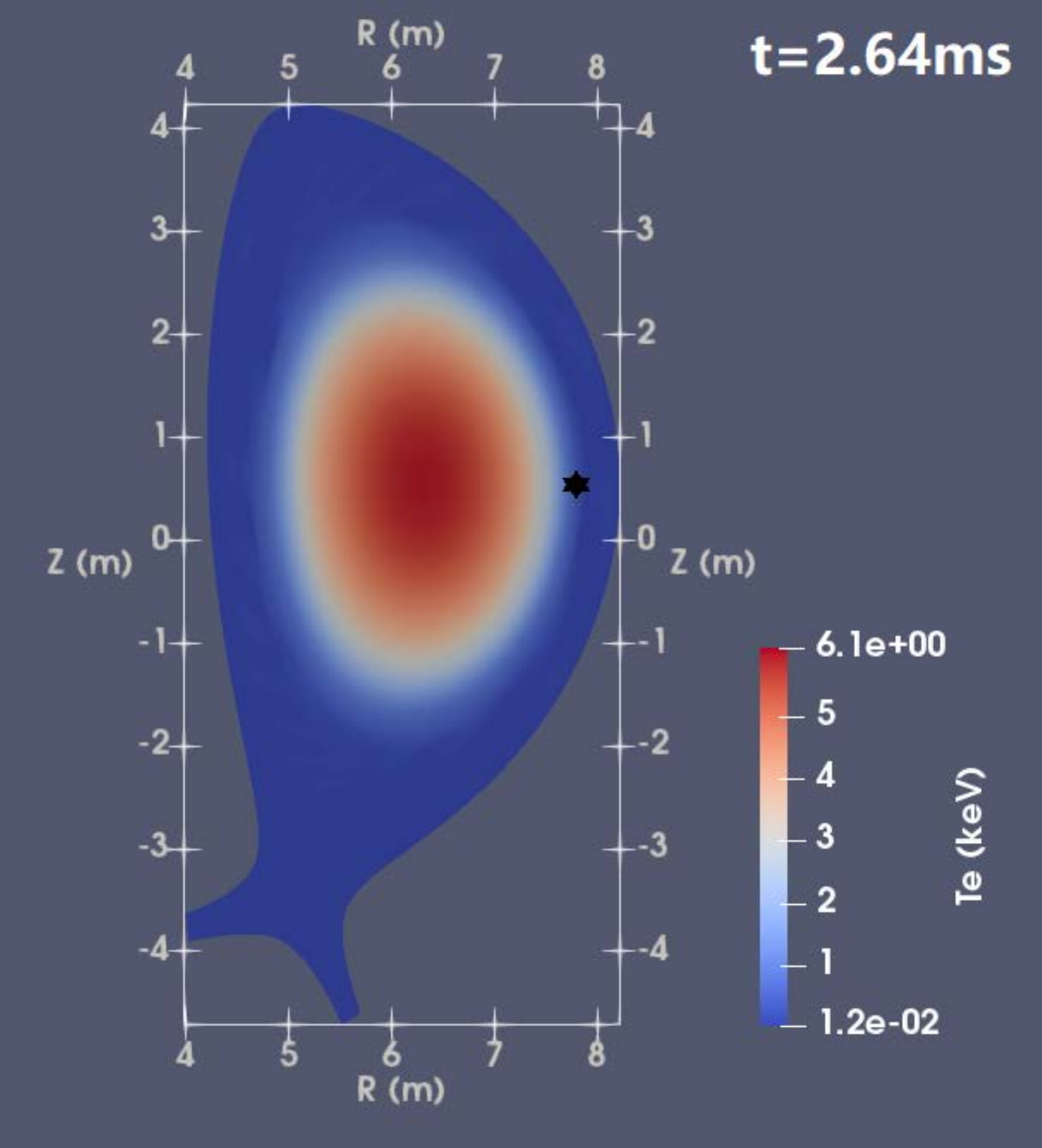}
}
&
\parbox{1.45in}{
	\includegraphics[scale=0.25]{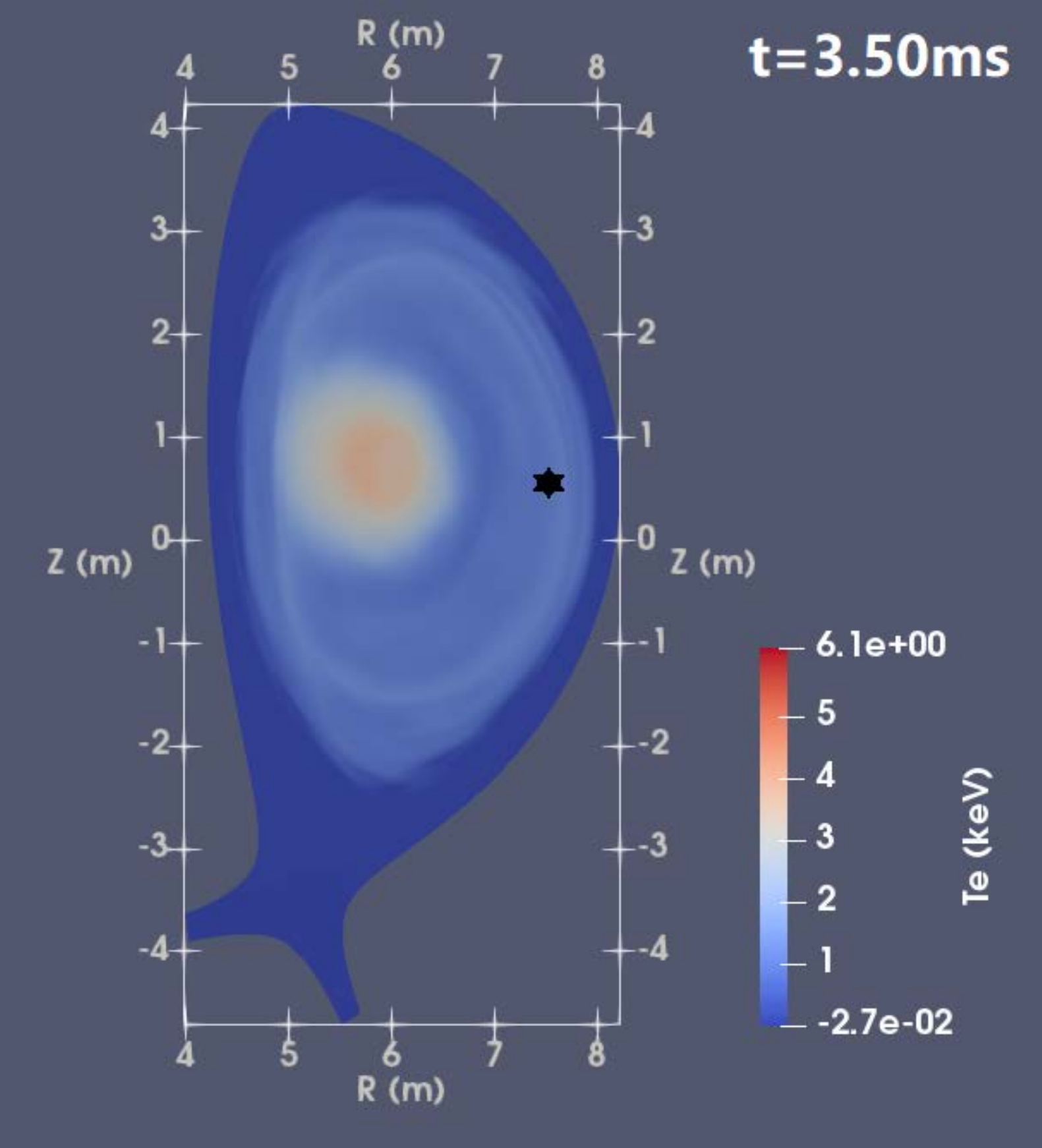}
}
&
\parbox{1.45in}{
	\includegraphics[scale=0.25]{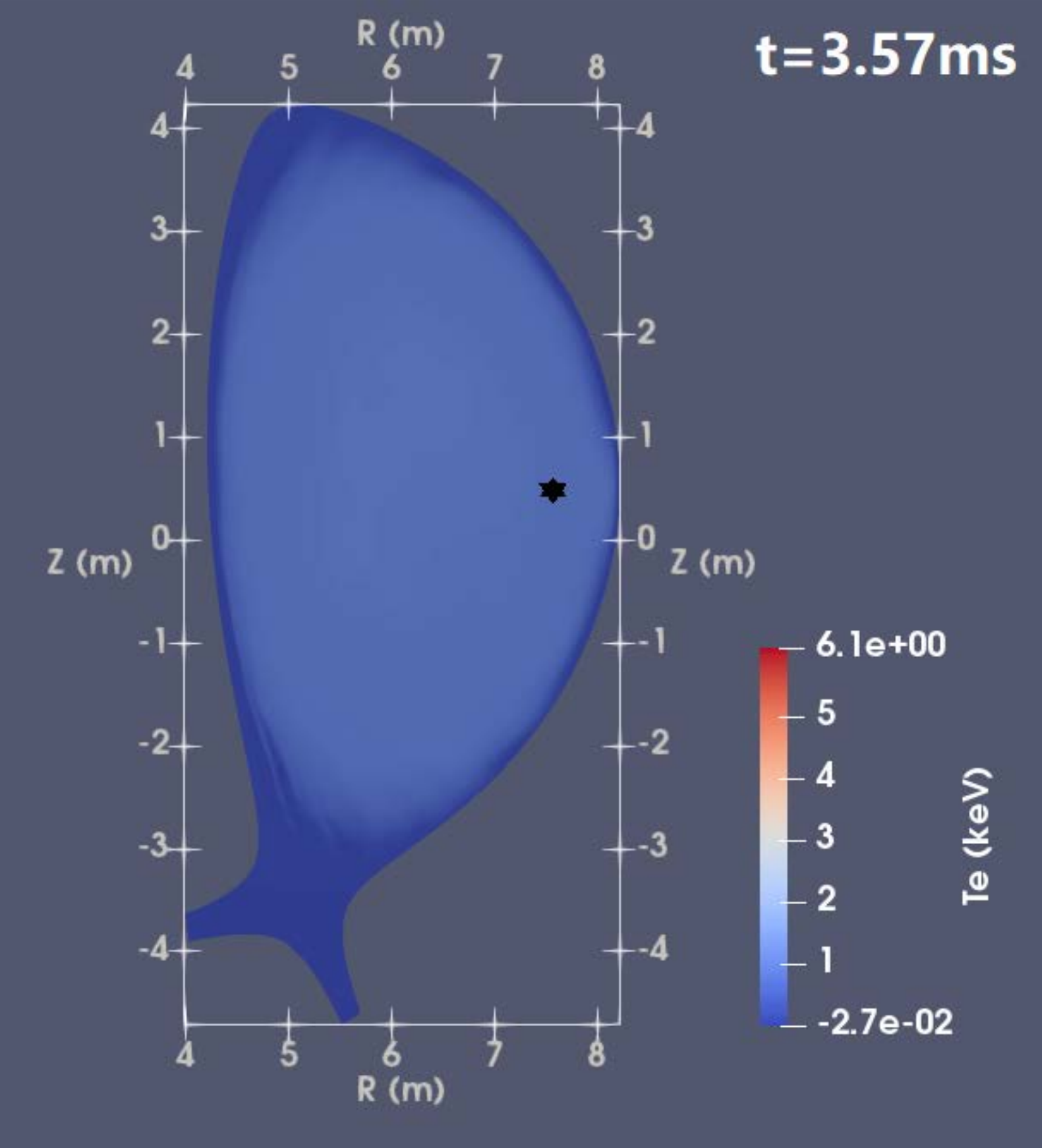}
}
\\
(a)&(b)&(c)
\etbl
\caption{The electron temperature profile for ``ITER shot 1'' at time (a) $2.64ms$, (b) $3.50ms$ (During the TQ) \& (c) $3.57ms$. The black stars mark the approximate position of the vanguard fragments.}
\label{fig:13}
\end{figure*}

Since the safety factor is around unity over a large portion of the radial domain and features two $q=1$ surfaces due to the non-monotonic profile, the $1/1$ resistive kink mode has a prominent role in the TQ dynamics. Indeed, the final core collapse, which marks the beginning of the TQ is triggered by the vanguard fragments entering the $q=1$ surface. As is shown in Fig.\,\ref{fig:13}, at the time of the TQ, apart from the diffusive confinement loss due to field line stochasticity, there is observable $1/1$ kink motion away from the position of the fragments, indicating the $1/1$ mode O-point coincides with the fragment position, signalling that the helical effect on the $q=1$ surface plays an important role in triggering the core collapse.

\begin{figure*}
\centering
\noindent
\btbl{ccc}
\parbox{1.4in}{
    \includegraphics[scale=0.2475]{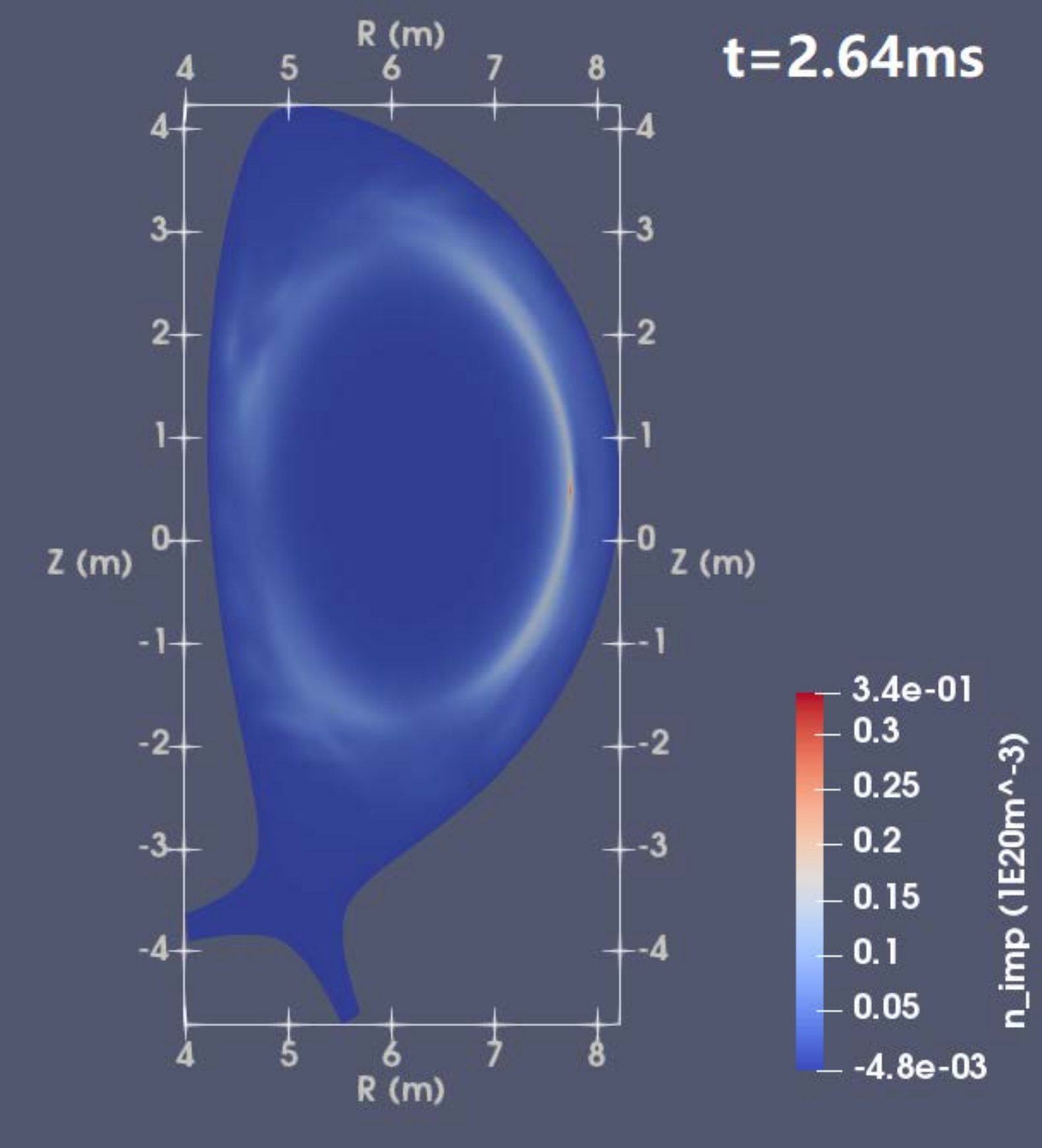}
}
&
\parbox{1.4in}{
	\includegraphics[scale=0.2475]{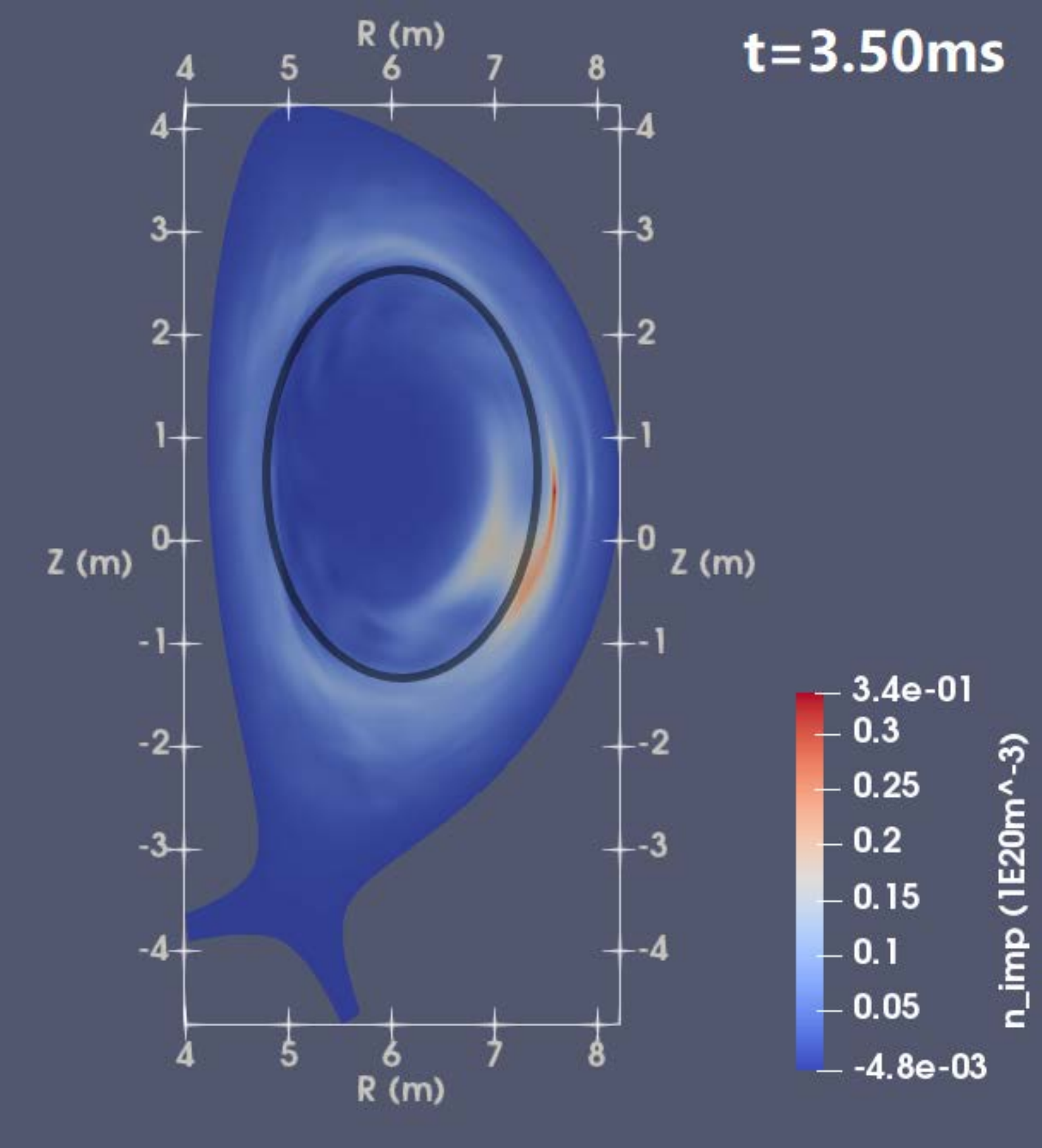}
}
&
\parbox{1.4in}{
	\includegraphics[scale=0.2475]{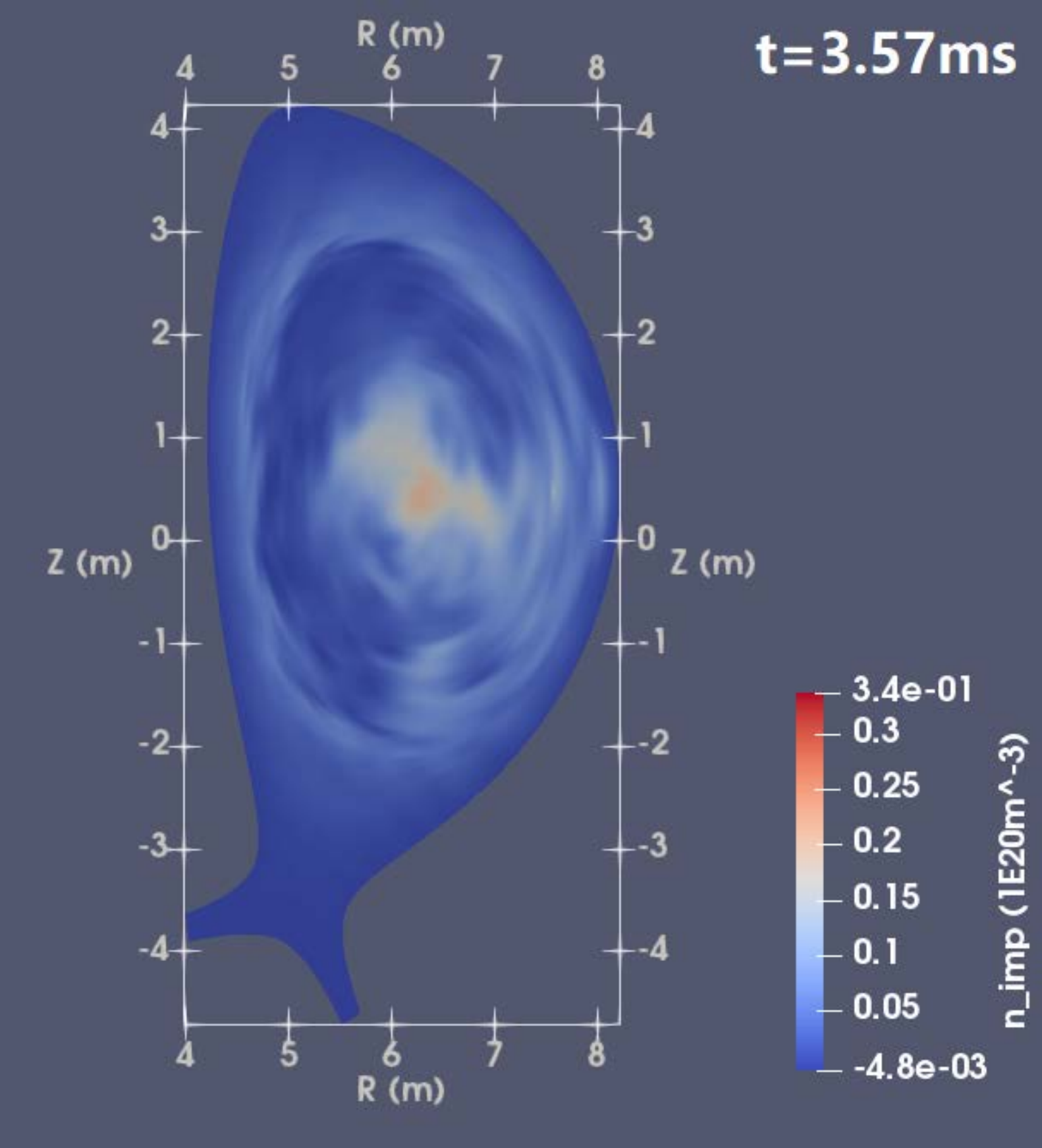}
}
\\
(a)&(b)&(c)
\\
\parbox{1.4in}{
	\includegraphics[scale=0.2475]{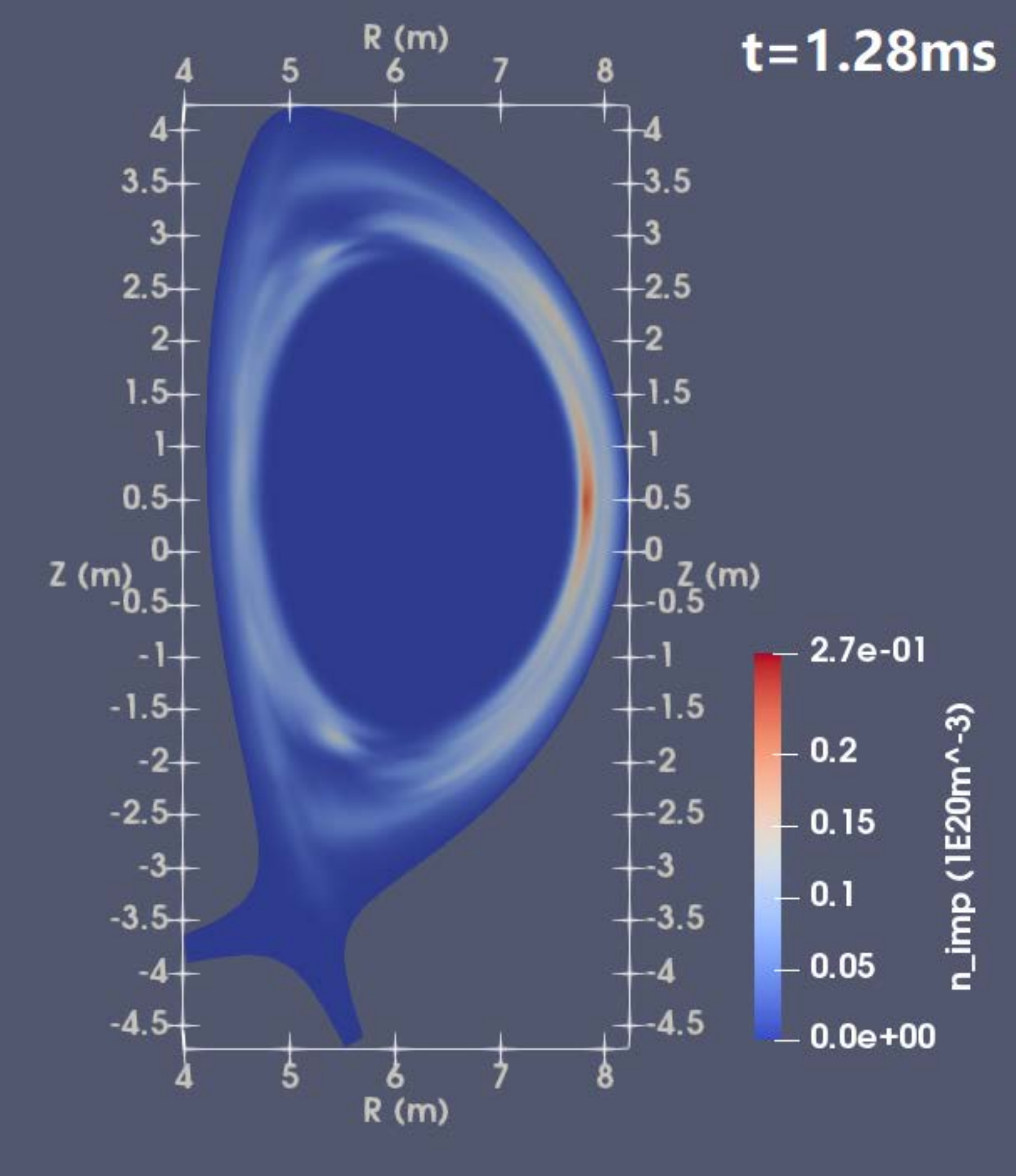}
}
&
\parbox{1.4in}{
	\includegraphics[scale=0.2475]{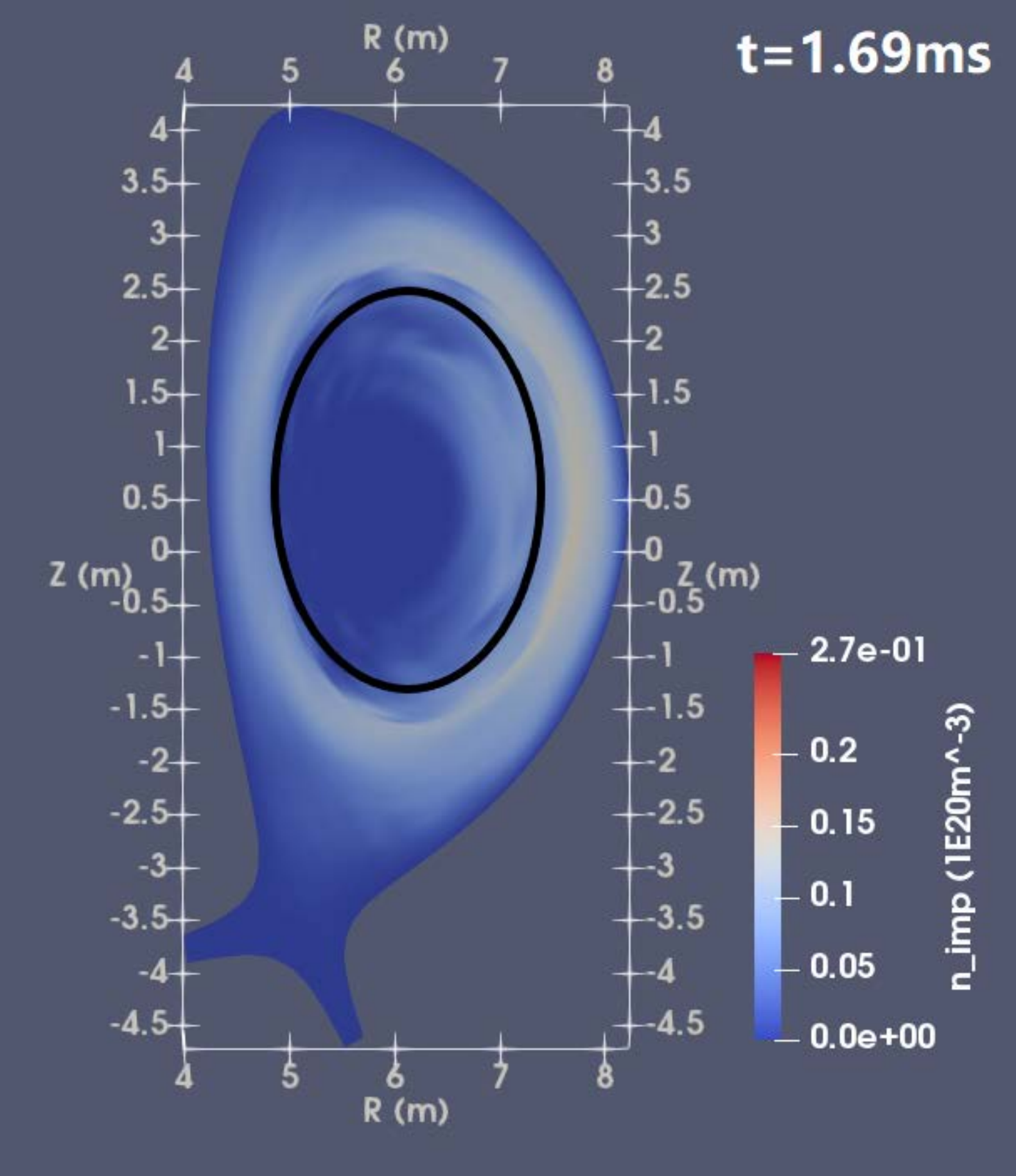}
}
&
\parbox{1.4in}{
	\includegraphics[scale=0.2475]{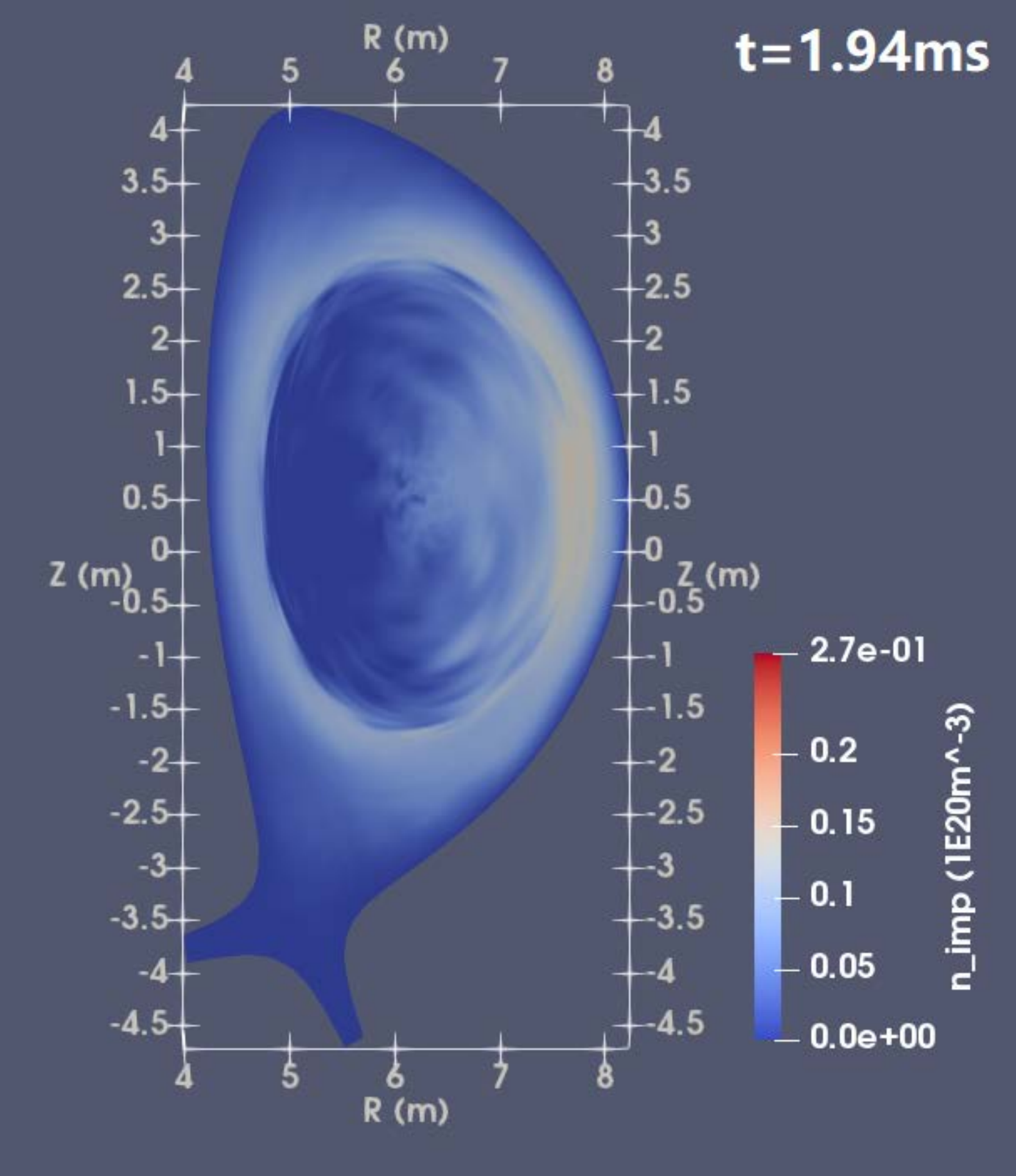}
}
\\
(d)&(e)&(f)
\etbl
\caption{The upper figures show the neon number density distribution (a) before the TQ at $t=2.64ms$, (b) at the time of TQ at $t=3.50ms$ (during the TQ) and (c) after the TQ at $t=3.57ms$ for ``ITER shot 1'' with mixed neon and hydrogen SPI. The lower figures show the neon number density distribution (d) before the TQ at $t=1.28ms$, (e) at the time of TQ at $t=1.69ms$ (during the TQ) and (f) after the TQ at $t=1.94ms$ for ``ITER shot 2'' with the same neon content but no hydrogen mixing. The black circles in (b) and (e) mark the approximate position of the $q=1$ surface.}
\label{fig:14}
\end{figure*}

We will now further show that utilizing this $1/1$ kink motion can be very beneficial to the core penetration of the injected material. For this purpose, we compare the neon transport of ``ITER shot 1'' with that of ``ITER shot 2''. The latter has exactly the same impurity content with the former, but without hydrogen in addition. As a consequence, more neon would be deposited on the edge area of the plasma compared with the ``ITER shot 1'' case according to our ablation model Eq.\,(\rfq{eq:NeD2AblRate})-(\rfq{eq:D2AblFrac}). Indeed, in the ``ITER shot 2'' case, the fragments barely reach the $q=1$ surface before being entirely ablated, depositing only a very small fraction of their content near or within the $q=1$ surface, as opposed to the ``ITER shot 1'' case, where a significant amount of neon is deposited close to the $q=1$ surface. This has substantial impact on the impurity transport as shown in Fig.\,\ref{fig:14}.

Both in Fig.\,\ref{fig:14}(b) and Fig.\,\ref{fig:14}(e), $1/1$ kink motion can be seen at the time of the TQ via the hollow region moving away from the fragment location towards the high field side. However, the core impurity transport for ``ITER shot 1'' case is remarkably better than that for ``ITER shot 2'' case. It should be noted that at the time of Fig.\,\ref{fig:14}(c) and Fig.\,\ref{fig:14}(f), the fragments (if they remain) are still far away from the axis. Such difference is the result of injection deposition relative to the mode structure. As the $1/1$ kink shows a broad displacement mode structure within the $q=1$ surface, particles deposited on or within that surface are expected to be carried deep into the core by mode convection. Otherwise, the dominant mode can not ``see'' the particles and the core penetration is not enhanced.
Hence we emphasize here the importance of utilizing the favorable mode structure for core penetration during SPI. Although such benefit is most obvious for the $1/1$ mode as we have shown, we will argue that such benefit can also be enjoyed for $m>1$ modes in the conclusion section, except that the exact physics differs from what we described above.

\subsection{The temperature deviation between the species}
\label{ss:TempDevSpecies}

\begin{figure*}
\centering
\noindent
\btbl{ccc}
\parbox{1.4in}{
    \includegraphics[scale=0.165]{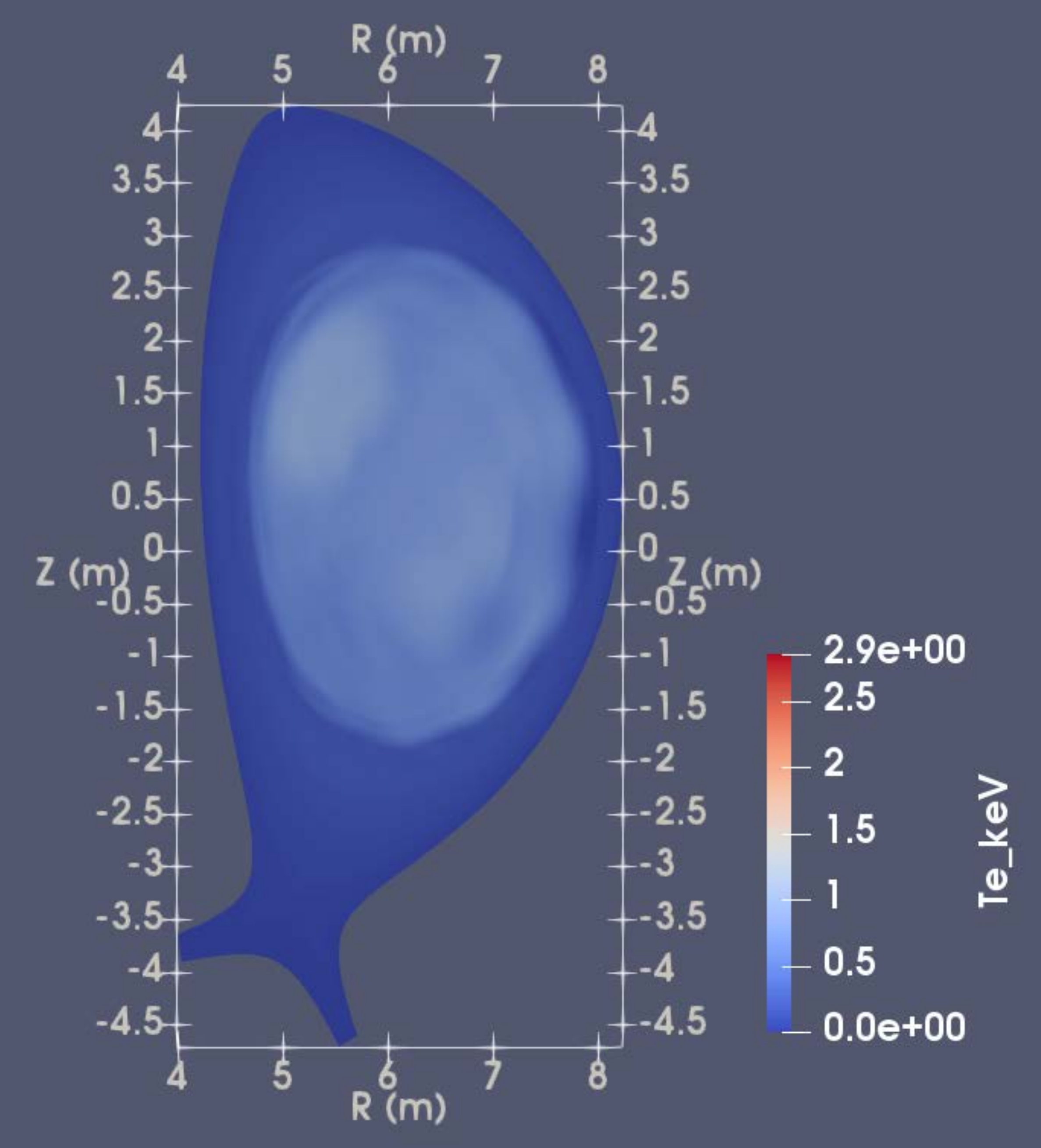}
}
&
\parbox{1.4in}{
	\includegraphics[scale=0.165]{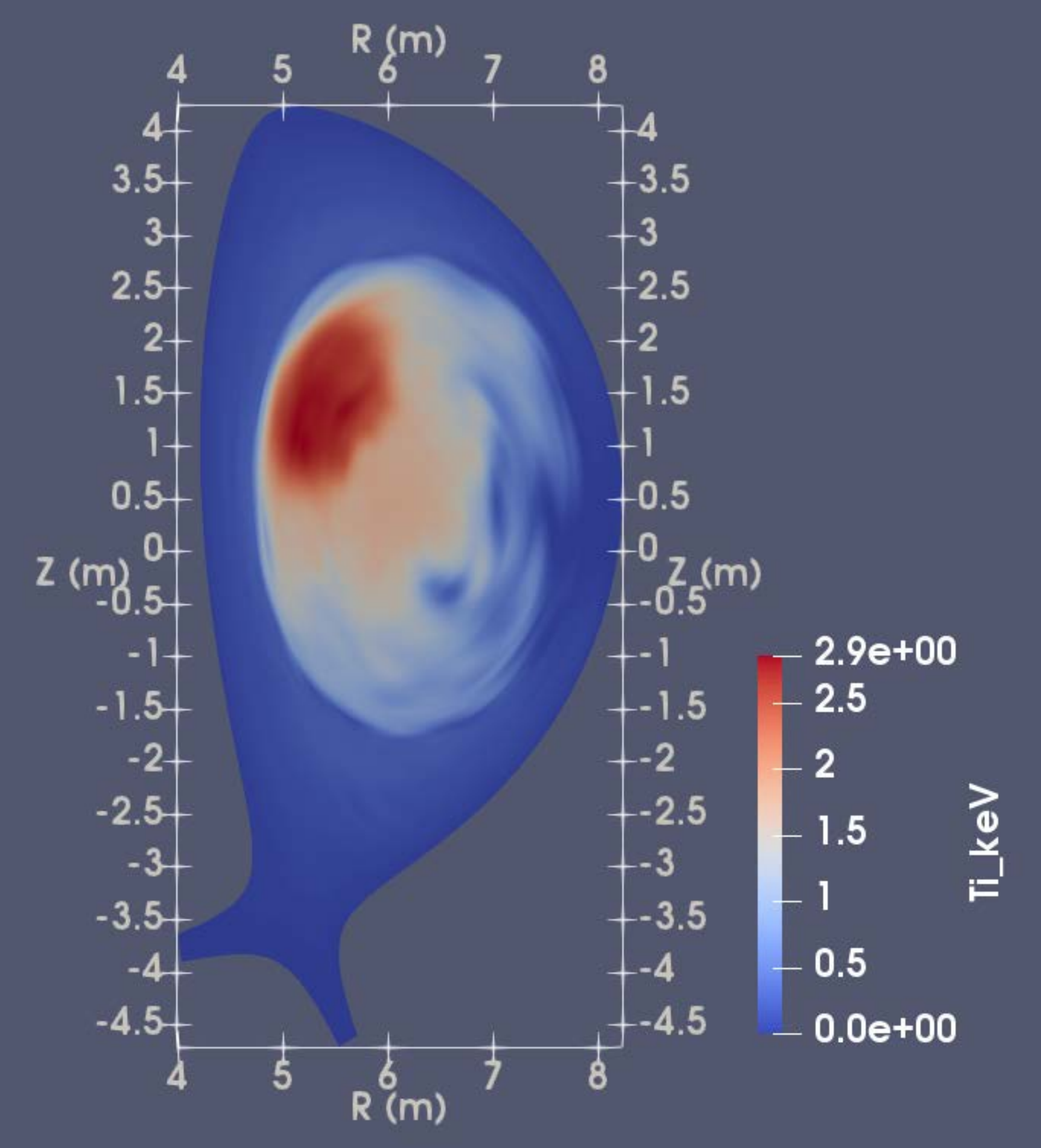}
}
&
\parbox{1.4in}{
	\includegraphics[scale=0.165]{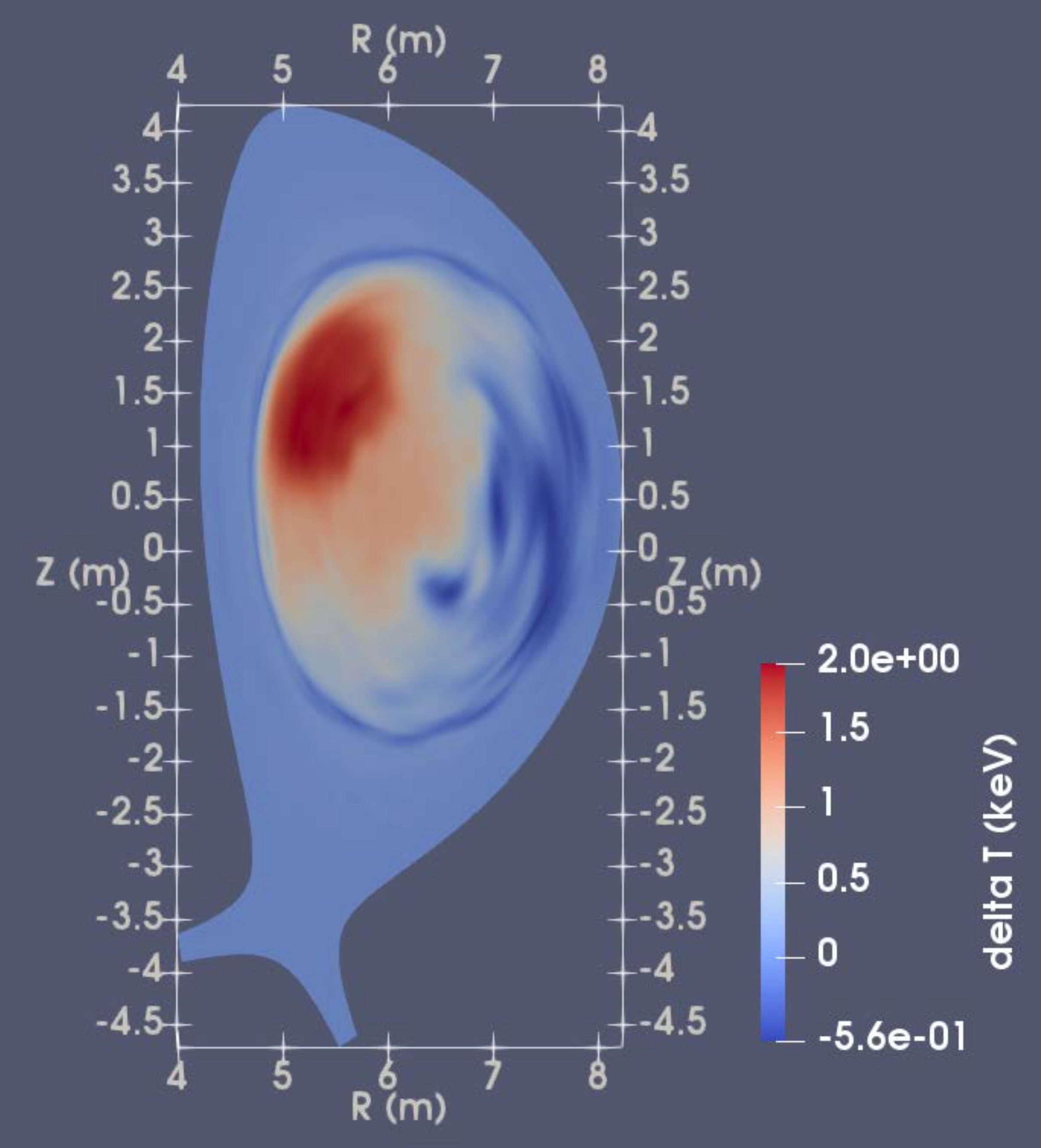}
}
\\
(a)&(b)&(c)
\etbl
\caption{The (a) electron temperature, (b) ion temperature and (c) their temperature difference during the TQ at $t=4.56ms$ for ``ITER shot 3''. }
\label{fig:15}
\end{figure*}

In this section, we discuss the temperature difference between the electron and ion species during SPI. We consider the ``ITER shot 3'' case with the two temperature model. As both the Braginskii parallel thermal conduction \cite{Braginskii1965RPP} and the free streaming thermal velocity scale like the inverse of the square root of the particle mass, it can be expected that the transport coefficients of the electron and ion temperature differ by a factor of the square root of the inverse mass ratio. During rapid confinement loss the different transport time scales of the species result in a strong decoupling of electron and ion temperatures as seen in Fig.\,\ref{fig:15}. Note that at this time the whole plasma is already stochastic, although the level of stochasticity may not be very strong in the core as demonstrated by the JET case shown in Fig.\,\ref{fig:10}.

At the time of the core collapse, the electron temperature profile is strongly flattened by the parallel conduction as shown in Fig.\,\ref{fig:15}(a), and only a vague shape of the kinking plasma remains. For the ion temperature however, the kinking motion is easily observable as it experiences much weaker conductive flattening. At this instance, the temperature difference between the two can be as large as $2keV$ in the core as is shown in Fig.\,\ref{fig:15}(c).

\subsection{The radiation asymmetry}
\label{ss:MonoSPIAsym}

Ideally, we would like the impurity radiation power density to be as uniform as possible after the injection, such that heat fluxes be distributed uniformly onto the PFCs. However, it will be seen that such uniform radiation is hard to achieve with a single SPI location, as the combination of unrelaxed impurity density distribution and the asymmetric outgoing heat flux can easily result in a toroidal radiation peaking factor $\baP_{rad}$ larger than 2 during the TQ when the radiation power is strongest. Here, the toroidal radiation peaking factor $\baP_{rad}$ is defined as the integrated radiation power within the poloidal planes normalized by the average radiation power.

\begin{figure*}
\centering
\noindent
\btbl{ccc}
\parbox{1.4in}{
    \includegraphics[scale=0.213]{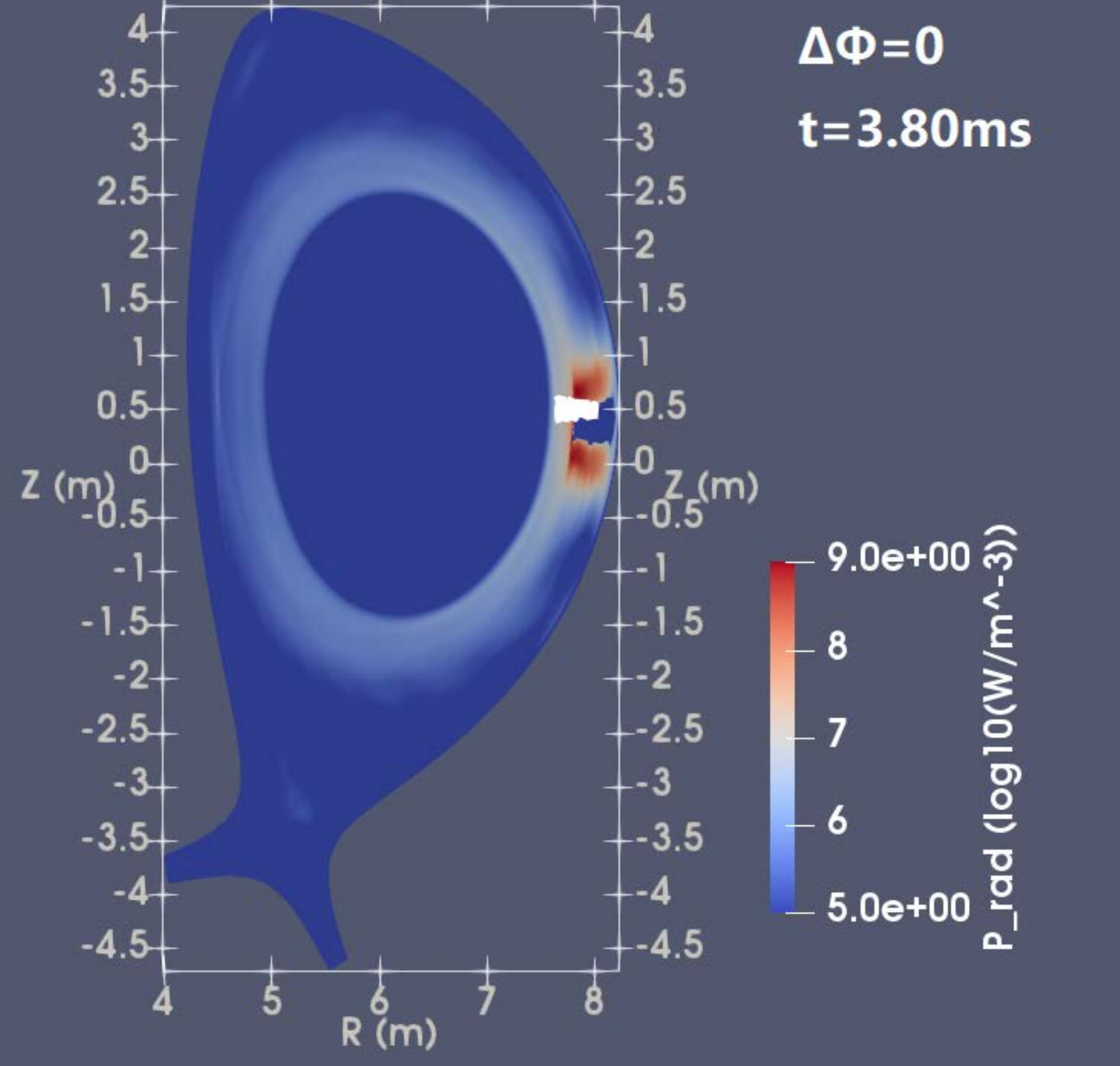}
}
&
\parbox{1.4in}{
	\includegraphics[scale=0.213]{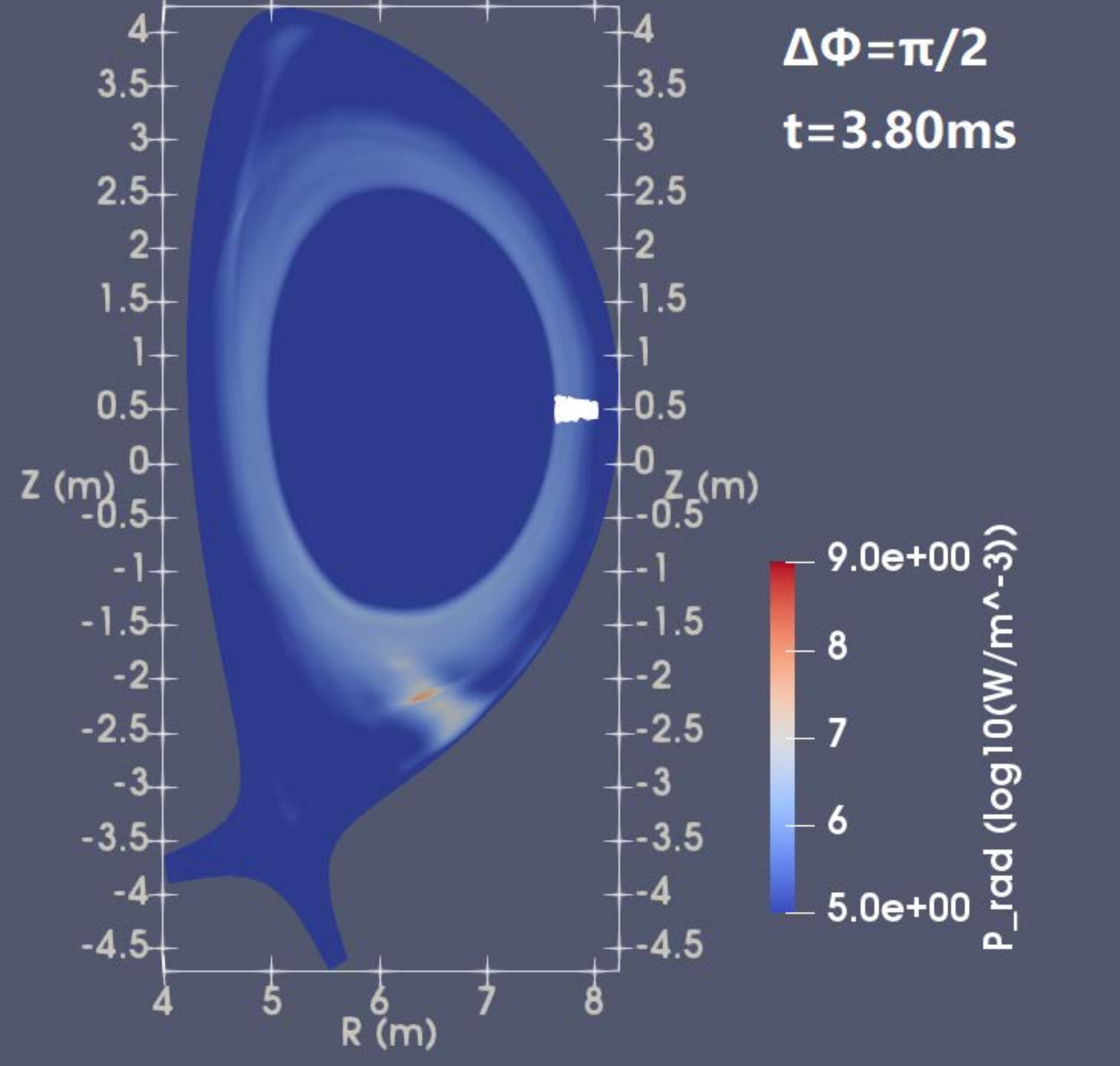}
}
&
\parbox{1.4in}{
	\includegraphics[scale=0.213]{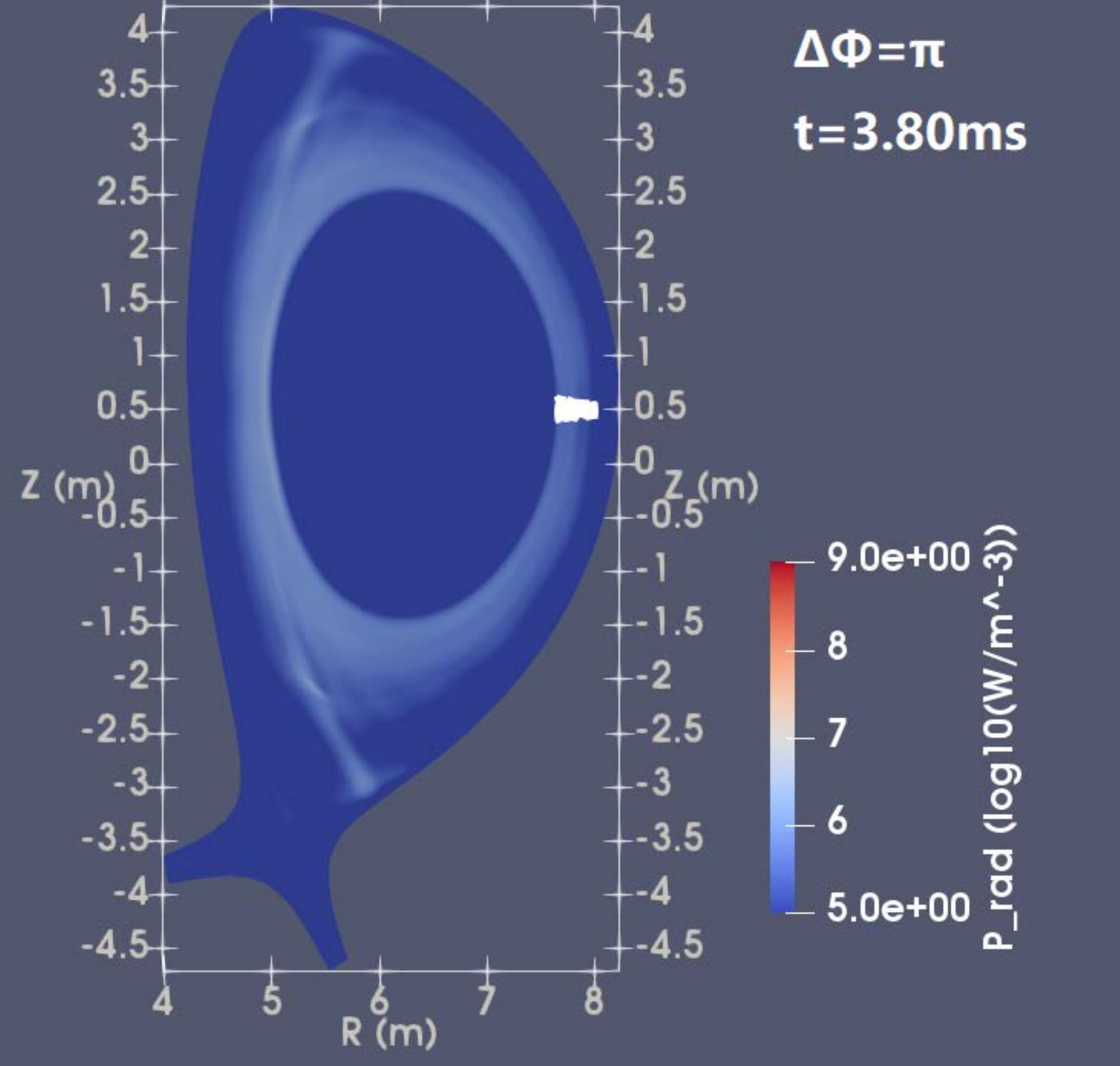}
}
\\
(a)&(b)&(c)
\\
\parbox{1.4in}{
	\includegraphics[scale=0.213]{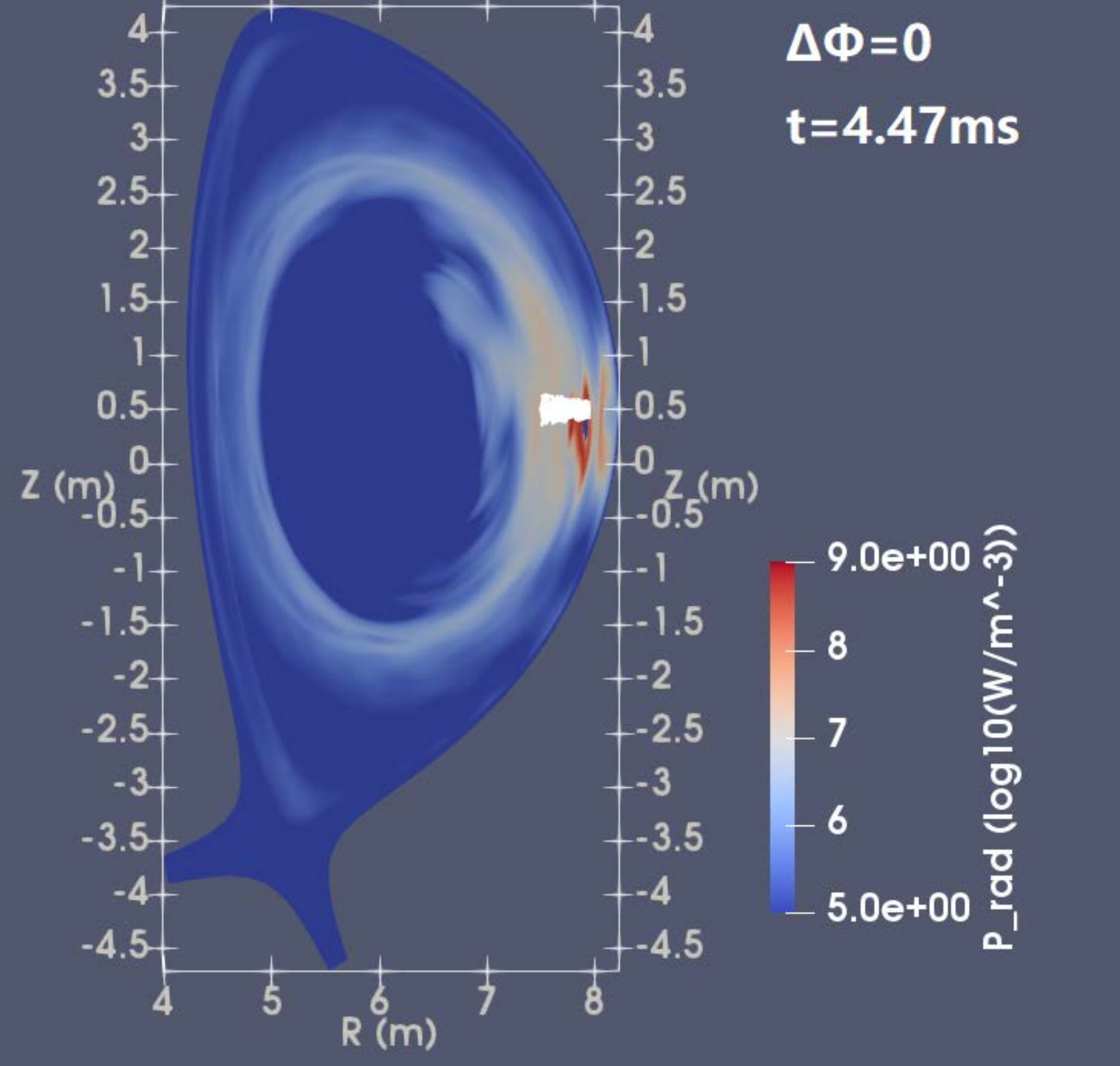}
}
&
\parbox{1.4in}{
	\includegraphics[scale=0.213]{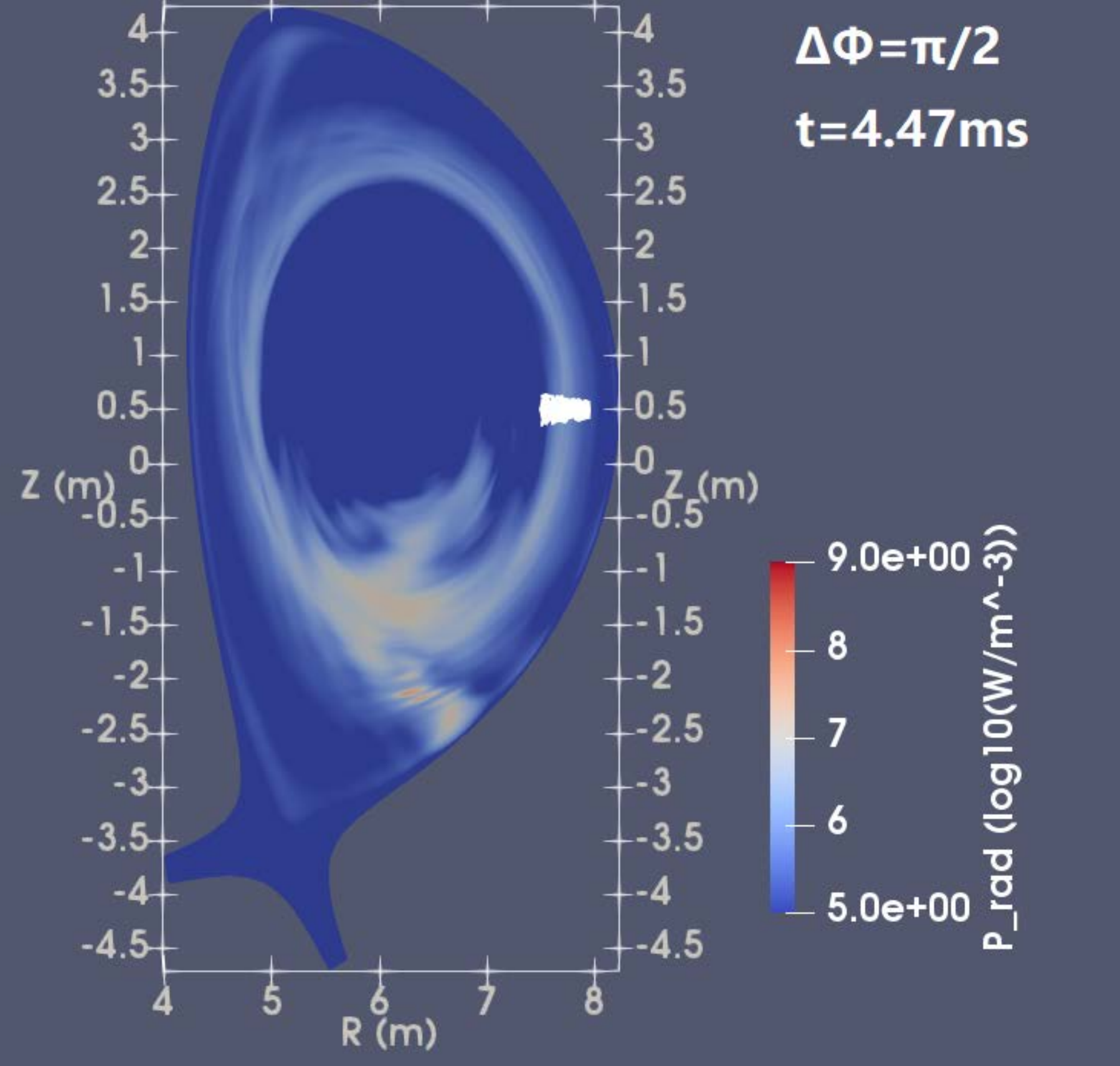}
}
&
\parbox{1.4in}{
	\includegraphics[scale=0.213]{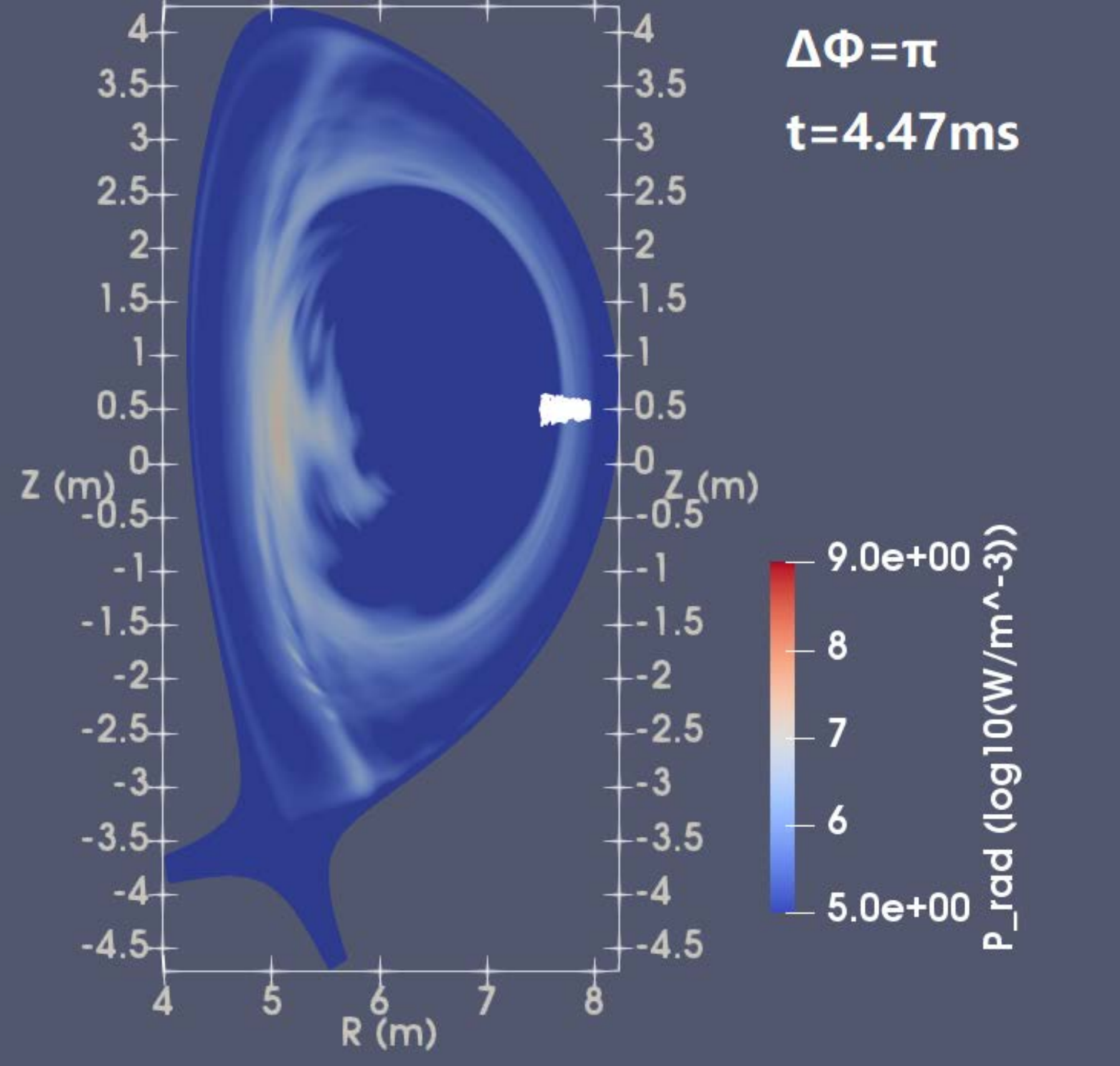}
}
\\
(d)&(e)&(f)
\\
\parbox{1.4in}{
	\includegraphics[scale=0.213]{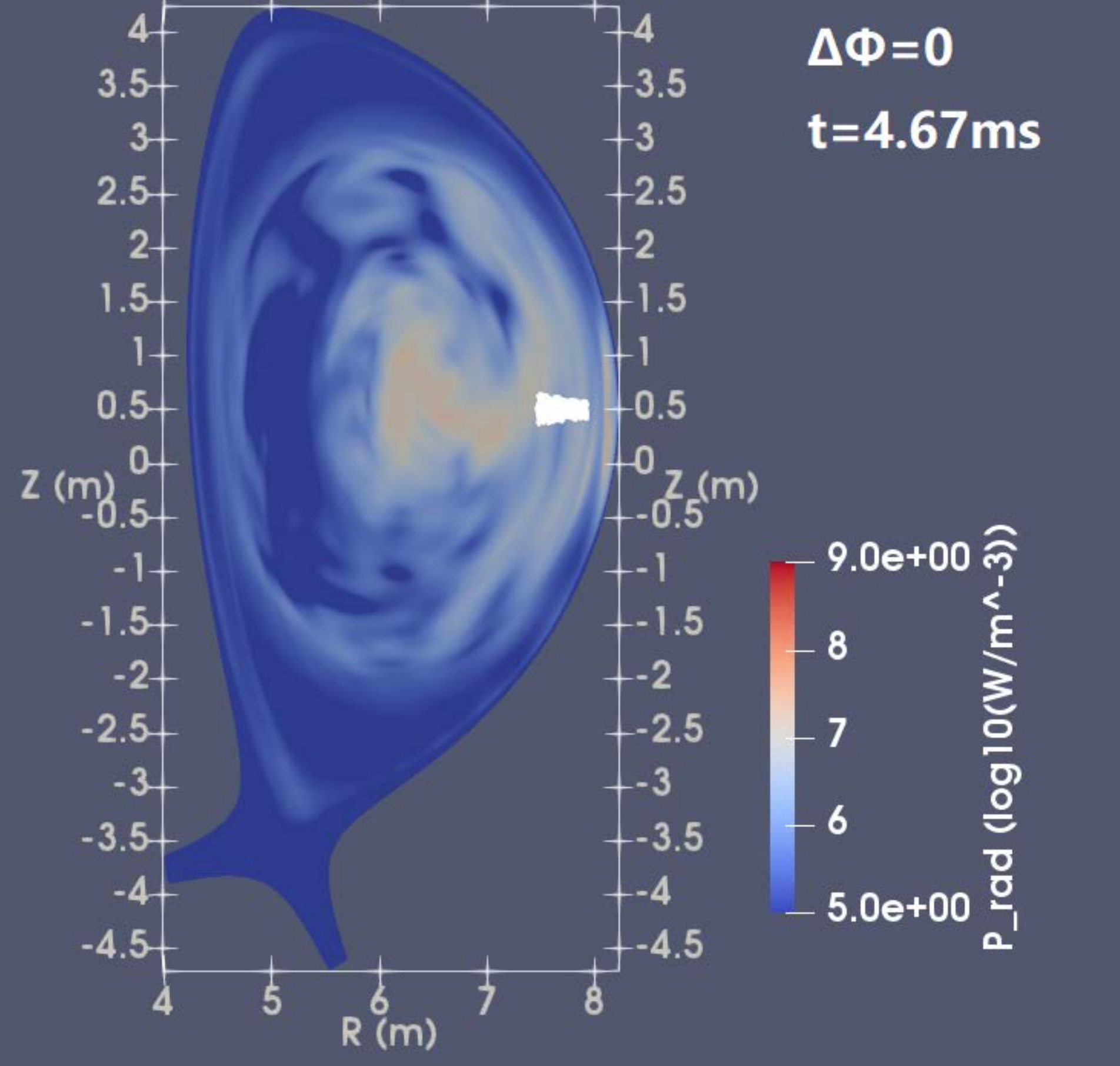}
}
&
\parbox{1.4in}{
	\includegraphics[scale=0.213]{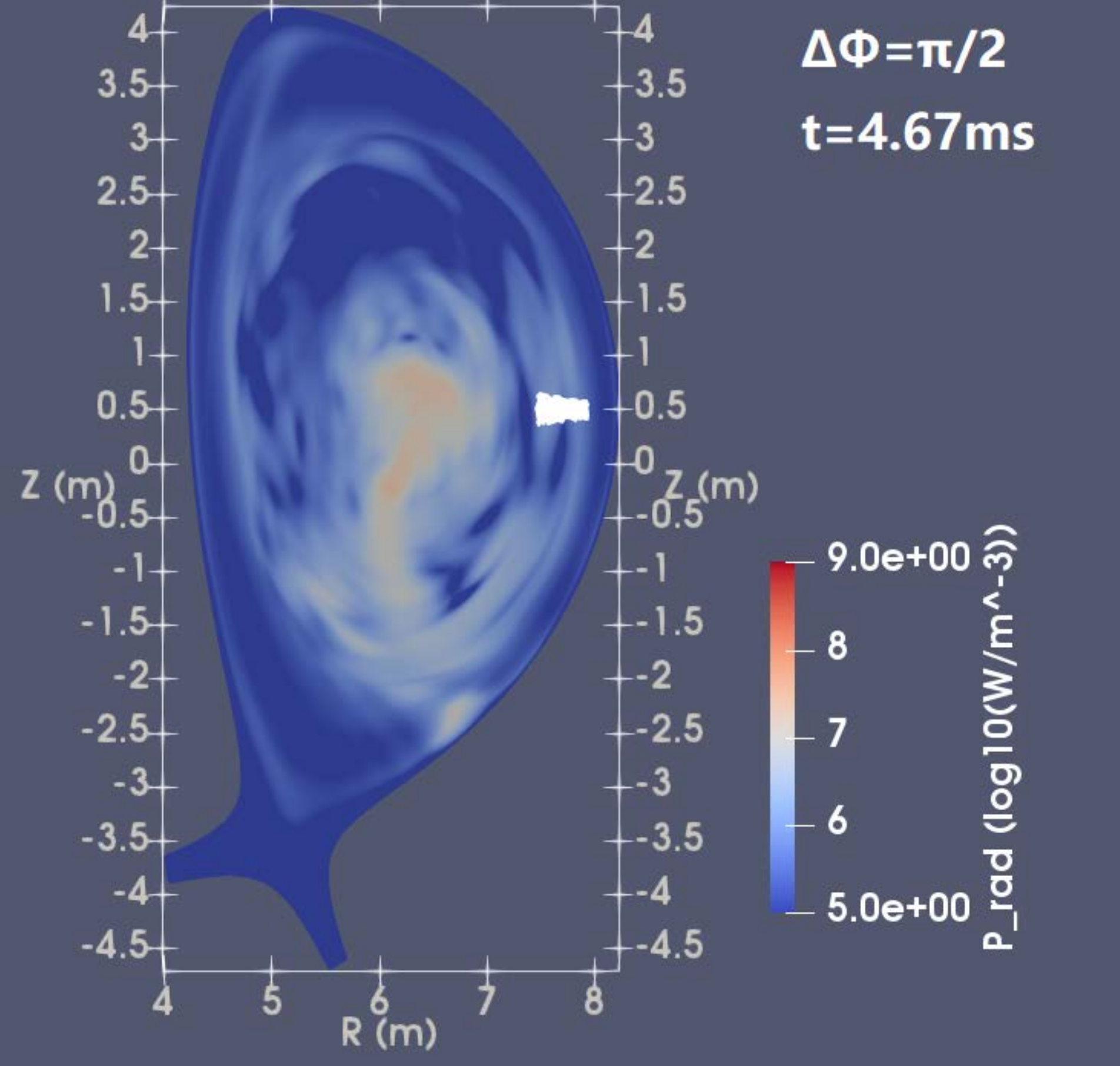}
}
&
\parbox{1.4in}{
	\includegraphics[scale=0.213]{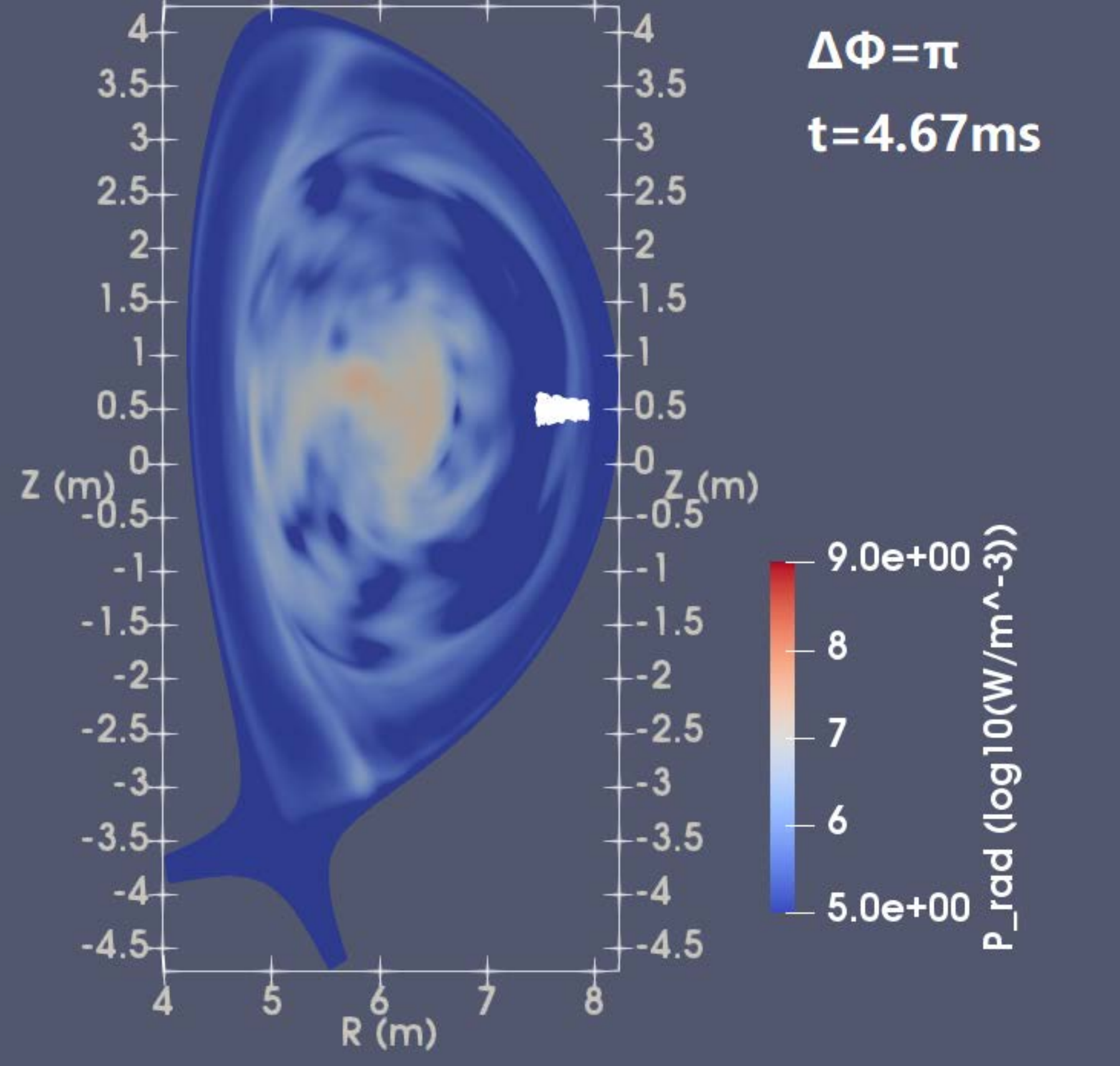}
}
\\
(g)&(h)&(i)
\etbl
\caption{The radiation power density for ``ITER shot 3''. The upper figures: The $\log_{10}$ of the radiation power density at time $t=3.80ms$ for toroidal location (a) $\gf=0$ (SPI location), (b) $\gf=\gp/2$ and (c) $\gf=\gp$; The middle figures: The $\log_{10}$ of the radiation power density at time $t=4.47ms$ for toroidal location (d) $\gf=0$ (SPI location), (e) $\gf=\gp/2$ and (f) $\gf=\gp$; The lower figures: The $\log_{10}$ of the radiation power density at time $t=4.67ms$ for toroidal location (g) $\gf=0$ (SPI location), (h) $\gf=\gp/2$ and (i) $\gf=\gp$. The white points indicate the projection of the fragment positions onto the poloidal plane.}
\label{fig:16}
\end{figure*}

Here, we will investigate such radiation asymmetry using the ``ITER shot 3'' case with the two temperature model. The toroidal location of injection is $\gf=0$. In Fig.\,\ref{fig:16}, the $\log_{10}$ of the radiation power density $P_{rad}$ is shown at times $t=3.80ms$, $t=4.47ms$ (around the time of the TQ) and $t=4.67ms$. The white points correspond to the projection of fragments onto the poloidal plane. One feature to note is that the radiation peak does not necessarily coincide with the position of the fragments nor the impurity density peak. More importantly, it can be seen that there are significant radiation asymmetries both poloidally and toroidally before and during the TQ, which is counterproductive to the goal of mitigating the radiation load during the TQ. Such asymmetry also shows distinctive helical structure corresponding to the $q=1$ surface, not unlike the helical radiation structure observed in DIII-D \cite{Sweeney2020NF}, although in that case the helicicty is $2/1$ since the TQ was triggered by fragments entering the $q=2$ surface. Ultimately, however, this asymmetry relaxes as the TQ proceeds as can be seen in Fig.\,\ref{fig:16}(g), (h) and (i).

We would like to mention again that the radiation asymmetry could be artificially reduced by our toroidally elongated deposition of the ablation cloud. Furthermore, the strong radiation asymmetry within the plasma does not necessarily result in a strong peaking in the radiative heat flux onto the PFCs which is the real indication of the thermal quench mitigation efficiency. Detailed analysis has to be carried out using simulation results and developing dedicated post-processing routines, and this is left for future works.

\section{The MHD response and radiation asymmetry for ITER dual-SPI}
\label{s:RadAsyDual}

The undesirable radiation asymmetry shown in Section \ref{ss:MonoSPIAsym} can be mitigated by conducting SPI at multiple toroidal locations simultaneously. To show this, we now use ``ITER shot 4'' and ``ITER shot 5'' with the two temperature model to investigate the MHD and radiation behavior after injecting from two toroidally opposite positions.

\subsection{The MHD response and radiation asymmetry for symmetric dual-SPI}
\label{ss:SymmDualSPI}

\begin{figure*}
\centering
\noindent
\btbl{cc}
\parbox{2.5in}{
    \includegraphics[scale=0.25]{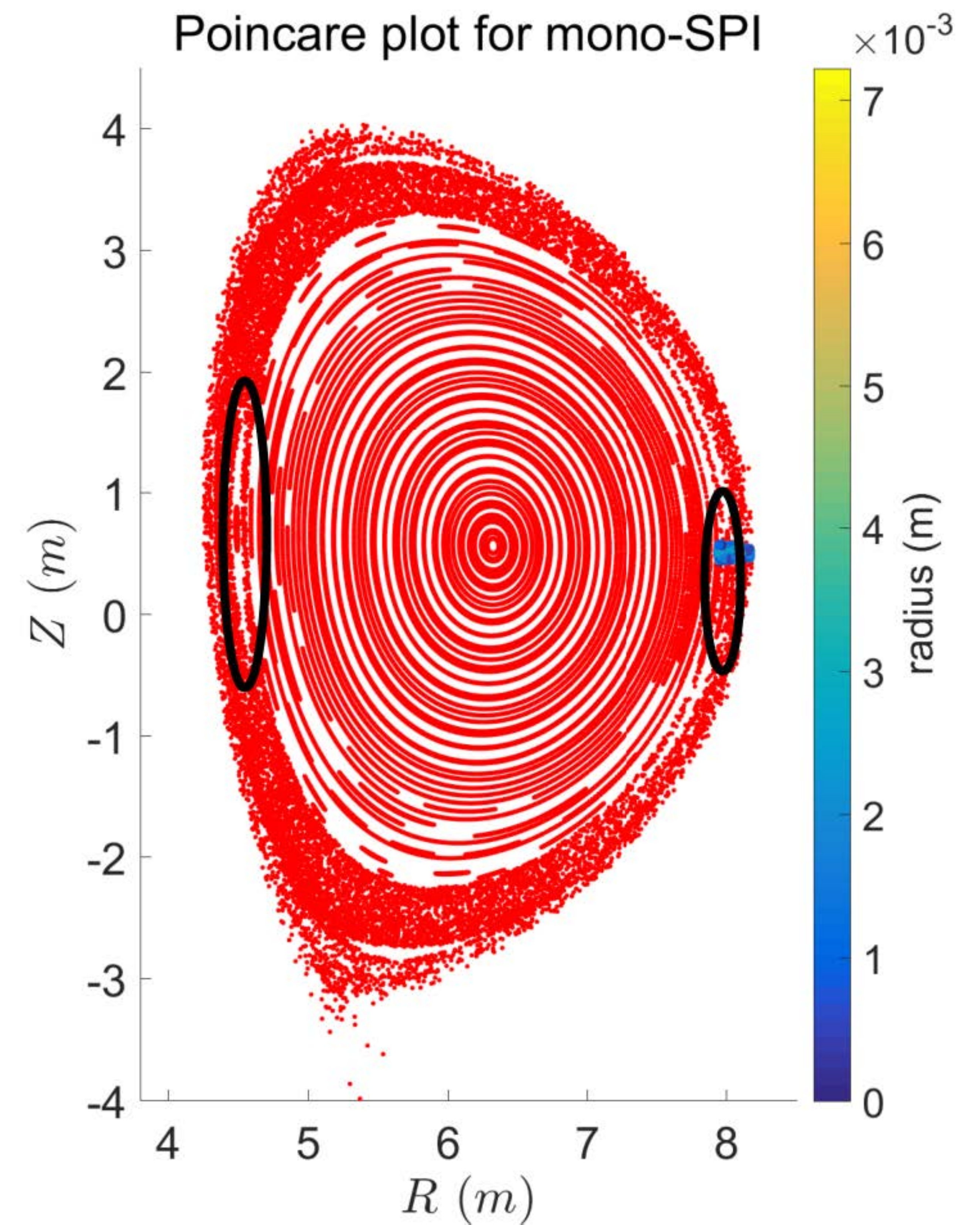}
}
&
\parbox{2.5in}{
	\includegraphics[scale=0.25]{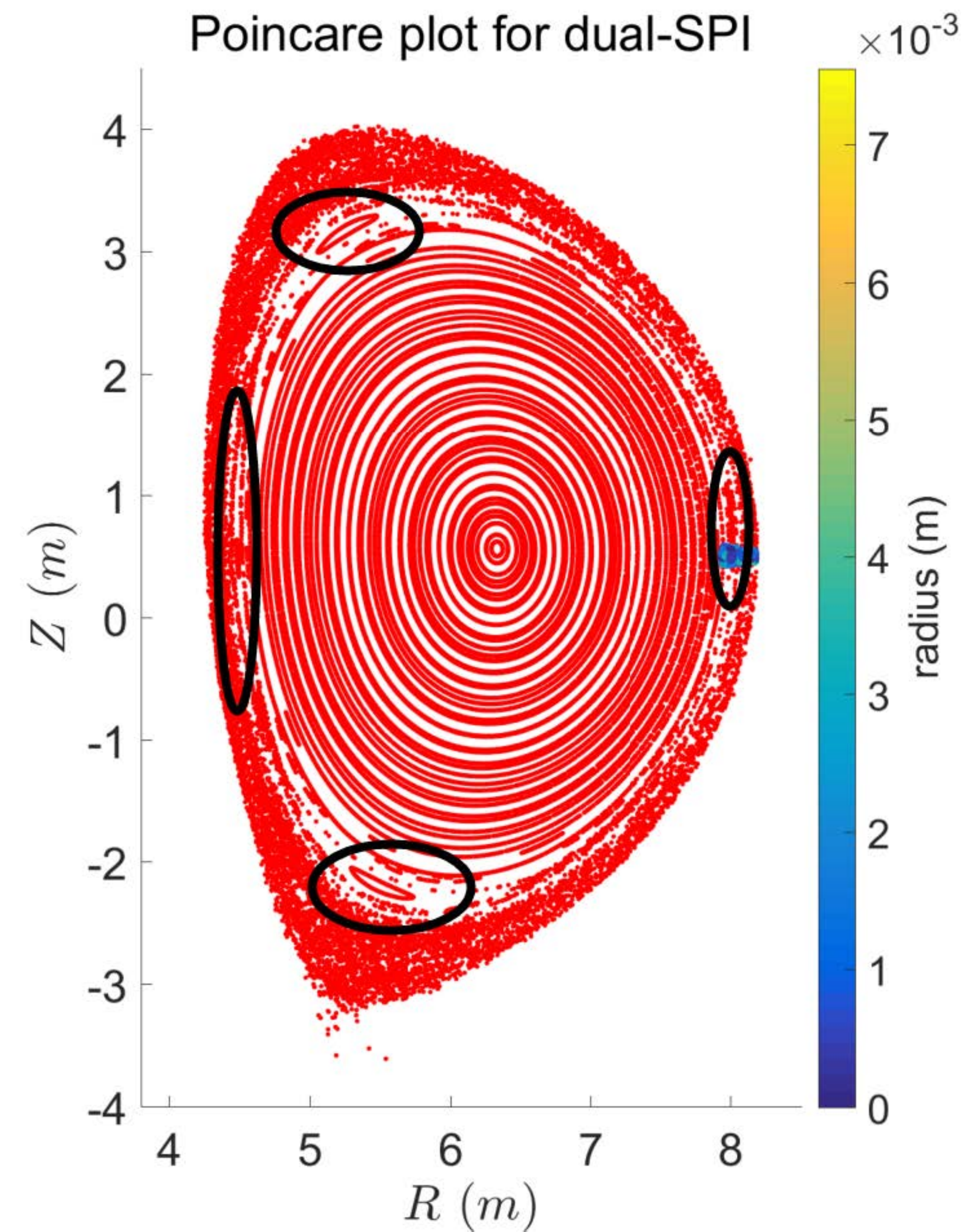}
}
\\
(a)&(b)
\etbl
\caption{The magnetic Poincar\'{e} plot for (a) the mono-SPI case and (b) the symmetric dual-SPI case. The remnant of the $2/1$ and $4/2$ islands, which are marked by the black circles, can be seen on the $q=2$ surface respectively.}
\label{fig:17}
\end{figure*}

We fist consider the case of symmetric dual-SPI ``ITER shot 4'' where the two injectors exactly mirror each other spatially and temporally, and compare it with the mono-SPI ``ITER shot 3'' case. The total injection amount between the two shots are the same as is shown in Table \ref{tab:1}. In such a scenario, if the helical effect dominates over the axisymmetric one, we expect that higher harmonics of the MHD modes compared with that of the mono-SPI case will dominate the MHD response. Otherwise, there should be little difference compared with the mono-SPI case. In our simulation, we observe that indeed the former case happens as can be seen in Fig.\,\ref{fig:17}. As the fragments arrive at the $q=2$ surface, the resonant $2/1$ island can be seen at the inner and outer mid-plane in Fig.\,\ref{fig:17}(a) for the mono-SPI case, with the outer one coinciding with the vanguard fragment location. On the other hand, for the dual-SPI case shown in Fig.\,\ref{fig:17}(b), the MHD activity at the $q=2$ surface is dominated by a $4/2$ island. Later in Section \ref{ss:AsymmDualSPI}, we will see that such good behavior of correspondence between the MHD parity and the injection symmetry is the result of the perfectly synchronized injections. In the scenario where fragments from one injector enter the plasma earlier than that from the other, the parity in MHD response is somewhat broken.

\begin{figure*}
\centering
\noindent
\btbl{cc}
\parbox{2.5in}{
    \includegraphics[scale=0.27]{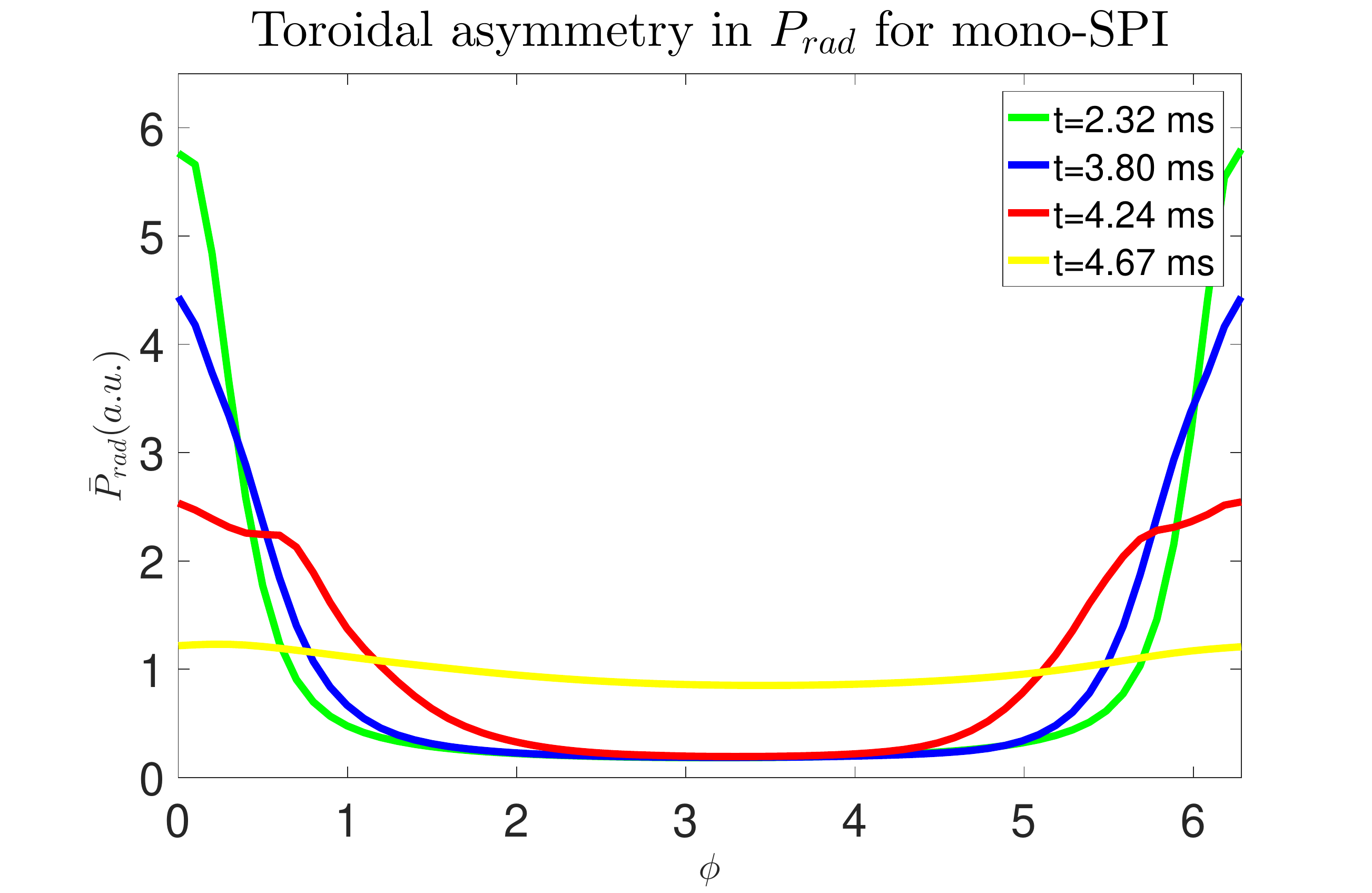}
}
&
\parbox{2.5in}{
	\includegraphics[scale=0.27]{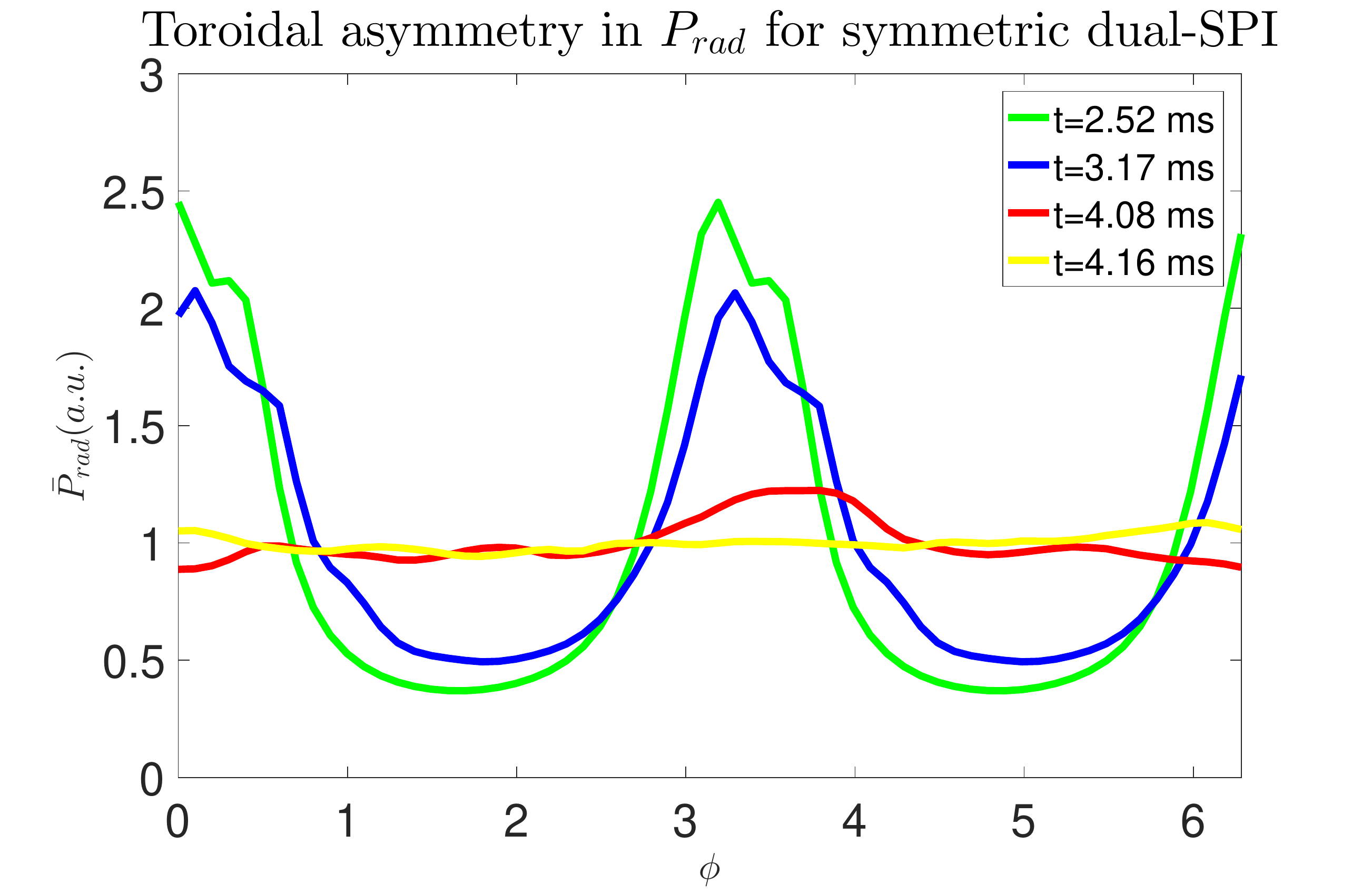}
}
\\
(a)&(b)
\\
\parbox{2.5in}{
    \includegraphics[scale=0.27]{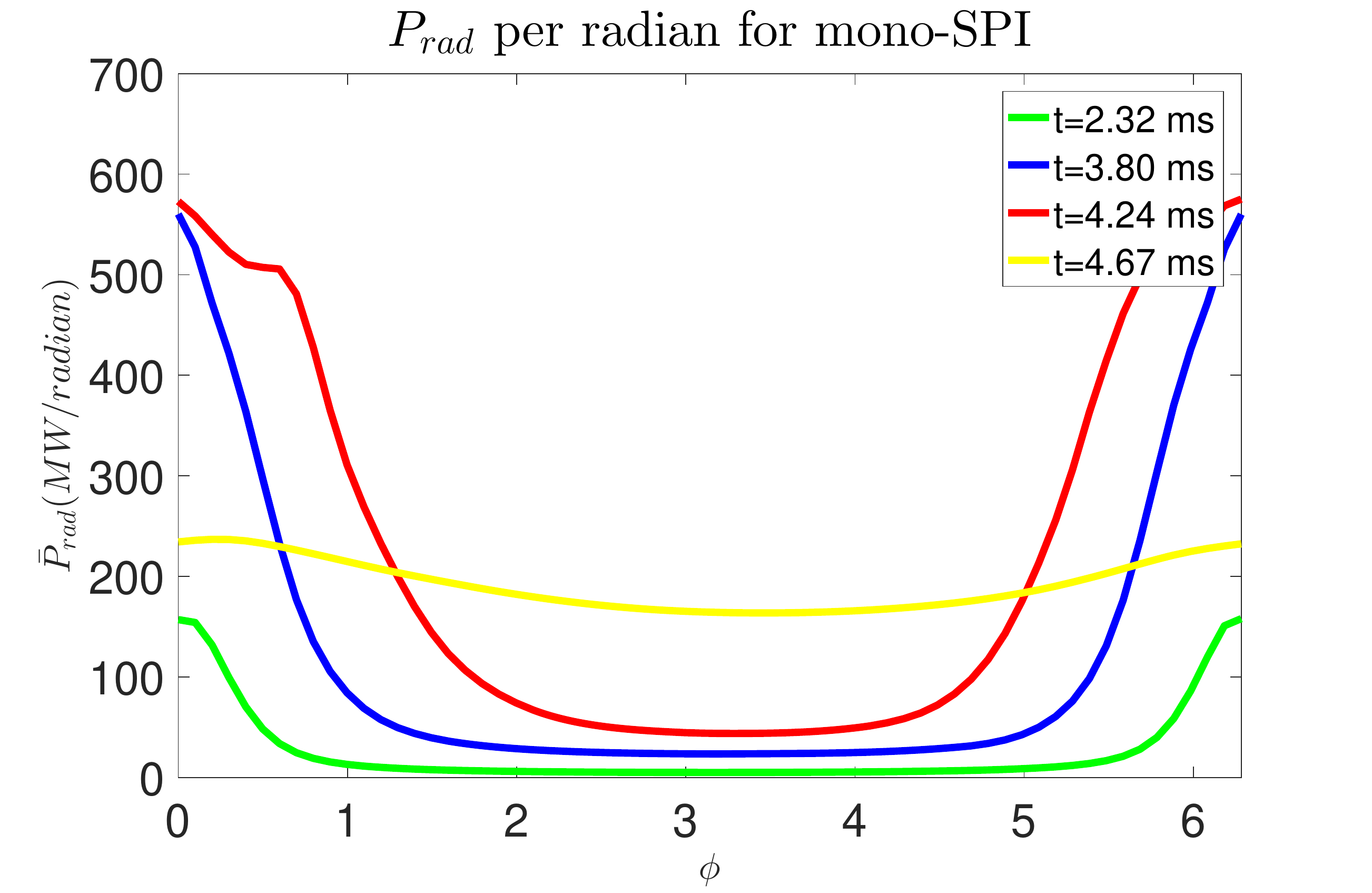}
}
&
\parbox{2.5in}{
	\includegraphics[scale=0.27]{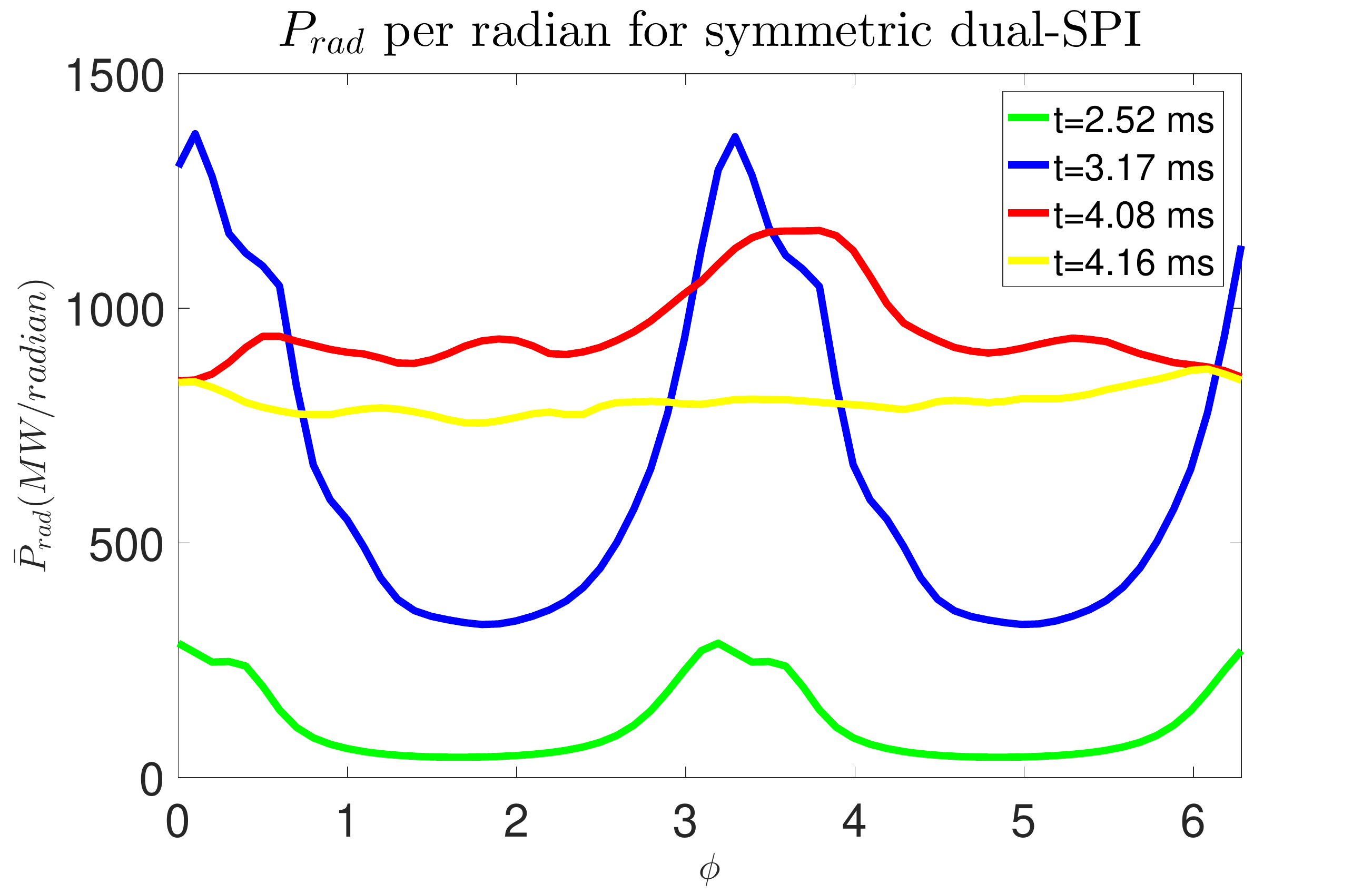}
}
\\
(c)&(d)
\etbl
\caption{The toroidal radiation peaking factor $\baP_{rad}$ of (a) the mono-SPI and (b) the symmetric dual-SPI, and the radiation power per radian for (c) the mono-SPI and (d) the symmetric dual-SPI. The green curves in both cases indicate the early periphery cooling by the SPI, the blue curves in both cases correspond to when the discharge is evolving towards the TQ, the red ones in both cases correspond to the early TQ, while the yellow ones correspond to hundreds of microseconds after the TQ.}
\label{fig:18}
\end{figure*}

The radiation asymmetry also bears the mark of the symmetry of the injection configuration in the early phase of the injection. We compare the toroidal radiation peaking factor $\baP_{rad}$ of the mono-SPI ``ITER shot 3'' with that of the symmetric dual-SPI ``ITER shot 4'', shown in Fig.\,\ref{fig:18}.
Since ``ITER shot 1'' and ``ITER shot 2'' case show similar radiation peaking with ``ITER shot 3'', we do not show them here.
For the mono-SPI case, the injection location is at $\gf=0$ and indeed we see the radiation peaking around that toroidal angle before the TQ ($t=2.32ms$ and $t=3.80ms$) and during the early TQ phase ($t=4.24ms$). Especially, a strong toroidal peaking is seen close to the TQ ($t=3.80ms$) and during the early TQ phase ($t=4.24ms$) as is shown by the blue and the red line in Fig.\,\ref{fig:18}(a), as well as those in Fig.\,\ref{fig:18}(c) where the actual numbers of $P_{rad}$ per radian are shown. This is partly due to a strongly unrelaxed impurity density distribution caused by the large ablation source as the outward heat flux increases during the TQ. It may also partly be caused by the asymmetric pattern in the outgoing heat flux itself. Later on, as the TQ proceeds, this peaking is ultimately relaxed. This behavior is in accordance with what we see in Fig.\,\ref{fig:16}. On the other hand, although the dual-SPI case also shows toroidal peaking corresponding to the injection parity in the pre-TQ phase, the peaking factor steadily decreases as the time approaches the TQ onset at $t\sim4.1ms$. Shortly after the TQ at $t=4.16ms$, the radiation peaking is already mostly mitigated as is shown by the yellow line in Fig.\,\ref{fig:18}(b) as well as that in Fig.\,\ref{fig:18}(d). Especially, the radiation peak right before the TQ at $t=4.08ms$ and that after the TQ at $t=4.16ms$ are no higher than that during the pre-TQ stage at $t=3.17ms$, despite the total radiation power is higher for the former two cases. Such behavior is very desirable since the total radiation power peaks shortly after the TQ is triggered \cite{Shiraki2016POP}, thus the toroidal peaking factor at that time is most important. Another noteworthy feature is that the radiation power is larger for the dual-SPI case compared with the mono-SPI case as can be seen from Fig.\,\ref{fig:18}(c) and (d). This behavior is qualitatively in agreement with recent K-STAR dual-SPI experiments \cite{Kim2020IAEA}, and could be partly because of the increased neon assimilation into the plasma for the dual-SPI case.

\subsection{The MHD response and radiation asymmetry for asymmetric dual-SPI}
\label{ss:AsymmDualSPI}

We have shown the good radiation peaking mitigation by perfectly symmetric dual-SPI in Section \ref{ss:SymmDualSPI}. However, such symmetry is not realistic, as there is no guarantee that the fragments would arrive exactly at the same time for both injectors. Here, we consider a more realistic case ``ITER shot 5'' where there is $1ms$ delay between the two SPIs and compare it with ``ITER shot 4'' to show the impact of imperfect timing.

\begin{figure*}
\centering
\noindent
\btbl{ccc}
\parbox{1.8in}{
    \includegraphics[scale=0.35]{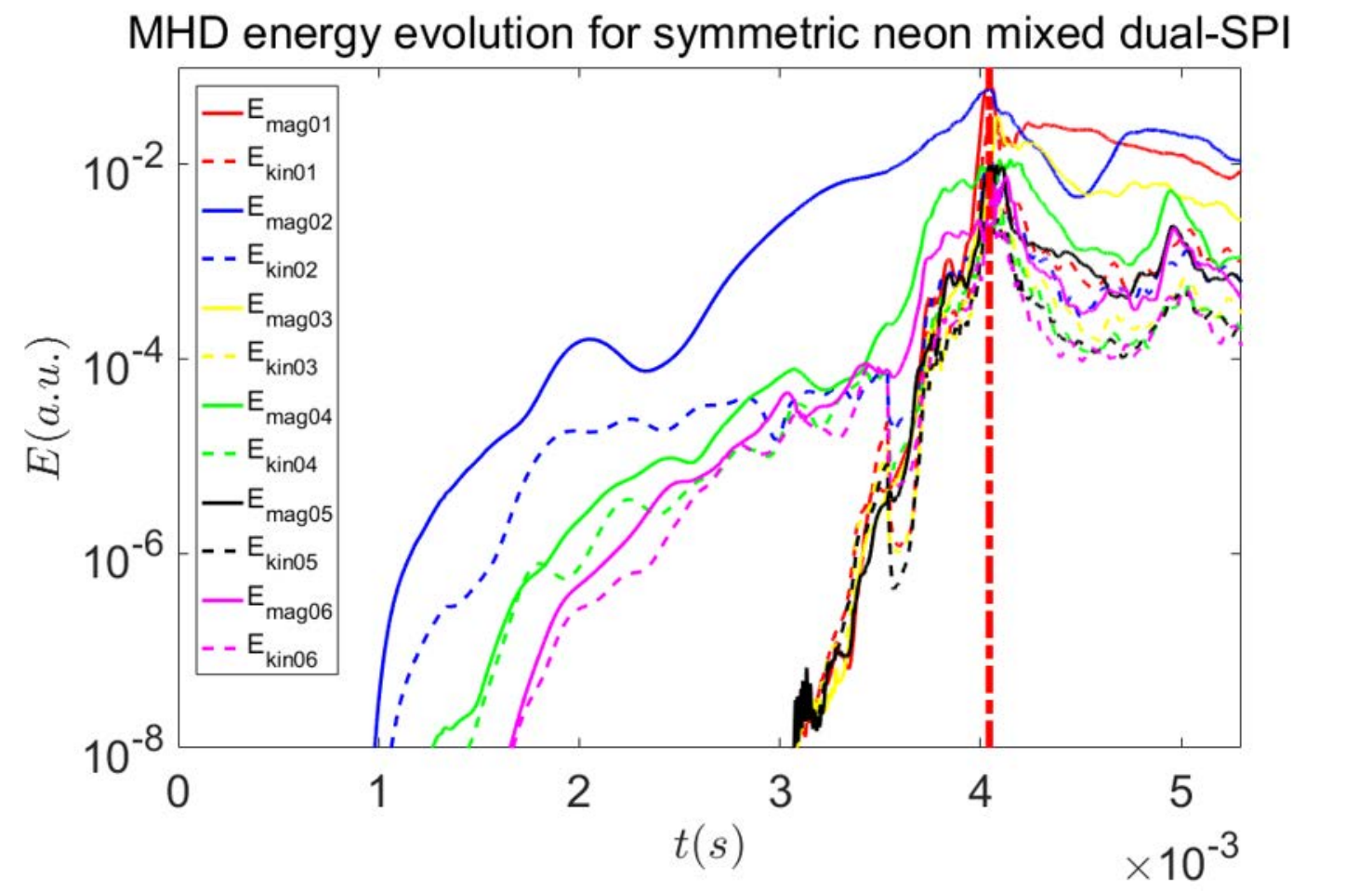}
}
&
\parbox{1.8in}{
	\includegraphics[scale=0.35]{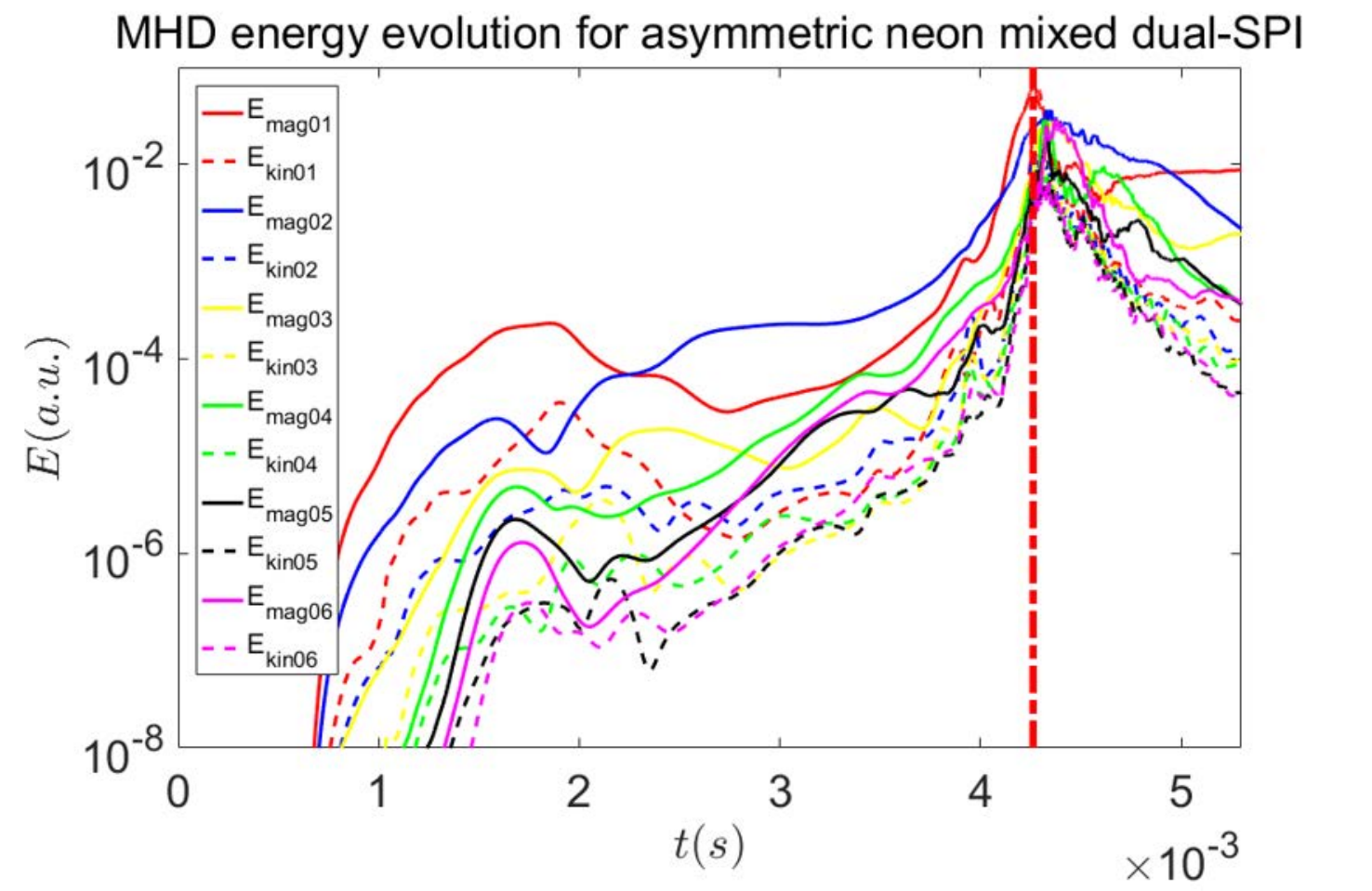}
}
&
\parbox{1.8in}{
	\includegraphics[scale=0.198]{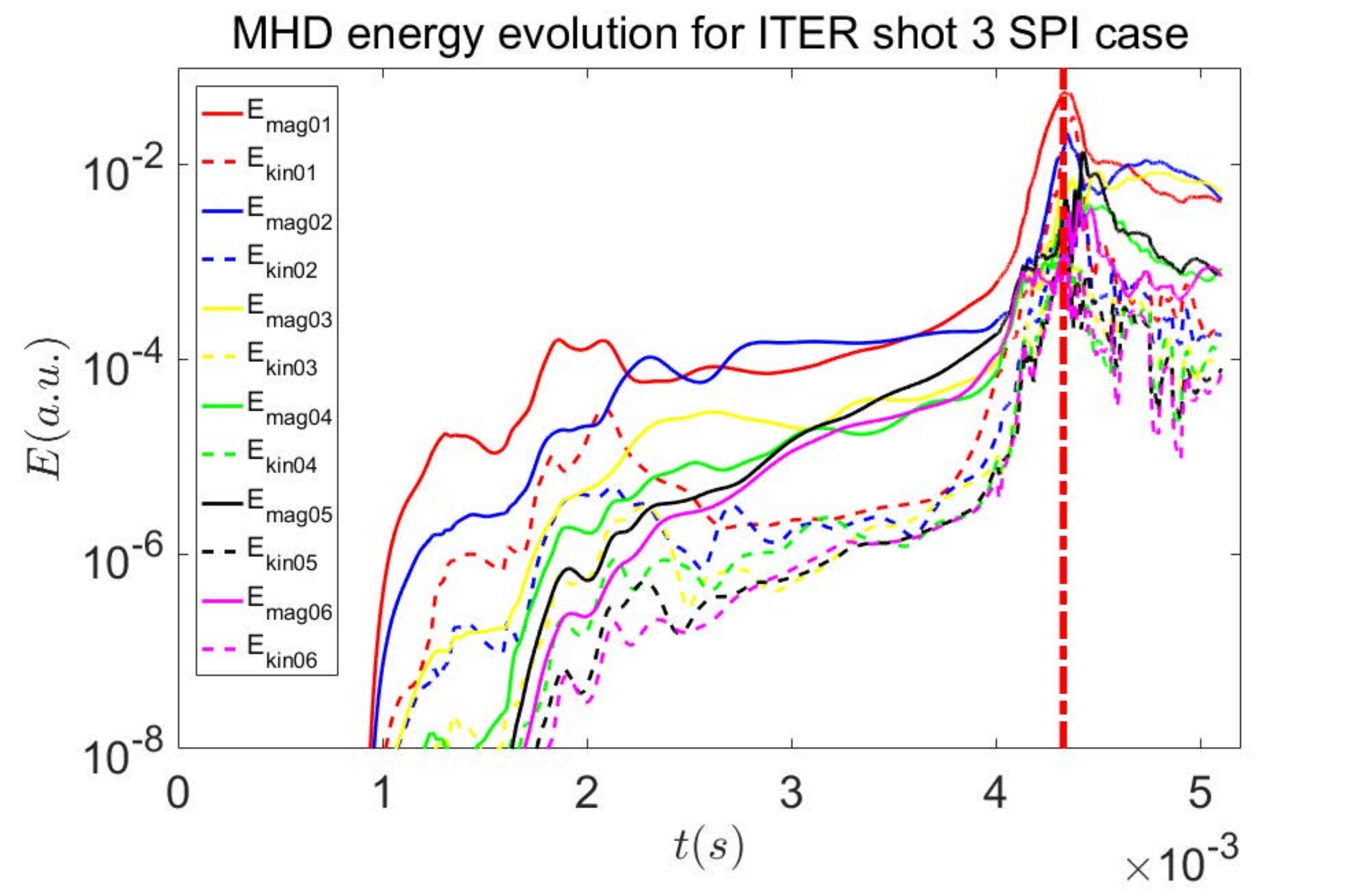}
}
\\
(a)&(b)&(c)
\etbl
\caption{The perturbed MHD energy for (a) the symmetric dual-SPI, (b) the asymmetric dual-SPI and  (c) the mono-SPI case ``ITER shot 3''. The vertical red chained line marks the onset of the TQ.}
\label{fig:19}
\end{figure*}

The most obvious difference is in the symmetry of the MHD response, as is shown in Fig.\,\ref{fig:19}. For the symmetric dual-SPI case, the dominant modes are the even modes as expected and in accordance with the Poincar\'{e} plot of Fig.\,\ref{fig:17}. In Fig.\,\ref{fig:19}(a), the $n=2$, $n=4$ and $n=6$ modes are much stronger than the odd modes, until we approach the TQ represented by the vertical red chained line when the whole plasma becomes turbulent. On the other hand, for the asymmetric case, initially the $n=1$ mode dominates over the even modes, although later on, the even modes also begin to grow and the $n=2$ amplitude becomes comparable with that of the $n=1$ mode. The continued growth of both odd and even modes collectively leads to the onset of the TQ at time $t\sim4.2ms$ as marked by the red chained line.
As a comparison to both dual-SPI cases, the MHD spectrum of the mono-SPI case ``ITER shot 3'' is shown in Fig.\,\ref{fig:19}(c). In this case, as the plasma evolve towards the TQ it can be seen that the $n=1$ mode dominate over the other modes. The comparison between the three cases show that the relative role of the $n=1$ mode increase gradually from the symmetric dual-SPI case to the asymmetric dual-SPI case, then further to the mono-SPI case which indeed can be seen as an asymmetric dual-SPI with extremely long delay between injections. Not shown in the figure, ``ITER shot 1'' and ``ITER shot 2'' case show similar MHD behavior with ``ITER shot 3'' as shown in Fig.\,\ref{fig:19}(c).

\begin{figure*}
\centering
\noindent
\btbl{ccc}
\parbox{1.4in}{
    \includegraphics[scale=0.160]{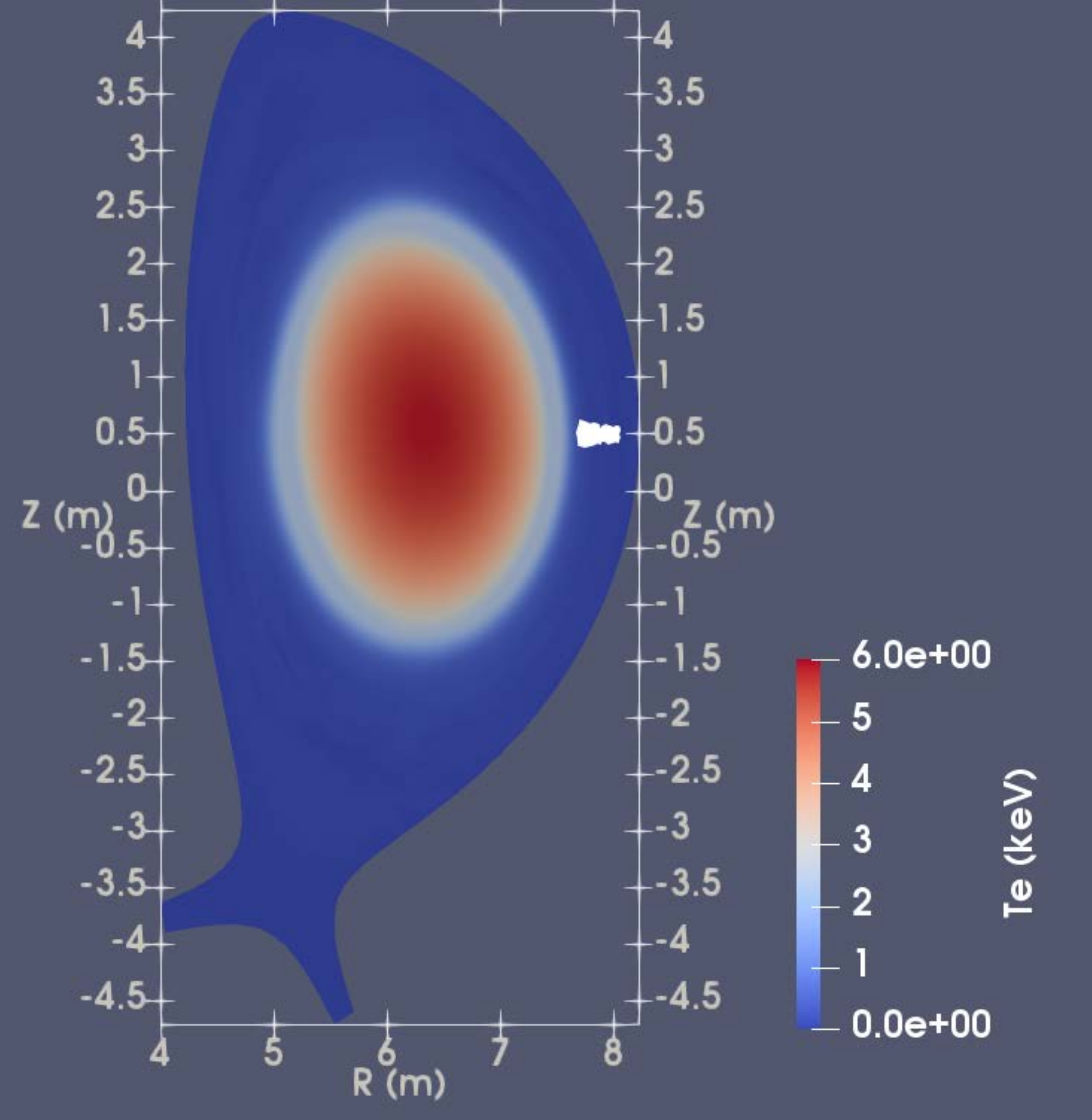}
}
&
\parbox{1.4in}{
	\includegraphics[scale=0.160]{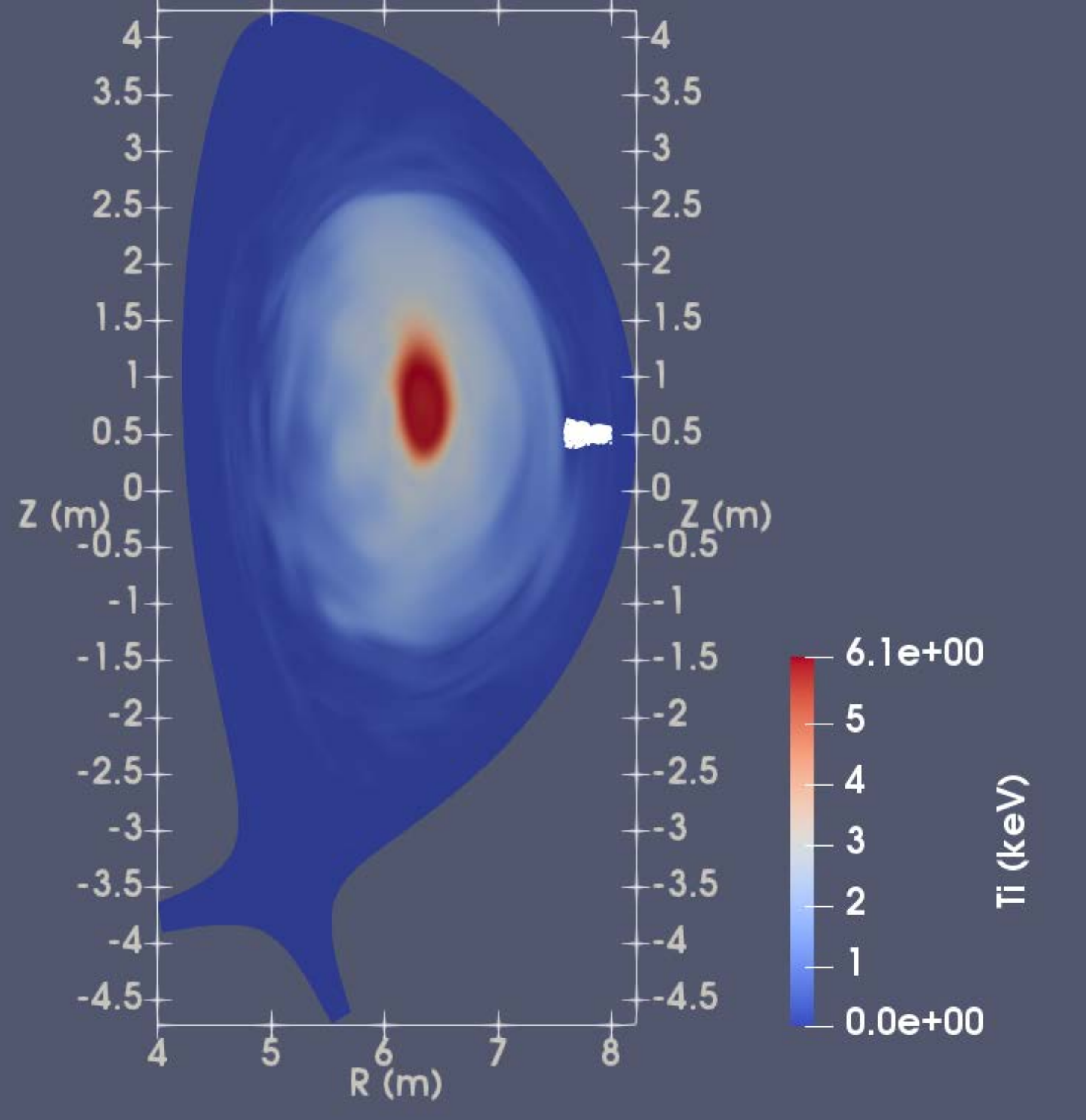}
}
&
\parbox{1.4in}{
	\includegraphics[scale=0.160]{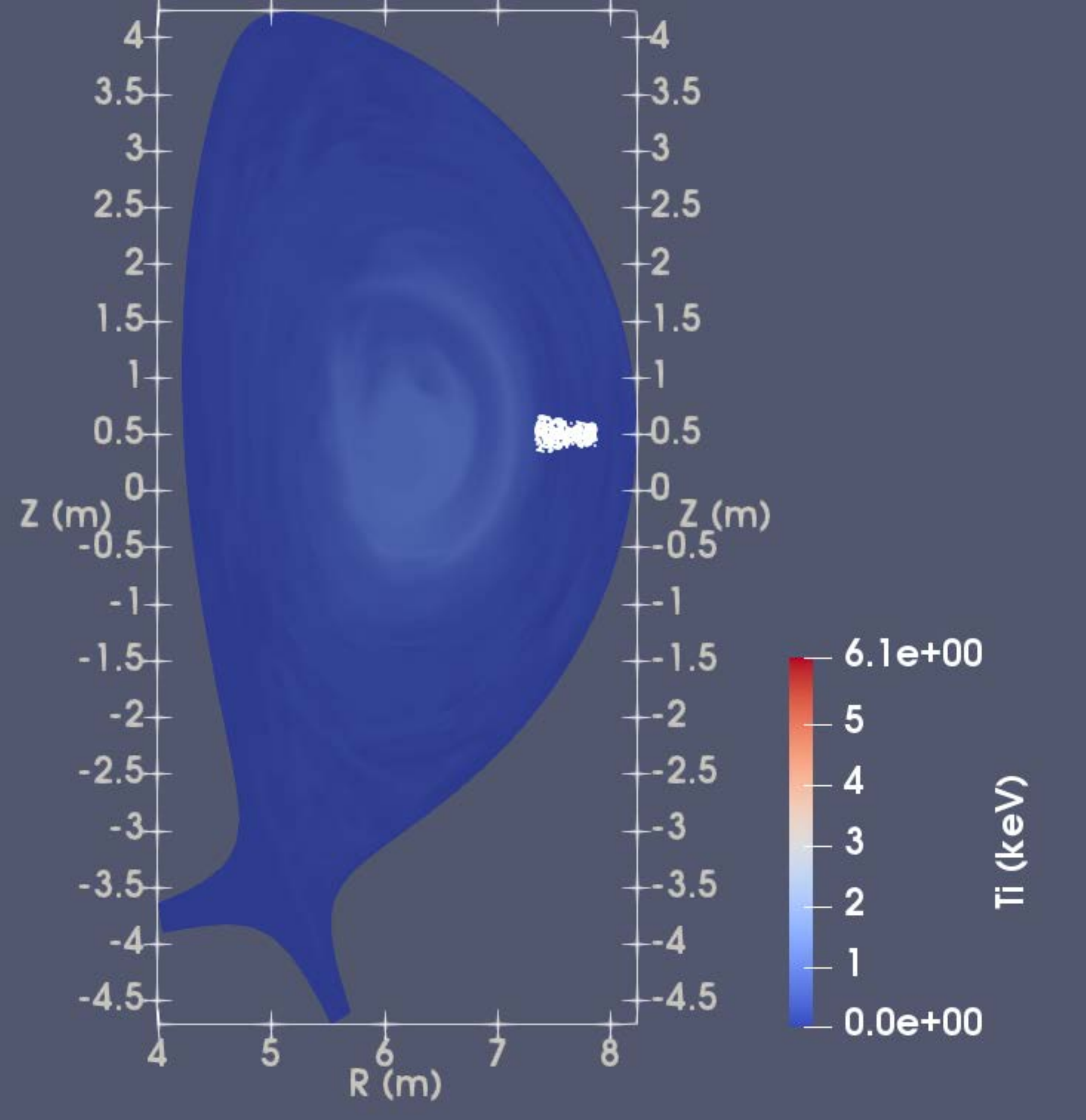}
}
\\
(a)&(b)&(c)
\\
\parbox{1.4in}{
	\includegraphics[scale=0.170]{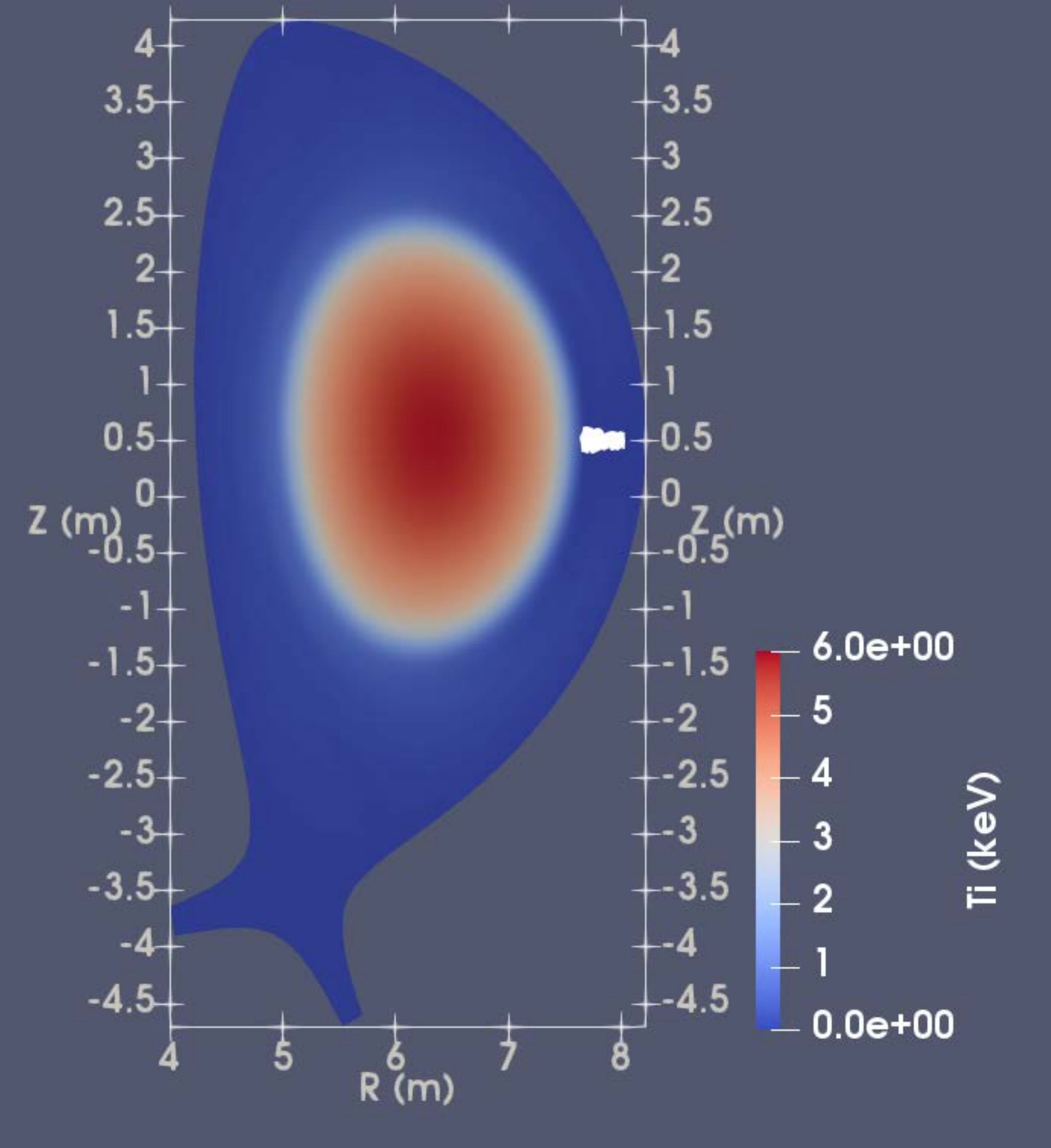}
}
&
\parbox{1.4in}{
	\includegraphics[scale=0.170]{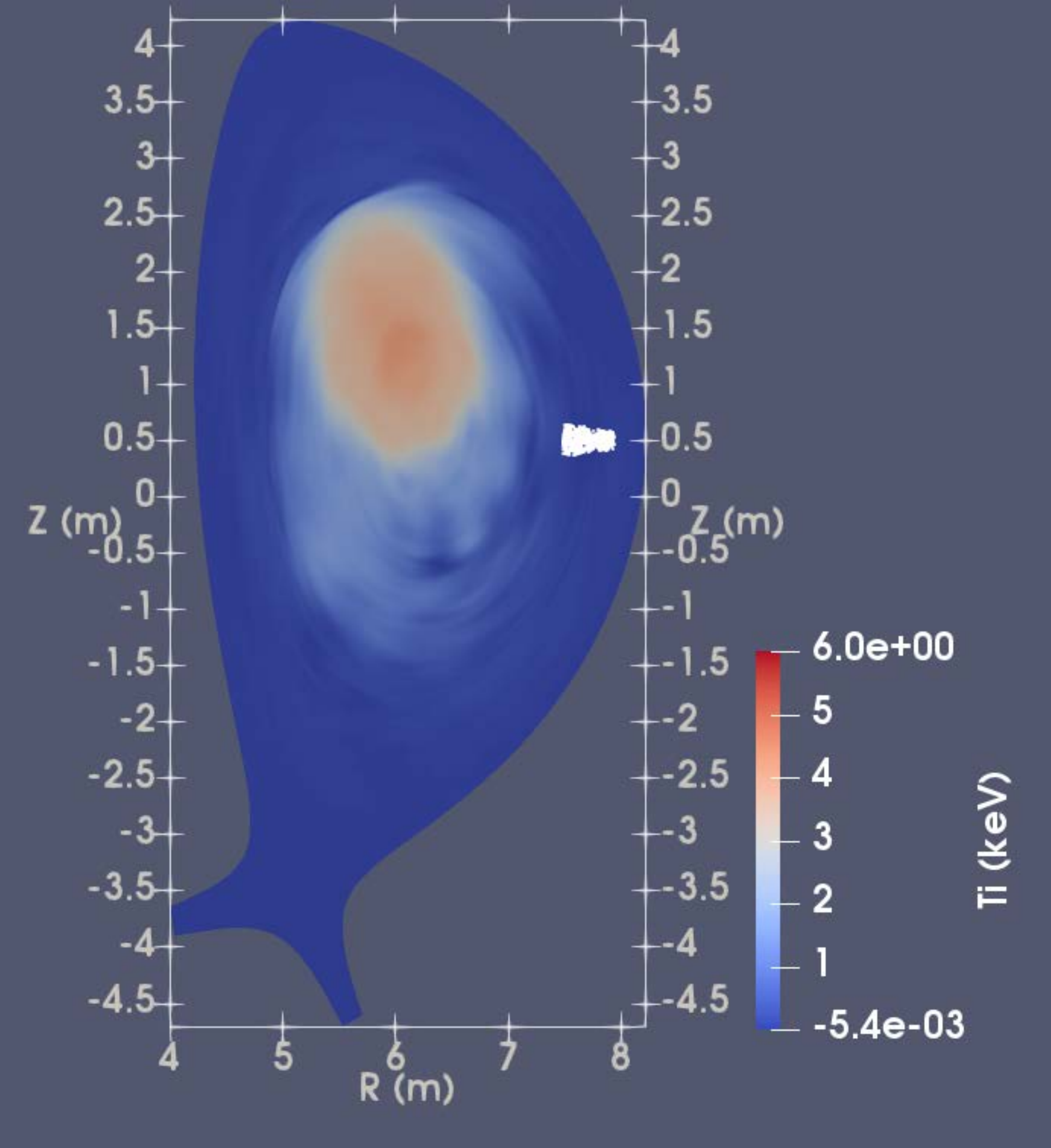}
}
&
\parbox{1.4in}{
	\includegraphics[scale=0.170]{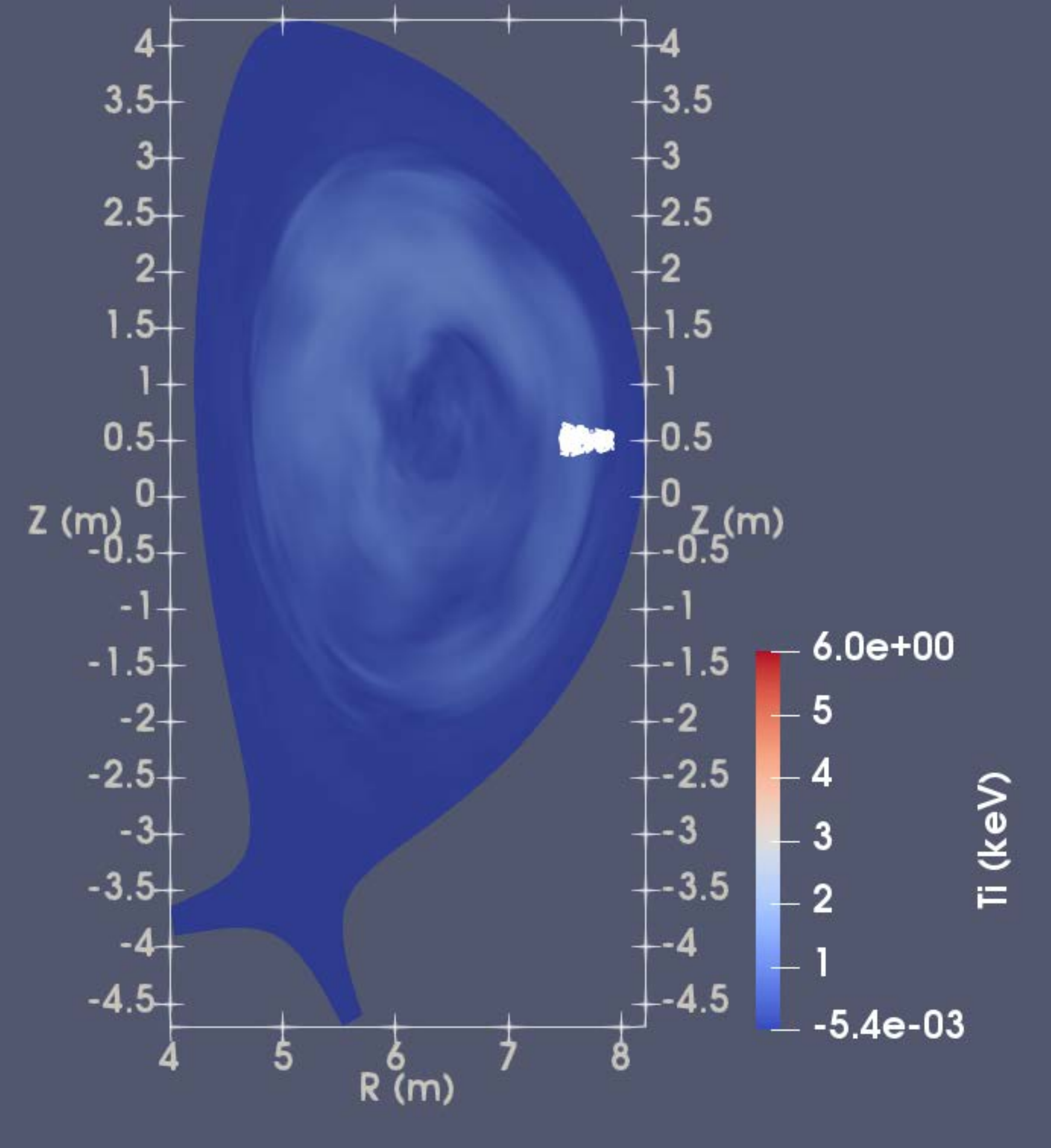}
}
\\
(d)&(e)&(f)
\etbl
\caption{The upper figures: the ion temperature profile (a) before the TQ, (b) at the time of the TQ and (c) after the TQ for the symmetric dual-SPI case; The lower figures: the ion temperature profile (a) before the TQ, (b) at the time of the TQ and (c) after the TQ for the asymmetric dual-SPI case. The dominant plasma motion is significantly different for those two cases. The white points are the projection of fragments onto the poloidal plane.}
\label{fig:20}
\end{figure*}

The evolution of the temperature distribution, given in Fig.\,\ref{fig:20}, shows more intuitively the dominant MHD response at the time of the TQ. The symmetric dual-SPI ion temperature evolution is shown in Fig.\,\ref{fig:20}(a), (b) and (c), where we can see that there is no observable $1/1$ kink motion during the core collapse; instead only a $m=2$ deformation exists. In contrast, for the asymmetric dual-SPI shown in Fig.\,\ref{fig:20}(d), (e) and (f), we see obvious $1/1$ kink moving away from the core, although its O-point does not exactly correspond to the position of the vanguard fragments, unlike that of the mono-SPI case shown in Fig.\,\ref{fig:13}. This means that despite the fact that the even and odd modes seem to collectively cause the onset of the TQ from the energy spectrum Fig.\,\ref{fig:19}, the $n=1$ mode may still play a more dominant role in the core collapse if perfect symmetry can not be achieved between the injectors.

\begin{figure*}
\centering
\noindent
\btbl{cc}
\parbox{2.5in}{
    \includegraphics[scale=0.27]{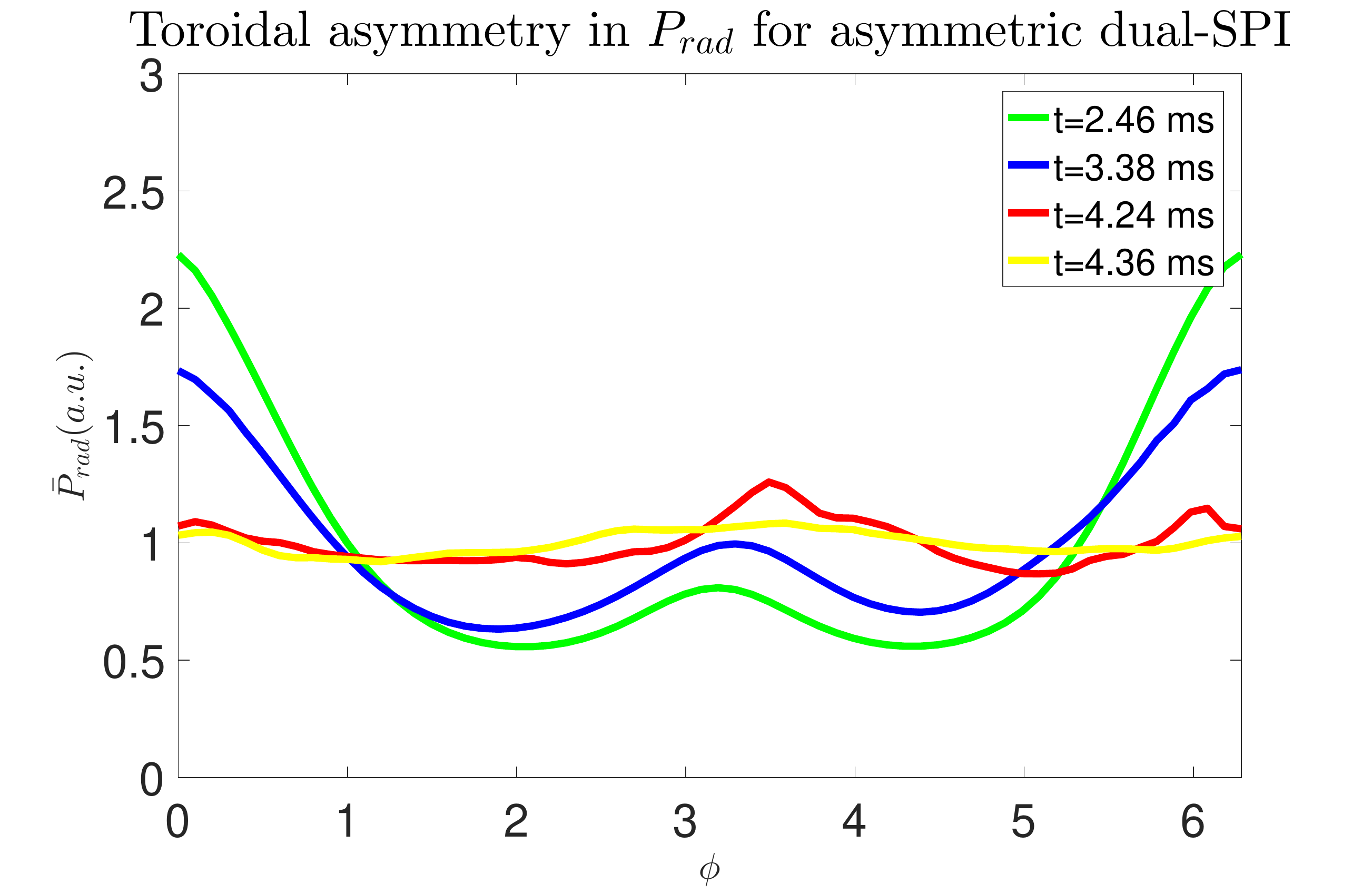}
}
&
\parbox{2.5in}{
	\includegraphics[scale=0.27]{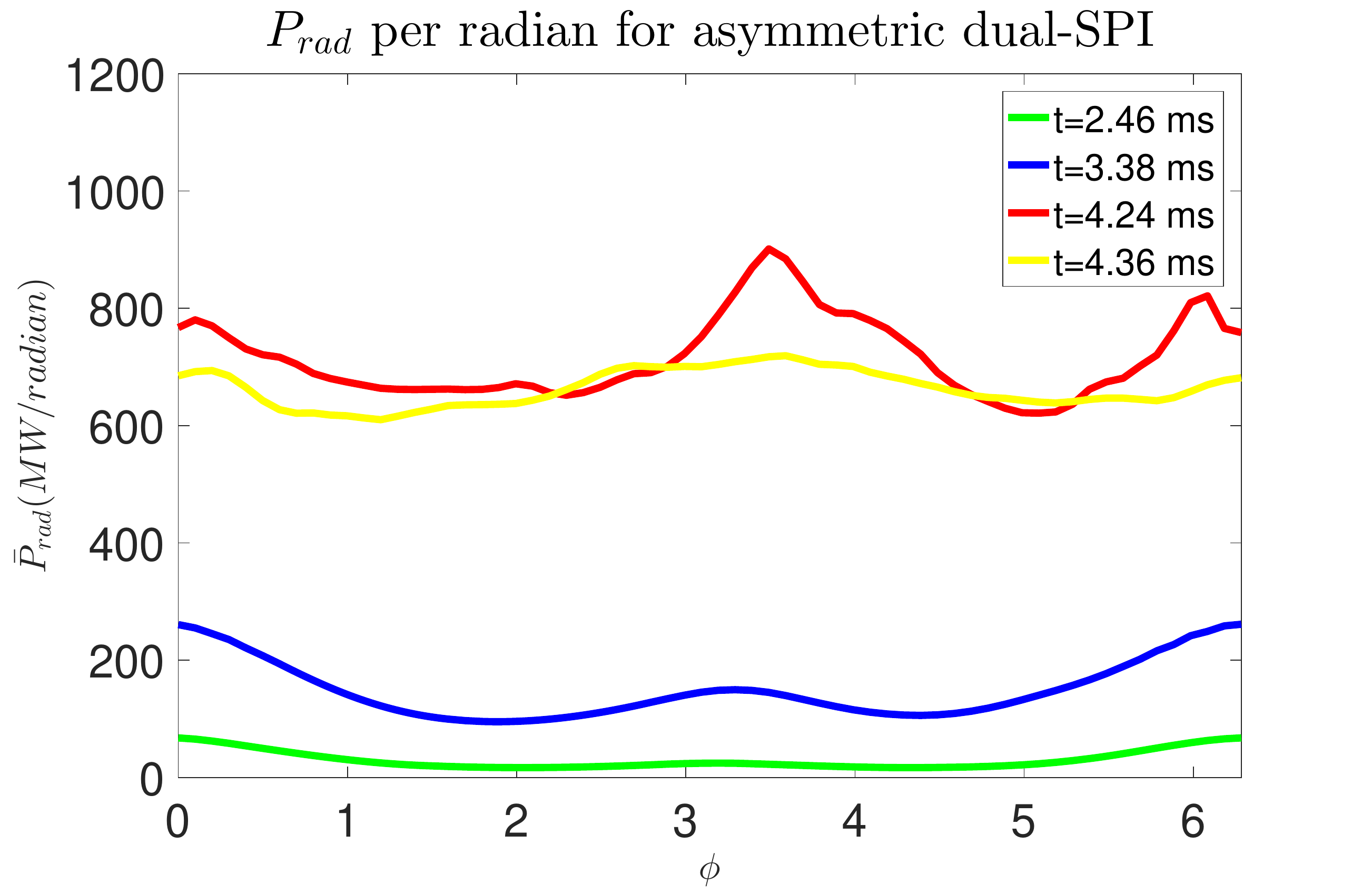}
}
\\
(a)&(b)
\etbl
\caption{(a) The toroidal radiation peaking factor for the asymmetric dual-SPI and (b) the radiation power density per radian. The TQ approximately happens at $t=4.2ms$.}
\label{fig:21}
\end{figure*}

Last, we examine the most important question: whether or not a good mitigation of the toroidal radiation peaking factor can still be maintained if there is no perfect symmetry between the injectors. This good mitigation is indeed maintained even with asymmetric dual-SPI, as can be seen in Fig.\,\ref{fig:21}. Before the TQ, there exists some toroidal peaking according to the injection asymmetry as is shown by the green and the blue curves. These two curves show similar maximum toroidal peaking amplitude with the ones in Fig.\,\ref{fig:18}(b). The latter exhibits strong symmetry between the two peaks, while in the asymmetric case, the $\gf=0$ peak, which corresponds to the injector which fired first, has a stronger peak than the $\gf=\gp$ one. Shortly after the time of the TQ at $t=4.24ms$ and $t=4.36ms$, however, the radiation peaking is mostly flattened, maintaining the same good behavior as the dual-SPI case. Especially, the radiation power density is already flattened when the total radiation power increased at the time of the TQ as is shown in Fig.\,\ref{fig:19}(b).

Naturally, the delay time between the two injectors must play an important role in how good this mitigation can be maintained, and in the case of a delay time comparable with the whole disruption mitigation timescale the radiation peaking would naturally approach that of the mono-SPI one. In this study, we merely conclude that good mitigation of toroidal peaking can be maintained with $\mathcal{O}\left(1ms\right)$ time delay, and leave more detailed analysis on this regard to future works.

\section{Summary \& Conclusion}
\label{s:Conclusion}

In this study we numerically investigated the MHD response and radiation asymmetry after full or mixed impurity SPI into JET or ITER plasmas by 3D nonlinear simulations. Our primary findings include a more complete picture of the MHD destabilization mechanisms and impurity transport in the strong cooling scenario, different TQ triggering surfaces depending on the plasma $q$ profile, insights into the importance of utilizing the favorable mode structure for better injection penetration, a demonstration of the temperature deviation between electron and ion species during the TQ, a detailed comparison between the radiation asymmetry characteristics for the mono-SPI and its mitigation for the dual-SPI cases, and lastly the correspondence between the MHD response symmetry and the injection synchronization for dual-SPI.

The rational surface upon which the core temperature collapse is triggered depends on the shape of the $q$ profile as would be expected. For our ITER cases with a large $q=1$ surface, flat $q$ profile in the core, but strong magnetic shear in the edge, the mode structure and island width are limited for the $2/1$ or $3/1$ mode, making it hard for them to nonlinearly couple with the core modes such as the $3/2$, thus the TQ is only triggered when the fragments enter the $q=1$ surface, destabilizing the $1/1$ kink. For the JET case with almost no $q=1$ surface and a lower shear near the $q=2$ surface, on the other hand, the $2/1$ mode is able to nonlinearly couple with the $3/2$ mode in the core thus triggering the TQ when the fragments arrive on the $q=2$ surface. The density transport mechanism also differs for the two cases. For the case with significant $q=1$ surface, the $1/1$ perpendicular convection is dominant. For the case without $q=1$ surface, convection along stochastic field lines also plays a significant role as the perpendicular transport from the $m>1$ mode is less efficient compared with the $1/1$ kink.
For both cases, the electron temperature profile show fast diffusive flattening at the time of the TQ due to conduction along the stochastic field lines.

For full argon SPI into a JET L-mode plasma which is compared with a deuterium SPI into another JET L-mode plasma with similar thermal energy content, our simulation confirms the argument that, for a strongly cooled scenario like the full argon one, the axisymmetric current contraction plays a significant role in the MHD excitation, while in the deuterium injection scenario, almost only the helical cooling effect on low order rational surfaces is important. The impurity density distribution and the radiation power density show distinctive helical feature with helicity corresponding to the TQ-triggering surface, similar to recent JET SPI experimental observations. The full argon SPI is also found to trigger the TQ when the fragments arrive on the $q=2$ surface, far away from the axis, resulting in a time delay between the core temperature loss and density increase, which may be unfavorable for runaway electron suppression, incentivizing more advanced approaches to efficiently achieve both the TQ and the CQ mitigation \cite{Nardon2020NF}. The evolution of the field line stochasticity can partly explain the aforementioned deviation in the temperature and the density response in the core. At the onset of the TQ, it is found that although global field line stochasticity exists, it still takes many turns to travel between the core and the edge region along the field lines. So that while the quick parallel electron thermal conduction is capable of collapsing the core temperature, the much slower parallel convection struggles to transport injected materials from the edge into the core. Until later on, as the stochasticity grows and the core becomes easily accessible from the edge, we see core density increase begin to manifest, and the previously hollowed density profile is flattened. Another interesting feature is that a slight change of origin position could result in remarkable better core accessibility as is shown by comparing Fig.\,\ref{fig:10}(d) against (e) and (f). Such behavior could be related to the perturbed flux mode structure of the dominant mode, as the anomalous diffusion within a sufficiently stochastic field is proportional to the summation of normalized magnetic perturbation squared \cite{Rechester1978PRL}. Thus, injection deposition on the large mode amplitude region would enjoy stronger core penetration along the stochastic field line.

Our ITER simulation with neon/deuterium mixed SPI shows efficient core penetration of injected materials can be achieved by taking advantage of favorable mode structures. Concretely, ablation deposition on and within the $q=1$ surface can result in strong convective transport into the very core of the plasma (prior to the TQ onset) due to $1/1$ kink's broad mode structure within the $q=1$ surface. In case of deposition outside the $q=1$ surface, such a convection into the core is not observed due to the smallness of the displacement mode structure. Even if the plasma does not have a $q=1$ surface to begin with, the above philosophy could still apply. First of all, in the presence of strong edge cooling, current induction in the core could push the safety factor below one if it is not far above unity initially \cite{Hoelzl2020POP}. Second, within the resonant surface, the $m>2$ displacement will also result in perpendicular convection away from the O-point towards the core, thus transporting the injected material inward which would then spread along the stochastic field lines, contributing to the core mixing. This would not be as efficient as the $1/1$ convection due to the decreasing displacement amplitude towards the axis, however. Last, as mentioned in the previous paragraph, the anomalous diffusion coefficient itself is dependent on the perturbed flux mode structure, thus deposition on the strong magnetic perturbation region would result in easier stochastic transport.

Furthermore, our two temperature model demonstrates that significant temperature deviation between electrons and ions can exist during the TQ.
Hence, in a single temperature treatment, the ions would artificially heat up the electrons compared with the two temperature case. Since the pellet ablation is dominated by the hot electrons and the ion contribution is small \cite{Sergeev2006PPR}, this would mean that the single temperature treatment could result in an artificially higher ablation rate compared with that from the two temperature one.
The impurity radiation would also be affected.

More importantly, even with artificially elongated impurity deposition, we find remarkable radiation asymmetry both in poloidal and toroidal direction, which threatens to undermine the TQ mitigation effort. Such asymmetry is found to be strongest at the time of the TQ when the radiation power is strongest and gradually relaxes over the course of the TQ. Moreover, the radiation structure at the time of the TQ shows distinctive helicity of the rational surface upon which the TQ is triggered, in agreement with both experimental \cite{Sweeney2020NF} and previous massive material injection simulation \cite{Izzo2013POP,Kim2019POP} results. The exact TQ-triggering rational surface differs depending on the q profile as we discussed above. In a further note, the SPI toroidal radiation peak location in our study as well as that of Ref.\,\cite{Kim2019POP} are in contrast with that of the MGI simulation done in Ref.\,\cite{Izzo2013POP}. Such difference is caused by a combination of unrelaxed impurity density profile and asymmetric outward heat flux pattern. The MGI case in Ref.\,\cite{Izzo2013POP} shows a much more toroidally uniform impurity distribution while retaining the poloidal asymmetry compared with the SPI ones, thus when the heat flux emerges from the X-point of the $1/1$ mode after the TQ, the radiation peak occurs at $\gp$ away from the injection location where both the heat flux and the impurity density are large. On the other hand, for the SPI case, the impurity density is much more toroidally localized due to increased local ablation source at the time of the TQ, thus the radiation peak occurs closer to the injection location toroidally. Nevertheless, injection from two toroidally opposite directions is found to effectively mitigate the aforementioned toroidal radiation peaking, even with imperfect timing between the injectors.
Another noteworthy feature of the dual-SPI is that they generally show larger radiation strength compared with the mono-SPI case. This implies that the impurity mixture ratio or total injection quantity have to be adjusted accordingly if one wishes to maintain the same post-SPI electron temperature with multiple SPIs.

Lastly, the MHD response is found to exhibit the same symmetry with that of the injection configuration, especially before the TQ and at the time of TQ onset. Even modes are found to dominate the pre-TQ MHD spectrum for symmetric dual-SPI, and the plasma shows no $1/1$ kink but instead a $m=2$ deformation as the core temperature collapses. On the other hand, imperfect synchronization between the SPIs causes dominant odd mode response early in the injection, much like the behavior of the mono-SPI case, although even modes also grow to comparable amplitude for asymmetric dual-SPI case later on towards the onset of the TQ. The plasma shows a significant $1/1$ motion at the time of the TQ, suggesting the core collapse is dominated by the $n=1$ mode as a result of imperfect synchronization, in contrast with the situation of the perfect one.
The comparison between the two dual-SPI cases and a mono-SPI case show that the relative strength of the $n=1$ mode gradually grow stronger as the time delay between the injections increases.

The above investigations provide insights into the characteristic MHD response and corresponding transport processes as a result of impurity SPIs. Above all, they improve the understanding for the mitigation of radiation asymmetry by multiple simultaneous SPIs, as well as the characteristic MHD behavior under such scenarios. Realistic impurity treatment considering a non-equilibrium model is currently under development and is of high priority for upcoming JOREK work, the analysis of line radiation photon mean-free-path based on this non-equilibrium is also planned. Furthermore, the proper treatment of the long tail hot electrons and their contribution to the ablation rate when conducting SPI into a high initial temperature H-mode plasma will also be pursued in the near future.
Last but not least, it be possible to use the small scale calculation of the expansion of the ablation cloud \cite{Aleynikov2019JPP} to provide a more accurate shape for the toroidally relaxed density and momentum source for our macroscopic simulation, this is also part of our future works.

\vskip1em
\centerline{\bf Acknowledgments}
\vskip1em

  The authors thank L. Baylor, P. Parks, L.E. Zakharov, R. Sweeney, D. Bonfiglio, B.C. Lyons and C.C. Kim for fruitful discussion. ITER is the Nuclear Facility INB no. 174. The views and opinions expressed herein do not necessarily reflect those of the ITER Organization. This publication is provided for scientific purposes only. Its contents should not be considered as commitments from the ITER Organization as a nuclear operator in the frame of the licensing process. Part of this work is supported by the National Natural Science Foundation of China under Grant No. 11905004. Part of this work has been carried out within the framework of the EUROfusion Consortium and has received funding from the Euratom research and training program 2014-2018 and 2019-2020 under grant agreement No 633053. The views and opinions expressed herein do not necessarily reflect those of the European Commission. This work is carried out partly on the supercomputer MARCONI operated by Cineca, and also partly on Tianhe-3 prototype operated by NSCC-TJ.

\end{document}